\documentclass[12pt,preprint]{aastex}

\received{}
\accepted{}

\slugcomment{}
\shorttitle{Is the Sun Lighter than the Earth?}
\shortauthors{T.R.\ Ayres et al.}

\begin{document}

\title{Is the Sun Lighter than the Earth?\\ Isotopic CO in the
Photosphere, Viewed through \\ the Lens of 3D Spectrum Synthesis}

\author{Thomas R.\ Ayres}
\affil{Center for Astrophysics and Space Astronomy,\\
University of Colorado,
Boulder, CO 80309; email: Thomas.Ayres@Colorado.edu}

\author{J.\ R.\ Lyons}
\affil{Department of Earth and Space Sciences, University of California, Los Angeles, CA}

\author{H.-G.\ Ludwig and E.\ Caffau}
\affil{Zentrum f\"ur Astronomie der Universit\"at Heidelberg, Heidelberg, Germany}

\author{S.\ Wedemeyer-B\"ohm}
\affil{Institute of Theoretical Astrophysics, University of Oslo, Oslo, Norway}

\begin{abstract}
We consider the formation of solar infrared (2--6 $\mu$m) rovibrational bands of carbon monoxide (CO) in CO5BOLD 3D convection models, with the aim to refine abundances of the heavy isotopes of carbon ($^{13}$C) and oxygen ($^{18}$O, $^{17}$O), to compare with direct capture measurements of solar wind light ions by the {\em Genesis}\/ Discovery Mission.  We find that previous, mainly 1D, analyses were systematically biased toward lower isotopic ratios (e.g., $R_{23}\equiv ^{12}$C/$^{13}$C), suggesting an isotopically ``heavy'' Sun contrary to accepted fractionation processes thought to have operated in the primitive solar nebula.  The new 3D ratios for $^{13}$C and $^{18}$O are: $R_{23}= 91.4{\pm}1.3$ ($R_{\oplus}= 89.2$); and  $R_{68}= 511{\pm}10$ ($R_{\oplus}= 499$), where the uncertainties are 1\,$\sigma$ and ``optimistic.''  We also obtained $R_{67}= 2738{\pm}118$ ($R_{\oplus}= 2632$), but we caution that the observed $^{12}$C$^{17}$O features are extremely weak. The new solar ratios for the oxygen isotopes fall between the terrestrial values and those reported by {\em Genesis}\/ ($R_{68}= 530$, $R_{67}= 2798$), although including both within 2\,$\sigma$ error flags, and go in the direction favoring recent theories for the oxygen isotope composition of Ca--Al inclusions (CAI) in primitive meteorites.  While not a major focus of this work, we derive an oxygen abundance, $\epsilon_{\rm O}\sim 603 {\pm}9$ ppm (relative to hydrogen; $\log{\epsilon}\sim 8.78$ on the ${\rm H}= 12$ scale).  That the Sun likely is lighter than the Earth, isotopically speaking, removes the necessity to invoke exotic fractionation processes during the early construction of the inner solar system.
\end{abstract}


\section{Introduction}

Isotopic ratios of the abundant light elements, especially carbon and oxygen, chart the history of
galactic chemical evolution (Langer \& Penzias 1993), and closer to home provide insight into
fractionation processes that shaped the primitive solar nebula
(Krot et al.\ 2005).  Geochemical mixing models that consider only bulk meteorites and terrestrial planets suggest that (inverse) isotopic 
ratios of the Sun --- expressed as, for example, $^{16}$O/$^{18}$O --- should be only
a few tenths of a percent higher than terrestrial standard values\footnote{Based on Vienna Peedee Belemnite [V-PDB] for $^{13}$C and the Vienna Standard Mean Ocean Water [V-SMOW] mixture for the oxygen isotopes: Gonfiantini, Stichler, \& Rozanski 1995;
$^{12}$C/$^{13}$C=\,$89.2{\pm}0.2$ (ibid., Table~3), $^{16}$O/$^{18}$O=\,$498.7{\pm}0.1$ (ibid., Table~1), and
$^{16}$O/$^{17}$O=\,$2632{\pm}5$ (ibid., Table~1).}  
(Wiens, Burnett, \& Huss 1997), representing an isotopic {\em deficit}\/ when expressed
per mil\footnote{In the
solar system context, isotopic
abundances conventionally are reported as differences with respect to the
standard terrestrial values in 
parts per thousand ($^{\circ}\!\!/\!_{\circ\circ}$)
according to, for example, $\delta{^{13}\mbox{C}_{\rm X}}\equiv [R_{23}^{\oplus}/
R_{23}^{\rm X})\,-\,1]{\times}10^{3}$, where $R_{23}\equiv \epsilon(^{12}{\rm C})/\epsilon(^{13}{\rm C})$ is
the {\em inverse}\/ isotopic ratio commonly quoted in solar work ($\epsilon$, for the Sun, is the abundance relative to hydrogen in parts per million [ppm]).}.  Lunar and Martian isotopic ratios are very close to Earth's; asteroids traced from
meteoritic compositions span a wider range, but still deviate from terrestrial
by less than 20\,$^{\circ}\!\!/\!_{\circ\circ}$ (Burnett et al.\ 2003); while comets have a mean $^{13}$C deficit of $\sim -45$\,$^{\circ}\!\!/\!_{\circ\circ}$ (Woods 2009).  To be sure, some
extraterrestrial materials show even larger isotopic
deficiencies, reaching $-$60\,$^{\circ}\!\!/\!_{\circ\circ}$ for the oxygen isotopes in 
Ca--Al inclusions (CAI) of
chondritic meteorites in the least altered CAIs (Wiens et al.\ 2004); while some nebular metal grains show isotopic enrichments in the other direction,
up to +180\,$^{\circ}\!\!/\!_{\circ\circ}$ in the primitive chondrite Acfer~094 (Sakamoto et al.\ 2007).  

One interpretation of the chondritic inclusions led to the prediction that the Sun should be similar to the isotopically lightest CAIs (Clayton 2002). Compositions of solar wind ions implanted in lunar metal grains covered the range from $^{16}$O enriched (rare isotopes depleted) and similar to the lightest CAIs (Hahsizume \& Chaussidon 2005), to $^{16}$O depleted (rare isotopes enhanced), more like atmospheric ozone (Ireland et al.\ 2006). NASA's {\em Genesis}\/ Discovery Mission resolved the conundrum by measuring solar wind isotopes to extremely high accuracy through direct capture of the light ions (Burnett et al. 2003).  Analysis 
of {\em Genesis}\/ collection plates found $\delta{^{18}\mbox{O}}= -102.3{\pm}3.3$\,$^{\circ}\!\!/\!_{\circ\circ}$ and $\delta{^{17}\mbox{O}}= -80.8{\pm}5.0$\,$^{\circ}\!\!/\!_{\circ\circ}$, which implied photospheric values of $-59$\,$^{\circ}\!\!/\!_{\circ\circ}$ for both isotopes, after accounting for fractionation due to inefficient Coulomb drag during solar wind particle acceleration (McKeegan et al.\ 2011).	

The {\em Genesis}\/ photospheric values are consistent with the isotopically lightest CAIs, in line with Clayton's prediction, and imply that a mass-independent fractionation (MIF) process occurred between the Sun and terrestrial planets during solar system formation.  Proposed mechanisms for the MIF include CO self-shielding in the solar nebula (Clayton 2002; Lyons \& Young 2005); CO self-shielding in the molecular birth cloud (Yurimoto \& Kuramoto 2004; Lee, Bergin, \& Lyons 2008); galactic chemical evolution of interstellar gas and dust with differing time scales, in this case with the dust younger than the gas (Krot, Nagashima, \& Ciesla 2010); and a chemical MIF accompanying O$_{3}$ formation in surface reactions on dust grains, which then is imparted to molecular cloud water (Dominguez 2010).  Large oxygen isotope effects in CO self-shielding have been measured in at least one molecular cloud (Sheffer, Lambert, \& Federman 2002), and in a proto-planetary disk as well (Smith et al.\ 2009), but with large enough error bars to preclude confirmation of a slope similar to the CAI mixing line.

Spectroscopic determinations of photospheric isotopic abundances offer a different, somewhat contrary, perspective.  An early study by Hall, Noyes, \& Ayres (1972) of the 2.3~$\mu$m $\Delta v=2$ overtone rovibrational bands of solar carbon monoxide at the McMath-Pierce telescope on Kitt Peak derived
a $^{12}$C/$^{13}$C ratio of 90 with an uncertainty of 15\% (at 95\% confidence level), consistent with the terrestrial value of 89.  The measurements were made in a sunspot, where thanks to the cooler temperatures,
the isotopomer transitions attained far greater strength than in the warm photosphere.  Overtone transitions of $^{12}$C$^{18}$O and $^{12}$C$^{17}$O were not discernible in the 2.3~$\mu$m
umbral spectrum, however.
A follow-up study by Hall (1973) extended the isotopic measurements
to the 4.6~$\mu$m $\Delta v=1$ fundamental bands, in the undisturbed photosphere as well as sunspots,
using a newly commissioned infrared grating spectrograph at the McMath-Pierce.  He derived a slightly lower $^{12}$C/$^{13}$C ratio of 84${\pm}$8, but still consistent with terrestrial.   
Hall also estimated values for $^{18}$O and $^{17}$O, again obtaining essentially terrestrial ratios, although with large error bars.

While the CO isotopomer bands are strongly enhanced in sunspots, so too are those of other molecules, leading to complex
spectra with substantial line crowding.  For this reason, photospheric observations are preferred for the isotope problem, because
the warmer temperatures favor durable CO over its more weakly bound cousins like OH and SiO, leading to cleaner, less crowded spectra.  Even so, from the
ground there are only a few usable windows in the infrared for the CO measurements, owing to heavy molecular blanketing by the Earth's atmosphere.
 
Addressing both these issues, Harris, Lambert, \& Goldman (1987) recorded the photospheric
CO fundamental bands with a Fourier transform spectrometer (FTS) during a
high altitude balloon flight.  Adopting the
Holweger--M\"{u}ller (1974) 1D reference temperature stratification (very similar to
contemporary 1D semi-empirical models), 
they derived $^{12}$C/$^{13}$C=~$84{\pm}5$ and
$^{16}$O/$^{18}$O=~$440{\pm}50$.  The $^{12}$C$^{17}$O
spectrum was too weak in their scans to measure.  
Again, the observed solar ratios were slightly lower than terrestrial, but
within the cited uncertainties.  

More recently, Ayres, Plymate, \& Keller (2006; hereafter APK) analyzed an extensive, high quality
record of the solar CO fundamental and first-overtone bands from the Shuttle-borne ATMOS FTS, supplemented
with ground-based disk center and limb scans obtained with the large 1~m FTS at the McMath-Pierce.  The
authors synthesized CO line profiles, and center-to-limb behavior, utilizing a variety of 1D photospheric reference models, and multicomponent variants intended to simulate at least the thermal impact of the
convective fluctuations that characterize the inhomogeneous solar plasma.  Their
recommended values of the isotopic ratios were: $^{12}$C/$^{13}$C=\,$80{\pm}1$, 
$^{16}$O/$^{17}$O=\,$1700{\pm}220$, and 
$^{16}$O/$^{18}$O=\,$440{\pm}6$, where the cited small uncertainties represented standard errors of the mean (1~s.e.)
over the large samples of parent ($^{12}$C$^{16}$O) and isotopic CO transitions
considered, but exclusive of possible systematic errors.  These ratios,
taken at face value, were significantly lower than terrestrial.  At the same time, a pioneering study by Scott et
al.\ (2006; hereafter SAGS) of solar CO and the associated isotopomers, again based on the ATMOS material but now using prototype 3D convection models, proposed isotopic ratios
$^{12}$C/$^{13}$C=\,$87{\pm}4$ and $^{16}$O/$^{18}$O=\,$480{\pm}30$; higher than APK, slightly lower than terrestrial, but
agreeing with both within (2\,$\sigma$) uncertainties.  

Notably, while the recent
studies found solar $^{12}$C/$^{13}$C ratios generally consistent with terrestrial,
when aggregated together, they all were systematically lower, rather
than, say, scattering uniformly around the standard value.  
Unfortunately, the uncertainties still were too large to distinguish 
between an isotopically ``heavy'' versus ``light'' Sun, especially for $^{18}$O and $^{17}$O, which as described earlier display a 6\% deficit relative to ocean water according to {\em Genesis.}\/  A more precise measurement of photospheric $^{12}$C/$^{13}$C, as well, might help to distinguish between the several proposed MIF mechanisms. For example, CO self-shielding should initially impart a large depletion of $^{13}$C$^{16}$O in the solar nebula, yet that would be removed at least partially, and possible entirely, by subsequent ion-molecule charge exchange reactions between C$^{+}$ and CO.  The potentially broad range of depletions has yet to be meaningfully constrained by model calculations, so an accurate measurement of the photospheric ratio could provide some needed guidance.

Here we reconsider the solar CO isotopic analysis, in light of a more recent generation of solar 3D
models, which now meet key observational tests that the initial complement of 
convection simulations was less successful matching.  We also have taken special precautions, on the one hand to control 
and narrow the random (mainly observational) uncertainties of
the analysis; and on the other to quantify a variety of systematic effects related to the models, and equally so the laboratory
molecular line parameters.  We stress that we are attempting to perform precision ``forensic'' spectroscopy of the solar plasma with
uncertainties ideally below 1\%, in order to compare to the very precise {\em Genesis}\/ findings
(with $\sim$1\% quoted errors).  To preview our conclusions,
that goal was frustrated by lingering uncertainties in the atomic physics ($f$-value scales) and more subtle aspects of
the modelization, even though state-of-the-art 3D convective snapshots were utilized.  Nevertheless, we
also demonstrate that especially for $^{13}$C, but $^{18}$O as well, the observational uncertainties can be
controlled to the desired level or better, so that in principle future comparisons of higher precision can be
carried out, once the external atomic physics and modelization issues are resolved.  We also confirm that the 1D spectrum synthesis approach is essentially useless for this particular molecular problem, although to be sure there are other less pathological cases where a careful 1D analysis can produce similar results to a full 3D study (see, e.g., Ayres 2008).

\section{Observations}

There is a clear advantage of infrared molecular spectra for tracing photospheric isotopic abundances, as noted in Hall, Noyes, \& Ayres (1972):
the isotopomer and parent ($^{12}$C$^{16}$O) rovibrational transitions are well separated in frequency, $\omega$ (in wavenumbers
[cm$^{-1}$]), owing to the
large influence of the different molecular weights on the rotational and vibrational properties of the
respective diatoms.  Isotopic shifts encountered in the electronic spectra of atoms and ions are minuscule in
comparison, and usually much smaller than the thermal line width at photospheric temperatures, rendering isolation 
of the isotopic component rather tricky, if possible at all.

At the same time, obstacles encountered in an isotopic analysis of solar molecular features
are myriad.  First, the heavy isotopes of C and O have very low abundances
and consequently their absorption lines tend to be weak.  These will be more
influenced by unrecognized blends, especially in telluric contaminated ground-based spectra, than
their much stronger $^{12}$C$^{16}$O counterparts.  Second, the strongest isotopomers are fundamental transitions at
the peak of the rotational distributions ($J_{\rm low}\sim 30$), but the equivalent
parent $^{12}$C$^{16}$O lines are too strong to
be useful in an abundance study (Hall 1973).  Consequently, the isotopic lines must be compared to 
intrinsically weak parent lines --- high excitation transitions in the case of $\Delta v=1$ --- thereby creating the possibility of biases in the highly temperature sensitive analysis.  Finally, historically there have been significant disagreements among proposed oscillator strengths for the ``hot'' CO bands (see, e.g., APK).

To be sure, many of these challenges have straightforward solutions.  For example, the dominant cause of blending in
the CO bands are other CO lines, whose frequencies are precisely known and whose relative strengths can be estimated
by spectrum synthesis.  Thus, minimally contaminated features can be readily identified.  Further, in
an isotopic analysis, the absolute accuracy of the oscillator strength scale is not as important as the
relative precision (e.g., $f_{36}/f_{26}$, where the subscript is the isotopomer
designation: $26\equiv ^{12}$C$^{16}$O, and so forth) over the rotational ladder, given that the relative scalings for the isotopomer $f$-values depend on
ostensibly simple atomic physics considerations, and should be calculated consistently even when the absolute parent strengths might be
in error.  Additionally,
one can consider features covering a range of lower level excitation energies, $E_{\rm low}$, to 
help identify temperature-dependent
trends, which might arise from systematic deviations of the oscillator strengths with $v_{\rm low}J_{\rm low}$,
which is correlated with $E_{\rm low}$, or 
from the solar model itself.  Even so, the analysis probably would not be possible if
it were not for the existence of very high-quality FTS scans of the average disk center solar
spectrum, free of telluric contamination and in the relevant portions of the infrared, from the space-based ATMOS FTS
experiment.

  \subsection{ATMOS}

As noted in APK, Shuttle-borne ATMOS was a high-resolution
($\omega/\Delta\omega\sim 150,000$) FTS
instrument designed to study trace molecular
species in the Earth's atmosphere backlighted
by the rising or setting Sun as viewed from low-Earth orbit.  Solar reference spectra,
obtained from zenith pointings, were free of terrestrial contamination.  The instrument recorded
a $\sim 0.1\,D_{\odot}$ diameter circular region at disk center, which 
corresponds to $\mu= 1$ for all intents
and purposes ($\mu\equiv \cos{\theta}$, where $\theta$ is the heliocentric angle).  The best results
were obtained from the final ATMOS flight in 1994 November as part of the ATLAS-3 payload
(Abrams et al.\ 1999).  The signal-to-noise of those data is extremely high, better than
$10^3$; well suited for a study like the present one, which must dig down to the very weak 
isotopic features.  Based on clean ATMOS CO lines in common to even higher-resolution McMath-Pierce FTS measurements,
we adopted a 2~km s$^{-1}$ FWHM Gaussian approximation to the ATMOS instrumental profile to smooth
synthetic CO spectra for comparison with observed line shapes.  Details concerning the reduction of the dearchived ATMOS scans can be found in the
earlier paper (APK).

  \subsection{Hybridization of ATMOS Spectral Features:\\ $\mathbf{^{12}}$C$\mathbf{^{16}}$O $\mathbf{\Delta v=1}$ and $\mathbf{\Delta v=2}$, and the
$\mathbf{\Delta v=1}$ Isotopomers}

As alluded earlier, the solar CO bands contain large numbers (literally thousands) of absorption lines,
many of which, especially $^{12}$C$^{16}$O, are minimally contaminated by extraneous features.  Further,
because of the modest, smooth change of $gf$-value and excitation energy along the rotational
ladders of each vibrational band, lines of similar rotational
numbers in the same band
are very similar to one another in absorption strength: spectral clones if you will.  Having a large sample of target
transitions potentially is a great benefit, especially to help isolate systematics.  In contrast,
abundance studies based on only a few accessible features --- like photospheric atomic 
oxygen --- become more
susceptible to accidental blends, distortions in the local continuum level, and NLTE excitation (e.g., Caffau et al.\ 2008).  Synthesizing absorption
profiles of hundreds of CO lines is a trivial matter in classical 1D spectral analysis, but is less attractive
for 3D work, where each temporal snapshot is equivalent to tens of thousands of 1D problems, and a dozen or more snapshots
well separated in time should be considered to ensure an adequate average.

We worked around the tension of wanting to consider as many CO features as possible, but at the same time
minimizing the number of transitions to synthesize in 3D, by combining groups of CO lines of similar excitation,
wavenumber range, and absorption depth together to yield ``hybrid'' features.  In this way, we could distill the
properties, both observational and laboratory, of perhaps a 150 separate absorptions into a smaller,
more tractable sample of a few dozen or so, with increased S/N and better defined continuum levels.  For this study, we restricted 
lower level energies to below 18,000 cm$^{-1}$, which characterizes the majority of the measurable 
isotopomer features.  The goal then was to develop a comparison sample of fundamental and overtone parent lines in this energy
interval, as well as the isotopic sample itself.  Initial line shape synthesis tests suggested that beyond an equivalent width,
$W_{\omega}$, of about
5~mKy (10$^{-3}$ cm$^{-1}$), a typical CO $\Delta v = 1$ line begins to saturate, and thus becomes less responsive for an
abundance analysis.   The saturation limit for $\Delta v = 2$ lines is about twice as high.
We thus also applied these respective $W_{\omega}$ cutoffs as part of our line selection criteria.

In practice, the $E_{\rm low}$ and $W_{\omega}$ cuts
severely limited the number of suitable $\Delta v=1$ features that could
be ``hybridized;'' many more $\Delta v=2$ lines were available, however.  This is an important issue because all of the
best (i.e., highest S/N) isotopomer features are of the $\Delta v=1$ type.  Ideally, parent lines
and the isotopomer counterparts should be compared on as close to the same basis as possible, to avoid potential
systematic errors that might creep into the analysis by, say, pitting $\Delta v=1$ isotopomers against $\Delta v=2$
parent lines (the overtone oscillator strengths are smaller by a factor of $\sim$100).  Because of
the relative scarcity of suitable $\Delta v=1$ target lines --- parents typically too strong, isotopomers typically too weak --- we were forced to collect them without paying too much
attention to the $v_{\rm low}J_{\rm low}$ combinations, but rather more closely to $E_{\rm low}$ (the fundamental
quantity that controls the number of absorbers through the Boltzmann factor).  

We identified suitable lines for hybridization by considering visualizations of all the CO features that
satisfied the $E_{\rm low}$ and $W_{\omega}$ cutoffs, and did not have another $^{12}$C$^{16}$O transition with any significant strength whose line center fell within
$\sim{\pm}10$ km$^{-1}$ of the target line (i.e., not close enough to disrupt the candidate line,
given the typical $\sim$4.4~km s$^{-1}$ line full width at half maximum intensity [FWHM]).  The visualizations
compared the observed line shape to simulations from the FAL-C 1D model (Fontenla, Avrett, \& Loeser 1993),
with abundances adjusted (as a function of $E_{\rm low}$, as described later) to achieve reasonable matches to known clean $^{12}$C$^{16}$O lines, and separated according
to isotopomer.  In this way, blends of $^{12}$C$^{16}$O target lines with, say, weak coincidental isotopic features
could be identified.  In order to increase the number of potential hybridization candidates, lines were included that
were affected in their far wings (beyond $\sim{\pm} 10$~km s$^{-1}$) by blends, as long as the nearest edge of the blend terminated
short of the target absorption, and as long as the local continuum level appeared to be adequately defined in at least one of the wings.  The affected parts of the target profile were ignored when the candidates subsequently were coadded.  A list of potentially suitable features was compiled in this way.  

That
list then was parsed into groups of transitions nearby in wavenumber, close in $E_{\rm low}$, and similar in 
absorption depth.
The hybridization candidates in each group were aligned by Gaussian centroiding, interpolated onto a finely sampled 
velocity scale (0.5~km s$^{-1}$ steps), and
coadded, evenly weighted, ignoring the portion(s) of the line profile previously flagged as
corrupted (trimming was limited to velocities outside the ${\pm}5$~km s$^{-1}$span of the line core).   

At the same time, the parallel collection of individual molecular parameters also was ``hybridized.''  The approach was to average the line strengths ($gf$-values) and
the temperature- and wavenumber-dependent components of the absorption --- stimulated emission and Boltzmann excitation factors --- for
a set of discrete temperatures between 4000~K and 5000~K, then fit the resulting average relation with an optimal
combination of $gf$-value and $E_{\rm low}$ (after setting the transition frequency to a weighted mean
of the group).  Note that the $E_{\rm low}$ 
for the R- and P-branch transitions of the same $J_{\rm low}$ are the same, as is the
statistical weight $g= (2\,J_{\rm low}\,+\,1)$, but in general the oscillator strengths will differ, favoring the R-branch, and the
difference increases up the rotational ladder (as illustrated later, in Fig.~1a).  The molecular parameters were hybridized for two independent oscillator strength scales, as decribed next,
and for a third scale --- $\langle{f}\rangle$ --- which was a straight average of the two.

  \subsubsection{Oscillator Strengths}

We considered two independent oscillator strength scales for the study: 
Goorvitch (1994: G94), based on experimentally derived transition matrix elements; and Hure \& Roueff (1996: HR96), who utilized theoretical electric dipole moment functions.  Figure 1a illustrates the 
dependence of $\Delta v=1$ and $\Delta v=2$ oscillator strengths separated by
vibrational band, as a function of lower rotational level, $J_{\rm low}$, as derived from the HR96 dipole matrix elements, and for
a version of the G94 $f$--values taken from the HITEMP section of the HITRAN database\footnote{see http://www.cfa.harvard.edu/hitran/}.  

We chose the HITRAN CO line strengths, which were derived from the G94 matrix elements, because 
the HITRAN values are given to higher precision than the original G94 electronic tabulations.  However, a detailed comparison
of HITRAN to the G94 scale showed that the latter was systematically very slightly higher (by 0.25\%), independent of
$J_{\rm low}$, for both
fundamental and first overtone; except for
the 2--0 band\footnote{The vibrational band designation is $v_{\rm up}$--$v_{\rm low}$, where the leading value is
the upper vibrational level, the trailing is the lower, and $\Delta v = $1,\,2 for the fundamental and first overtone, 
respectively.   Note, also, that the rotational lines in
a given fundamental or overtone band can be in one of two branches: ``P'' if the upper state $J_{\rm up}$ is one
lower than the initial state $J_{\rm low}$ ($\Delta J = -1$); and ``R'' if the upper state is one higher ($\Delta J = +1$).} which showed a much larger shift ($\sim 5$\%), with a strong increase for higher $J_{\rm low}$ in the
P branch, but not the R branch.  The small systematic offset undoubtedly can be traced to a slightly different numerical value for the
combined physical constants that appear in the conversions from matrix elements to Einstein $A$--values and ultimately
$f$-values; but the larger
$J_{\rm low}$-dependent offset for the 2--0 band likely signals a more significant numerical issue.  Ratios of oscillator strengths
between neighboring bands (e.g., ${[3-1]}\,/\,{[2-0]}$) at the same $J_{\rm low}$ show very systematic behavior. On this
basis, the HITRAN 2--0 band appears to be anomalous.  We thus corrected the HITRAN 2--0 band to the G94 scale, and
adjusted the rest of the HITRAN $\Delta v = 1$ and $\Delta v = 2$ values for the slight deficit mentioned earlier.  
  
Figure~1b compares ratios of the G94 to HR96 oscillator strengths on an expanded scale to illustrate the differences more clearly.  The
G94 $\Delta v=1$ $f$-values are systematically a few percent lower than those of HR96, but
independent of vibrational band or rotational number.  In contrast, the $\Delta v=2$ overtone bands tend to have larger oscillator strengths on the G94 scale,
at least for the lower vibrational bands, and
the deviations relative to HR96 vary strongly and systematically with both vibrational band and rotational number.  

Figure~1c compares $f$-values of
$^{13}$C$^{16}$O to $^{12}$C$^{16}$O on the G94 and HR96 scales separately.  The isotopomer strengths are 
smaller by about 4\%, essentially identical on the two scales, and show minimal dependence on $v_{\rm low}$ or $J_{\rm low}$.  A similar deficit is seen for the oxygen isotopes: $^{12}$C$^{17}$O is down by 2\% and $^{12}$C$^{18}$O by 4\%.  The consistent behavior between the two oscillator strength scales supports the
premise stated earlier that the relative precision of the isotopomer $f$-values should be better 
determined than the absolute parent values.

  \subsubsection{Equivalent Widths}

After the hybrid profiles were constructed, equivalent widths were measured.  First, a continuum level
was established by considering the intensities on either side of, 
but well away from, line center (in practice, 6~km s$^{-1}$ $\leq |{\upsilon}| \leq 10$ km s$^{-1}$).  An ``Olympic'' filter
(throwing out the two highest and two lowest values) was applied to the collection of (nine) points in each flanking continuum band
separately, and the larger of the medians of the two sets of surviving points was adopted as
the continuum level.  The dispersion of the normalized
points in the flanking reference continuum windows served as an empirical estimate of the photometric
noise, which later was utilized to assign an uncertainty to the equivalent width according to the
FTS noise model of Lenz \& Ayres (1992).  

The equivalent width, itself, was
measured by fitting a Gaussian profile to the normalized intensities within ${\pm}5$~km s$^{-1}$ of line
center (which now was located very close to zero velocity owing to the initial centroiding of the constituent
profiles).  Although the solar CO lines display a systematic convective blueshift of around
350~m s$^{-1}$, the profiles nevertheless are very close to Gaussian in shape (with a FWHM of around 4400~m s$^{-1}$ for the weak fundamental and overtone lines, but
C-bisector amplitudes of only a few hundred m s$^{-1}$ [SAGS]).  The Gaussian fitting approach was adopted here
(and in the previous work of APK) because as applied (later) to the synthesized profiles, it allows the
use of a sparsely sampled frequency grid.  This is advantageous in the computationally challenged 3D setting.

The initial sample of hybrid features was subjected to an abundance analysis in a representative 3D snapshot using the average oscillator strengths, to derive an $\epsilon_{\rm O}$ consistent with the observed equivalent width for each line separately (see below for details).  The resulting distributions of 
inferred $\epsilon_{\rm O}$ with both $E_{\rm low}$
and $W_{\omega}$ were examined, and any deviants from the expected smooth relations were culled out.  The offending
hybrids were vetted to see whether one of the constituent lines was anomalous with respect to the others (affected perhaps by
an unrecognized non-CO blend), and if eliminating the offending line could improve the behavior.  If not, the outlier hybrid was simply discarded.  Ultimately, through this
iterative, somewhat arbitrary, process a final sample of 36 hybrid parent lines was developed, 10 $\Delta v=1$
and 26 $\Delta v=2$, from about 150 input transitions.  Table~1 summarizes the constituent lines and final hybrid parameters (for the $\langle{f}\rangle$ scale).  Figure~2a illustrates the procedure for
four representative $\Delta v=1$ hybrids, and Figure~2b similarly for four $\Delta v=2$ cases.  Note in Table~1 that 
the hybrid designated 1J26 is the trivial case of a single line, because only one suitable $\Delta v=1$ could be found for the (key) energy range
below 7000~cm$^{-1}$.  Also, we caution that the Gaussian parameters ($W_{\omega}$, FWHM) listed for the individual contributing lines in a hybrid group should be taken only as a guide, especially for those features with distortions in one wing or the other: the continuum level in the pre-coadded profiles is less well defined than for the final hybrid owing to higher photometric noise and the possible influence of intensity deviations, due to blends, outside the line core.  This caution particularly extends to the isotopic sample described next.

The hybridization scheme was applied in a similar fashion to $^{13}$C$^{16}$O, $^{12}$C$^{18}$O,
and  $^{12}$C$^{17}$O (9, 4, and 4 hybrids, respectively, representing nearly 70 input transitions).  For the isotopomers, the Gaussian fitting imposed a fixed FWHM of
4.14~km~s$^{-1}$, as inferred from the average of the measured widths of the higher-S/N $^{13}$C$^{16}$O hybrids.  The apparent decrease with respect to the FWHM$\sim 4.4$~km s$^{-1}$ of the narrowest parent $\Delta v=1$ profiles is somewhat more than expected from
the influence of the larger molecular weights on the thermal broadening, but the decrease is replicated in the
3D line shapes of the isotopomers.
In addition,
the Gaussian centroiding step, prior to coadding, was omitted in
favor of accepting the laboratory line centers on the ATMOS frequency scale.  (An empirical zero-point calibration of the ATMOS velocity scale was determined for each isotopomer separately according to the mean shift of the
hybrid lines relative to 1D synthesized profiles, determined in an initial fitting step.)  Examples of the isotopomer hybrids are illustrated in Figures~2c ($^{13}$C$^{16}$O), 2d ($^{12}$C$^{18}$O) and 2e ($^{12}$C$^{17}$O).  The calculated hybrid line parameters are summarized
in Table~2, again for the reference $\langle{f}\rangle$ scale.

\section{Analysis}

In overview, our strategy was first to derive an oxygen abundance
for a given solar model using the $^{12}$C$^{16}$O ``parent'' hybrid line sample, then calculate the isotopic ratios
that reproduced the equivalent widths of the weak isotopomer absorptions for that oxygen abundance.
This was done in practice by synthesizing
the hybrid line equivalent widths for a small set (4--6) of discrete values [of the oxygen abundance, $\epsilon_{\rm O}$; or scale factors, $s_{\rm ISO}$, relative to reference isotopic ratios, $(R_{\rm ISO})_{\rm STD}$, where ${\rm ISO}= 23$, e.g., refers to
$^{12}$C/$^{13}$C] spanning the full range of potential sample variation, and then
interpolating with the observed $W_{\omega}$ to deduce the corresponding target value (of $\epsilon_{\rm O}$
or $s_{\rm ISO}$).  For convenience, we assumed
$\epsilon_{\rm C}= \onehalf\,\epsilon_{\rm O}$, so that the CO concentration effectively becomes quadratically dependent on
the oxygen abundance (see APK for the physical motivation behind this choice).  

The derived $\epsilon_{\rm O}$'s in the first step typically were found to
depend systematically 
on $E_{\rm low}$ (see SAGS; APK), and usually there was an
offset between the $\Delta v=1$ and $\Delta v=2$
hybrids as well.  The abundance gradients and fundamental/overtone shifts could be due to: (1) systematic errors in the
$f$-value relations as a function of $v_{\rm low}J_{\rm low}$ (which correlates with $E_{\rm low}$); or (2) small thermal differences between the 3D convection models and the real Sun
in the CO-absorbing layers (the middle photosphere, well above the continuum-forming zone: see APK and Fig.~3b, below).  That the 
{\em ab initio}\/
3D models might depart somewhat from the true Sun in the mid-altitudes of the photosphere would not be Earth-shattering, because
the underlying simulations typically are purely radiation-hydrodynamic, lacking the small-scale magnetic fields of
the true photosphere (Stein 2012), and usually also missing
high-frequency acoustic waves (with wavelengths not resolved by the grid spacing in the computational box), which are strongly damped in the outer photosphere (Ulmschneider 1974).  Both of
these neglected phenomena potentially could produce additional heating
there (e.g., Ayres 1975).

We thus also considered slightly perturbed models for which the thermal structure above the continuum-forming
layers was systematically raised by a small amount ($\lesssim 100$~K) so that the resulting mean temperature stratification
matched that of a semi-empirical 1D reference model.  
In these temperature-enhanced models, we usually found a significantly different 
slope of $\epsilon_{\rm O}$ against $E_{\rm low}$, and that the $\Delta v=1$ and $\Delta v=2$
samples displayed an opposite offset compared with the baseline.  
We then imposed an intermediate temperature correction chosen
specifically to force the $\Delta v=1$ and $\Delta v=2$ samples to agree.  In principle, this middling ``Goldilocks'' solution represents an atmospheric model whose thermal structure is in harmony
with respect to the
CO fundamental and overtone bands; but, as we show later, the specific temperature correction does depend on the
adopted oscillator strength scale.

We pursued this tripartite strategy for the G94 and HR96 CO oscillator strengths 
separately, to
test for potential systematic errors from that source.  When calculating the isotopomer hybrid samples, we
imposed the specific $\epsilon_{\rm O}(E_{\rm low})$ relation inferred for the $\Delta v=1$ parent hybrids (since
all the isotopic lines are of the $\Delta v=1$ type).  We strove to attain as much consistency at each level
of the analysis as possible, but also to explore as much of the potential parameter space as was permitted (by other
constraints) to judge by how much the final results might be influenced by the models and/or molecular parameters.  The individual
pieces of the strategy are described in more detail next.

  \subsection{CO5BOLD 3D Snapshots and Continuum Tests}

We adopted the so-called ``CO5BOLD'' radiation-hydrodynamic (RHD) scheme to
provide the baseline 3D models for the CO infrared spectral synthesis.  The methodology behind the CO5BOLD 
3D stellar convection simulation code has been described by Freytag et al.\ (2012), who have compared results from this and other contemporary 3D RHD simulations, with excellent agreement achieved for key tests such as visible continuum center-to-limb behavior (see, also, Beeck et al.\ 2012).  Wedemeyer-B\"ohm \& Rouppe van der Voort (2009) have examined photospheric intensity distribution functions in blue, green, and red
continuum bands, at disk center and near the limb, from {\em Hinode}\/ Solar Optical Telescope images (corrected
for point response and stray light), and concluded that the CO5BOLD 3D simulations well reproduced the
various observables that characterize the granular fluctuation patterns.  This agreement gives us
confidence for the infrared problem, since the 2.3~$\mu$m and 4.6~$\mu$m continua bracket the formation depths of
the visible radiation.

We utilized sixteen independent 3D snapshots, well separated in time, from a fully-relaxed CO5BOLD run.  This is a slightly reduced set of those employed by
Caffau et al.\ (2008) in their analysis of photospheric atomic oxygen: see that study for more modelization details.  
Figure~3a illustrates schematic
temperature and velocity maps for the snapshots collectively, for a constant height slice (median
$z$ of the $p=10^5$ dyne cm$^{-2}$ pressure surface) characteristic of the deep
photosphere where the visible continuum arises.  There are a dozen, or so, identifiable
granules in each snapshot, which have the familiar broad, warm upwelling centers, surrounded by narrower, cool subducting lanes. The
area asymmetry largely is responsible for the convective blueshifts seen in most photospheric lines.

Each snapshot has 40~km steps
in the two horizontal directions, covering a 5.6${\times}$5.6~Mm$^{2}$ patch of the surface, and 15~km grid steps in the vertical direction, with an extent of 2.3~Mm; altogether a resolution of
140$\times$140$\times$150.  The top boundary of the snapshots was very high ($< 10^{-6}$ in Rosseland optical depth), well above the CO-absorbing layers in the middle photosphere; and for convenience we truncated the vertical grid at a maximum
depth such that $\tau>> 1$ at any relevant frequency, thus ignoring the deepest layers that are important for the deep-seated granulation dynamics,
but not the CO spectrum synthesis confined to the surface layers.  

Twelve opacity bins were utilized for the radiation transport in the RHD simulations, a significant improvement over earlier generations of such models that were derived using only one or a few radiative bands.  Abundances were taken from Grevesse \& Sauval (1998), except for CNO where values closer to Asplund, Grevesse, \& Sauval (2005) were adopted.  The fact that the 3D time series was calculated with CNO abundances that we later show are inconsistent with what we find from CO is not vital, because: (1) the CNO abundances are not critical to the model pressure structure (the metals that provide the electrons for the H$^{-}$ opacity are far more important); and (2), we apply
a ``calibration'' adjustment to the model pressure scales to compensate for any slight such inconsistencies, as
outlined next.

As described in a previous 3D study (of [\ion{O}{1}] $\lambda$6300; Ayres 2008), an important practical consideration in
utilizing ``imported'' RHD models is that there might be slight inconsistencies between the opacities and
equation of state employed in the original simulation, and the (usually more detailed) ones embedded in a
{\em post facto}\/ spectrum synthesis code.  Such inconsistencies could lead, for example, to the prediction of slightly
different continuum intensities (say, in the visible) than were inherent in whatever effective temperature constraint
was imposed at the lower boundary of the 3D simulation.  We mitigated against such inconsistencies by requiring that the full 3D model (consisting of the 16 independent snapshots)
successfully predict absolute continuum intensities (see, e.g., APK) over the visible range 0.44--0.68~$\mu$m.  We forced such agreement by
slightly adjusting the pressures in the model by a uniform multiplicative factor.  The slight shift of the pressure scale
moves the steep part of the deep-photosphere temperature profile, where the visible continuum forms, inward or outward in optical depth space,
thereby changing the temperatures where the continuum becomes optically thick, and thus
the emergent intensity field.  For this particular multi-snapshot CO5BOLD model,
the derived scale factor was 1.033: a small correction, to be sure, given the potentially large
inconsistencies that in principle could be found in such a comparison.  The pressure adjustment has
much less impact on the middle photosphere, where the temperatures have
flattened out compared to the deeper layers.  

Also highlighted in Fig.~3a are three snapshots selected for
a more detailed analysis to examine the role of systematic effects, for example arising from the stochastic nature of
the independent thermal profiles (i.e., how many snapshots are needed for a reliable average) or from
the oscillator strength scales (which have a subtle impact on the ``Goldilocks'' balance temperatures mentioned earlier).  Snapshot ``B'' has an average upper photospheric temperature profile very similar to
that of the full model; snapshot ``A'' is cooler than average at high altitudes; and snapshot ``C'' is warmer than average.  In general, these temperature deviations are reversed at depth.  Also, the pressure scale factor derived from
the full 3D model was applied to the individual reference snapshots, rather than deriving a specific scale factor for
each, in order to preserve the stochastic intensity fluctuations inherent in such snapshots (although it turned out, by accident, that the three reference snapshots have nearly identical visible
continuum intensities, all slightly above the full 3D model).
 
Figure~3b compares a spatially averaged (on surfaces of constant pressure) version of the full 3D model to the
FAL-C semi-empirical 1D stratification.  Also shown is the result of applying a uniform
temperature shift of +90~K to the outer layers of the 3D model, so that the pressure-surface average more closely resembles
the temperature profile of the semi-empirical model.  (We subsequently refer to such a temperature-enhanced 3D model as the
``maximally perturbed'' scenario, or MAX for short.)  The figure includes depth distributions of CO densities, calculated in
the models for the same $\epsilon_{\rm O}$ (600~ppm), and again averaged
on constant pressure surfaces; together with formation depths of the visible continuum (0.5~$\mu$m),
and those at the CO overtone (2.3~$\mu$m) and fundamental (4.6~$\mu$m).  Although the 1D FAL-C and the MAX variant of the 3D model share the same
average $T(p)$ stratification in the CO layers, MAX evidently produces significantly more CO, presumably because the
gain in molecular formation in the downward temperature fluctuations more than compensates for the loss in the
upward ones.  This 1D--3D difference is exaggerated for the CO problem owing to the large, non-linear temperature sensitivity of the molecular chemistry.

Figure~4 illustrates predicted absolute disk center continuum intensities and 
center-to-limb behavior at $\mu=0.2$, quite close to the limb, for the three representative
snapshots.  The underlying observations were described previously
in APK, and the 3D continuum synthesis approach in Ayres (2008).  In short, Planck functions (thermal emission term) and continuum opacities for each vertical column of the
3D model
were interpolated from precalculated tables, according to the local temperature and pressure.
Next, the continuum source functions (including Thomson and Rayleigh scattering) were calculated.  For simplicity,
we implemented a ``1.5D''\footnote{1.5D means treating each column of the 3D model as an
isolated, laterally homogeneous atmosphere to solve for the angle-dependent radiation fields in the local scattering term.  This approximation is valid for the visible continuum, for which the scattering
component is small, and the scattering mean-free-paths are short compared to the horizontal variations
of the convective structures.  In the IR, the approximation is irrelevant, because the scattering
term is insignificant, and the source function equals the local Planck function.}
partial coherent scattering (PCS) formalism, on the native optical depth scale of each vertical column,
using the Feautrier-based  Hermitian method of Auer (1976). 
Given the continuum source functions, $S(\tau)$, for each column, 3D specific intensities then were calculated
along rays (vertical for disk center, inclined for the limb) according to the formal solution of the transport equation.
This was accomplished in practice by interpolating the vertical source functions and opacities onto the rays (which intersect
multiple columns for the limb viewing case), and then interpolating the newly derived ray source function
onto a fixed optical depth grid, with associated precalculated Feautrier coefficients,
to achieve a fast but accurate formal
solution for the emergent intensity.

To achieve higher accuracy for the vertical transport calculations,
the columns were interpolated onto a finer, 5~km grid.  For the limb sightlines, the models were resampled onto
a finer, 20~km horizontal grid, and the views from the four cardinal azimuthal directions (${\pm}x$, ${\pm}y$) were averaged (in
practice, the differences between the independent views were minimal).  

Referring back to Figure~4, there are two sets of center-limb curves depicted.  Generally, the lower ones are for the baseline 3D snapshots,
while the slightly higher ones are for the temperature-enhanced MAX versions.  It is clear that not only
do the baseline snapshots reproduce the continuum center-limb behavior reasonably well (and much better than the initial
generation of 3D models, which solved the radiation transport utilizing only a few 
opacity bins), but also
the high-altitude temperature perturbation has essentially no influence on the disk center continuum intensities
and minimal
impact on the center-limb behavior.  This is a consequence of
restricting the temperature rise to the higher, transparent layers.
  
We consider the MAX option to be an extreme
case, because the contrived match of the mean 3D model to the 1D $T(p)$ profile glosses over important
aspects of how spatial averages of different spectral diagnostics feed into the practical construction of
a semi-empirical model.  Nevertheless, it is encouraging that the simple high-altitude temperature enhancement does not
fundamentally violate the ``continuum test.''  We included the temperature-enhanced snapshots in the
CO modeling as a hedge against the possibility that the purely RHD 3D simulations might be slightly too
cool in their outer layers, owing to lack of magnetic heating effects and/or damping of high-frequency acoustic waves, as
noted earlier.  The derived MAX temperature enhancements for the three reference snapshots, treated as independent models, ranged from 75~K (``C'') to 110~K (``A''), with ``B'' --- and the full 3D model --- in the middle at 90~K.  Under other circumstances these temperature
enhancements would be considered minimal, but are significant for the highly temperature-sensitive
molecular problem.

  \subsection{Oxygen Abundances from $\mathbf{^{12}}$C$\mathbf{^{16}}$O $\mathbf{\Delta v=1}$ and $\mathbf{\Delta v=2}$}

The initial part of the analysis involved utilizing the parent $\Delta v=1$ and $\Delta v=2$ 
hybrid samples to derive the oxygen abundance, and any dependence on excitation energy, over the same range of
$E_{\rm low}$ that the strongest isotopomer lines span, as a surrogate for the ideal direct 
comparison between parent and isotopomer lines
of the same $v_{\rm low}J_{\rm low}$ (which is not feasible here either due to saturation of the parent lines when the isotopomers are
strong, or too weak isotopomers when the parent lines are unsaturated).  We emphasize that the main role  of the derived oxygen abundance is as a ``transfer standard'' by which to calibrate the relative isotopic abundances.  

As alluded
earlier, line shapes of each hybrid transition were synthesized for a set of discrete oxygen abundances,
typically four evenly spaced in the log (by 0.03~dex = factor of 1.07) to span the full range of expected values in
that particular snapshot (which varied from model to model, and from unperturbed to MAX).  The CO number density was calculated in the instantaneous chemical equilibrium approximation (ICE), again assuming $\epsilon_{\rm C}= \onehalf\,\epsilon_{\rm O}$, and parsed into the various isotopomer contributions (more details are provided below).  Wedemeyer-B{\"o}hm et al.\ (2005) simulated a detailed chemistry network for CO in 2D dynamical models of the photosphere and chromosphere, and concluded that the ICE approximation
was accurately obeyed in the dense middle photosphere, where the chemical time scales are very short, although the authors did identify strong departures from equilibrium in the higher, more tenuous layers, especially near chromospheric shock fronts.  

The line opacity was
determined assuming purely thermal Doppler broadening, compensating for the molecular weight of
the particular isotopomer (G94), and shifting the resulting Gaussian profile by the line-of-sight component of the
local 3D velocity field (i.e., $v_{z}$ for the disk center simulations).  The line source function was taken to be
purely thermal (LTE), an excellent approximation for the middle photosphere where the CO rovibrational collisional
rates are very high (Ayres \& Wiedemann 1989). The total source function weighted the continuum and line components according to the respective monochromatic opacities.

The line profile was calculated at 26 frequencies.  One was a pure continuum point
for normalization purposes.  The other 25 were distributed symmetrically in velocity: 0.5~km s$^{-1}$ spacing
over the inner ${\pm}5$~km s$^{-1}$ core of the profile, with a few additional points at 1~km s$^{-1}$ spacing
in the line wings.  To ease the CO computations, we selected only every other column from the full 3D model, and the
three reference snapshots, to carry out the disk center profile synthesis.  This is justified because the horizontal sampling of the convective pattern 
in the CO5BOLD snapshots is, by numerical necessity,
much finer than the typical scale lengths of the dominant granular structures.

The
equivalent width of each theoretical profile (after smoothing to ATMOS resolution) 
was measured by a Gaussian fit over the ${\pm}5$~km s$^{-1}$ core of
the line (as for the observed features), after compensating for the average convective blueshift.  The calculated $W_{\omega}$ versus $\epsilon_{\rm O}$ distribution then was modeled by a parabolic function,
and the fitted relation was inverted for the observed $W_{\omega}$ to yield the oxygen abundance appropriate
to the particular hybrid transition.  

This first
phase is illustrated in Figures~5a (theoretical and observed profiles) and 5b (abundance fits) for the full 3D model
and the average oscillator strengths.  Fig.~5a shows that the (smoothed) synthesized profiles very closely match the observed ATMOS hybrid profiles, as was demonstrated
by SAGS in more detail for individual CO rovibrational
lines (especially by consideration of the C bisectors).  The calculated convective blueshifts of the fundamental and overtone
lines are about 300~m s$^{-1}$ and 380~m s$^{-1}$, respectively, about half the maximum shifts of weak \ion{Fe}{1} and \ion{Fe}{2} absorptions in the visible (e.g., Asplund et al.\ 2000), which tend to form in deeper, more dynamic layers.  (The observed profiles were matched to the simulated ones
by adjusting the ATMOS frequency scale by a constant velocity offset, separately for $\Delta v=1$ and $\Delta v=2$, so that the
O$-$C velocity difference was zero, averaged over all the hybrid transitions of the specific type.  The velocity zero point of the
ATMOS observations is not known well enough to carry out a direct comparison of observed and calculated convective blueshifts.  Furthermore, the CO fundamental and overtone lines form in slightly different
layers owing to the displaced continuum ``horizons'' [see Fig.~3b], so the average convective
shifts can be somewhat different.)

The second step in this part of the analysis was to examine the derived $\Delta v=1$ and $\Delta v=2$ oxygen abundances (e.g., from Fig.~5b) as a function of excitation energy,
$E_{\rm low}$, and equivalent width, $W_{\omega}$, to uncover any trends for the particular model.  Figure~5c depicts the procedure for the same case illustrated in Figs.~5a and 5b (full 3D model, $\langle{f}\rangle$ scale).  One sees that the derived $\Delta v=2$ oxygen abundances (in lower portions of the panels) average just below 600~ppm; the $\epsilon_{\rm O}$'s have a
slight but systematic dependence on $E_{\rm low}$; the $\Delta v=1$ and $\Delta v=2$ samples are separated; but neither group
displays a significant trend with equivalent width.

Also included in Fig.~5c are analogous results for the FAL-C 1D model (upper portions of the panels), also using the $\langle{f}\rangle$ scale, and with all the other assumptions the same (except that a depth-independent microturbulence parameter of 1.7~km s$^{-1}$ was
introduced so that the 1D profiles matched the observed line widths).  Now,
the derived oxygen abundances are much higher (e.g., APK); the excitation slope is much steeper; the 
$\Delta v=1$ and $\Delta v=2$ samples still are separated, and in the same sense as for the 3D model; but also the $\Delta v=1$ hybrids show
increased scatter.

The slope of the
$\epsilon_{\rm O}$ versus $E_{\rm low}$ distribution was measured by a linear least squares fit to the (more numerous and
more diverse in energy) overtone sample, and an offset (specified by the ratio $\rho\equiv <\epsilon_{\rm O}>_{\Delta v=1} / 
<\epsilon_{\rm O}>_{\Delta v=2}$) was inferred by forcing the (smaller) sample of fundamental lines to fit the same slope.  As is clear
from the figure, the main dependence is on $E_{\rm low}$ rather than $W_{\omega}$, consistent with the
selection criterion designed to minimize inclusion of partially saturated lines (which would stand out in the
$W_{\omega}$ part of the diagram as anomalous).  Note that the scatter of 
individual values about the linear relations is small for the 3D model.   

Figure~5d is similar to 5c, but now compares results from the baseline ($\Delta{T}\equiv 0$~K) full 3D model to those of Goldilocks ($\Delta{T}= 34$~K) and MAX ($\Delta{T}= 90$~K), again for the $\langle{f}\rangle$ scale.  The model variants show
a systematic increase in $\epsilon_{\rm O}$ with $\Delta{T}$; a smooth change in the abundance/excitation slope
from positive to negative; and a reversal of the $\Delta v=1$ and $\Delta v=2$ separation (and, of course,
$\rho\equiv 0$ for the Goldilocks option).  Significantly, the
excitation slope is nearly zero for the Goldilocks case, an outcome that is not a foregone
conclusion from simply forcing the $\Delta v=1$ and $\Delta v=2$ samples to agree.

Figure~6 summarizes the oxygen abundance exercises for snapshots A--C, provided
separately for the independent oscillator strengths, G94 and HR96; the full 3D model for the
$\langle{f}\rangle$ scale; and the 1D model, also for the average $f$-values.  

Although the details differ between the different oscillator strength scales, the overall behavior is
similar: the baseline models predict lower oxygen abundances, significant $\epsilon_{\rm O}$/$E_{\rm low}$ slopes,
and separated $\Delta v=1$ and $\Delta v=2$ distributions (dots mark the $\Delta v=1$ values), whereas the maximally-perturbed models yield higher 
oxygen abundances, more negative $\epsilon_{\rm O}$/$E_{\rm low}$ slopes, and $\Delta v=1$ and $\Delta v=2$ separated
in the opposite sense.  The Goldilocks snapshots
have intermediate values of $\epsilon_{\rm O}$ and intermediate excitation slopes.  A subtle consequence of the 
differences between the G94 and HR96 $f$-values is on the balance temperatures (the $\Delta{T}$ mentioned earlier) to
force $\rho\equiv 0$ for the different models.  The G94 scale picks a lower balance temperature, typically closer
to the baseline ($\Delta{T}\equiv 0$), whereas HR96 picks a higher value, closer to the MAX option.  Not surprisingly,
the average $f$-value scale selects intermediate balance temperatures.  The effect on the balance temperatures
is important given the highly temperature sensitive nature of the problem.  Also note that the
G94 ``bow ties'' scatter around zero excitation slope, whereas the HR96 symbols appear to be offset from the zero line.  This could be taken as some support for the G94 scale over HR96 (although we will see later that other evidence points in the opposite direction).

The $\Delta v=1$ oxygen abundances of the baseline models (550--580~ppm) are higher than the ``low'' oxygen abundances (460--490~ppm)
derived from atomic oxygen lines initially by Allende-Prieto, Lambert, \& Asplund (2001) and subsequently Asplund et al.\ (2004), which originally inspired the so-called ``Oxygen Crisis'' conflict with helioseismology (e.g., APK; and references to previous work therein).  In fact, the
higher values from the ``extreme'' MAX perturbed snapshots fall around the $\epsilon_{\rm O}\sim 680$~ppm favored by 
helioseismology (e.g., Ayres 2008).  The Goldilocks values are close to the intermediate $\epsilon_{\rm O}$
range proposed in a more recent study of atomic oxygen by Caffau et al.\ (2008).  Again, this comparison assumes that
$\epsilon_{\rm C}$ is exactly one-half $\epsilon_{\rm O}$: the inferred oxygen abundance would have to be modified to the extent that the true abundance ratio deviates from that value. 

The 1D model illustrated here shows a starkly different behavior than the 3D snapshots (as anticipated in
SAGS and APK): the inferred $\epsilon_{\rm O}$ are much larger, and the excitation slopes
are steeper.  To the extent that shallow slopes indicate thermal harmony in the models, the 1D example
clearly is far away from that ideal; almost certainly because it must represent at each pressure level the wide diversity of
temperatures in the real inhomogeneous atmosphere by only a single compromise value.  The spatial averages of ultraviolet and visible spectral diagnostics inherent in forming the 1D model clearly
fail miserably to capture the average behavior of the CO $\Delta v=1$ and $\Delta v=2$ bands, undoubtedly because
the UV/optical has a natural bias toward hotter temperatures, while molecules have the diametrically opposite bias.  We can safely conclude that the CO problem is the gold-standard
example where a 1D model provides a very misleading --- in fact completely useless ---
viewpoint compared to 3D (as ventured earlier by SAGS).

As a tracer of the oxygen
abundance, the CO bands clearly are very model dependent, and thus are more susceptible to systematic errors in
the thermal properties of the 3D models than typical atomic lines.  Personally we favor the Goldilocks
scenario over the unperturbed baseline snapshots, or the MAX variants.  We
consider the latter to be an extreme case, for the reasons outlined earlier.  At the same time, the baseline model very likely is slightly too cool
compared with the real photosphere, by virtue of the additional heating process not currently included
in the simulations.  Thus
our natural preference is for the intermediate Goldilocks case: not too cool, not too warm.  Our discussion of the oxygen abundance here mainly is for
sake of completeness, since we utilize $\epsilon_{\rm O}$ essentially as a transfer standard, as cautioned earlier.  Thankfully, the rather broad range of derived $\epsilon_{\rm O}$ values has a smaller impact on the
isotopic analysis, owing to its differential nature (although we will see that part of the anticipated cancellation
is not realized owing to the influence of the different $f$-value scales on the Goldilocks balance temperatures, which feeds back
into the isotopic problem).

  \subsection{Isotopic Ratios}

The final piece of the puzzle was to determine the isotopic ratios for
the reference snapshots, the full 3D model, and their temperature-perturbed cousins, given the oxygen abundances and
excitation slopes derived in the first phase.  

Owing to the different molecular weights of the isotopic diatoms, not only are the rovibrational properties different,
but so too is the chemistry.  The dissociation potential, for example, is slightly affected by the different ground level energies
of the isotopomers (Morton \& Noreau 1994), as is the kinetic factor in the dissociative equilibrium.  Furthermore, the
partition functions of each isotopomer are affected by the slightly compressed rovibrational energy ladders
(G94).

In the equation of state, we accounted for the ICE formation of the significant
diatomic molecules in a solar mixture, based on Grevesse \& Sauval (1998) abundances for all elements except O, which
we varied; C, which we took as $\onehalf$O; and N, which we set at $\frac{1}{8}$O.  In the ICE solution, for simplicity, we included only the dominant isotope
of each element (e.g., 
$^{12}$C for carbon and $^{16}$O for oxygen), with the appropriate molecular formation
parameters for that combination.  This is a good approximation for CO because the contribution by the rare isotopes is
only about 1\%.

We then treated the CO isotopomer abundances as fractions of the total concentration obtained
from the dominant-isotope ICE solution.  To derive the relevant ratios, we
simulated the full isotopic chemistry, but restricted to CO itself (and the main isotopomers 26, 27, 28, and 36;
ignoring the minute concentrations of 37 and 38).  We expressed the isotopomer 
abundance for, say, $^{13}$C$^{16}$O, relative to the calculated total density as 
\begin{equation}
n_{36}/n_{\rm CO}= \gamma_{36}\,[ \frac{R_{23}^{-1}}{(1 + R_{23}^{-1})} ] [\frac{1 }{(1 +  R_{67}^{-1} + R_{68}^{-1})}]\,\,\,\, ,
\end{equation}
where $n_{36}$ is the $^{13}$C$^{16}$O number density, $n_{\rm CO}$ is the total CO density from the ICE solution, $R_{23}$ is the $^{12}$C/$^{13}$C ratio, $R_{67}$ is 
the $^{16}$O/$^{17}$O ratio, and so forth.  The leading term in square brackets simply reflects the fraction of
carbon that is $^{13}$C, and the trailing term in square brackets reflects the fraction of oxygen that is $^{16}$O.  The product of the two
would be the isotopomer abundance ratio if the chemistry were independent of the molecular weights
and dissociation constants.  Finally,
$\gamma$ is a correction factor, derived from the CO
isotopomer network, which captures the impact of the more subtle chemistry issues.  In practice, $\gamma$ is temperature and 
density dependent, especially for $T< 4500$~K and
high densities ($n_{\rm H}\gtrsim {\rm few}{\times}10^{16}$ cm$^{-3}$); and 
exhibits asymptotic behavior below 3500~K and above
5000~K.  However, it also happens that the dominant altered factors in the dissociative equilibrium nearly
cancel, so $\gamma$ is close to one, differing from unity at most by only about half a percent in the worst 
\{$T,\,n_{\rm H}$\} case.  That $\gamma\approx 1$ implies that the isotopomer chemistry is close to that of the parent CO, on a per molecule basis, which conveniently justifies the dominant-isotope approximation in the global ICE chemistry. 

We calculated the $\gamma$'s in detail
for medium snapshot ``B,'' and determined representative
values:  1.0010, 1.0007, 1.0016, and 1.000 for isotopomers 36, 28, 27, and 26
respectively.  We introduced these as temperature- and density-independent constants 
in the isotopomer synthesis.  This is reasonable because the peak
of the CO density function occurs at temperatures much warmer than the low-$T$ limit where the maximum departures
of $\gamma$ from unity occur. 

SAGS performed a similar type of analysis, although they also included the factor
of the partition function that appears in the line opacity, together with the factor of
the partition function in the dissociative equilibrium, because in principle the two cancel, at least if the molecular formation is completely
unsaturated (in the sense that $n_{\rm CO}\sim \epsilon_{\rm C}\,\epsilon_{\rm O} n_{\rm H}^2$; i.e., only
a small amount of the free carbon and oxygen has associated into CO).  We chose to retain the partition function factors explicitly, and separately, in the line opacity and chemical equilibrium to allow for
partial saturation effects in the lower temperature overshooting
convective plumes of the 3D models, which are not accommodated by the SAGS strategy.
Furthermore, the SAGS ``opacity scale factors'' were derived taking the
isotopomer oscillator strengths equal to those of the equivalent parent transitions.  However, the isotopomer
strengths on either the G94 or HR96 scales are $\sim$4\% less than for the equivalent parent transition (for 36 and 28; e.g., Fig.~1c),
leading to a systematic error of at least that magnitude (isotopic abundances underestimated, and isotopic ratios, $R$, overstated),
all else being equal.

We adopted reference $(R_{\rm ISO})_{\rm STD}$'s of 90, 500, and 2650 for ${\rm ISO}= $ 23, 68, and 67,
respectively (essentially the terrestrial standard values); 
then calculated equivalent widths for multiples ($s_{\rm ISO}$, as described earlier) of these, from $-$0.050~dex to +0.075~dex, in steps of 0.025~dex (six total).
As we did for the oxygen abundances from the parent sample, we fitted a parabolic relation to the calculated
equivalent widths and solved for the isotopic ratio scale factor that matched the observed $W_{\omega}$.  Figures~7a--c illustrate
the procedure for the Goldilocks version of the full 3D model, utilizing the $\langle{f}\rangle$ scale.  We adopted a specific oxygen abundance, $(\epsilon_{\rm O})_{\rm ref}$, for each simulation based on the $\Delta v=1$ sample, but
adjusted the oscillator strengths of the isotopomer transitions (all $\Delta v=1$ type) by a multiplicative factor $\beta$ to compensate for any slope in $\epsilon_{\rm O}$ versus $E_{\rm low}$:
\begin{equation}
\beta= (\frac{{\rm c}_{0}}{(\epsilon_{\rm O})_{\rm ref}} + \frac{{\rm c}_{1}}{(\epsilon_{\rm O})_{\rm ref}}\times E_{4})^2\,\,\,\, ,
\end{equation} 
where $E_{4}\equiv (E_{\rm low}/10^4)$.  The intercept, $c_0$, and slope, $c_1$, of
the linear $\epsilon_{\rm O}/E_{\rm low}$ relation initially were derived from the
$\Delta v=2$ sample, but then adjusted to $\Delta v=1$ according to the $\rho$ factor mentioned earlier;
note also that $(\epsilon_{\rm O})_{\rm ref}\equiv {\rm c}_{0}\,+\,{\rm c}_{1}$.  The $\beta$-scaling of the $f$-values simulates the effect of a gradient in the oxygen abundance with $E_{\rm low}$, at least in the regime
where the CO formation is unsaturated.  

Figure~7c shows that for the full 3D model Goldilocks option, and the average oscillator strengths,
the isotopic abundance ratios $^{12}$C/$^{13}$C and $^{16}$O/$^{18}$O
are slightly higher than terrestrial (by about 2\%), with only small random errors due to the dispersion of values
within each sample, particularly if we are permitted to describe these as standard errors of the mean.  The $^{16}$O/$^{17}$O
ratio is slightly higher still, although now the dispersion among the four independent hybrid points is large,
paralleling the large individual uncertainties.  The $^{12}$C$^{17}$O features are extremely weak, and therefore
especially susceptible to errors in the local continuum level, so any conclusions drawn from
that isotopomer will be much less secure than for the stronger $^{13}$C$^{16}$O and $^{12}$C$^{18}$O systems.

The derived isotopic ratios are depicted schematically
in Figure~8, as a function of $\epsilon_{\rm O}$ ($\Delta v = 1$ values) and the $\rho$ ratio described earlier, for the same set of models
illustrated previously in Fig.~6.

On the G94 scale, the isotopic ratios are systematically $\sim 2$\% lower than those derived from the HR96 scale,
for the baseline and MAX models.
Going from the baselines to the MAX models, the isotope ratios increase systematically for both oscillator strength
scales, by nearly same amount for each: 8\% higher for 36 and 10\% higher for 28 and 27.  The dispersion in $R_{\rm ISO}$ among the
reference snapshots is 2--3\% for the baselines, and about half that for the MAX variants (undoubtedly because
the latter are closer
together in temperature than the baselines thanks to the temperature-correction procedure). 

On the G94 scale, the $\Delta v = 1$ $\epsilon_{\rm O}$ values are systematically $\sim 1$\%  higher than those from the HR96 scale, for the baseline models and the MAX variants, due to the systematically slightly lower G94 $\Delta v = 1$ 
$f$-values.  For the baseline models,
the average $\epsilon_{\rm O}$ is about 570~ppm with a $\sim$2.6\% snapshot dispersion.  For the MAX models,
the $\epsilon_{\rm O}$ dispersion shrinks to about 1\%, again likely because of the temperature confluence effect, but the absolute oxygen abundance rises a dramatic 21\% to about
685~ppm.  The large increase in the derived oxygen abundance with only a small $\Delta T= 75$--110~K change from the
baseline models demonstrates the strong temperature sensitivity of CO.  Note, however, that
the rise in the isotopic ratios from baseline to MAX is only about half that of the oxygen abundance itself,
demonstrating some significant cancellations in the differential analysis.

Now consider the Goldilocks models that force the $\Delta v=1$ and $\Delta v=2$ 
parent samples to agree in oxygen abundance space.  The differences between the results for the two oscillator strength
scales are more pronounced, because the systematic offsets between the G94 and HR96 $f$-values
for $\Delta v=1$ and $\Delta v=2$ went in 
opposite directions (at least for the lower vibrational bands that are relevant in this
analysis),
so the balance temperature to force them into agreement was different.  In fact, the HR96 Goldilocks
temperatures are closer to the MAX values, averaging 50\% of the
maximum $\Delta{T}$'s; while the G94 temperatures are closer to the baselines: only
1~K higher in the case of snapshot ``C,'' and only 7~K higher for ``B.''  
As before, the dispersions of the isotopic ratios
among the three Goldilocks snapshots for a given oscillator strength scale are small ($\lesssim$1\%), but now the HR96 values are 
systematically higher by about 5--6\% than those for G94.  The average oxygen abundances are about 600~ppm for G94 and 630~ppm
for HR96, with a small ($\sim$2\%) dispersion among the snapshots in both cases (although with snapshots ``B'' and ``C'' giving
nearly identical results, but ``A'' almost 25~ppm higher). 

We now are in a position to recommend isotopic ratios based on the global analysis.  We choose the Goldilocks option as our preferred temperature correction, because these are the models that achieve agreement between the $\Delta v=1$ and $\Delta v=2$ bands.  This would be the desired outcome for a model that has an optimum thermal structure as far
as the molecules are concerned, if the oscillator strength scale is reliable.  We chose recommended values
from the full 3D model, and the average $f$-values, because we have no reason (yet) to prefer one of the oscillator strength scales over the
other.  Further, within the set of reference snapshots
we can identify the sources of uncertainty devolving from: (1) the reference $\Delta v=1$ oxygen abundances
(small, contributed by the $\rho$ parameter derived from the ten $\Delta v=1$ hybrids); (2)
``noise'' in the isotopomer hybrid line samples (generally small, except
for isotopomer 27); (3) variations
among the snapshots (also small, when considering the full set of 16); and (4) differences imposed by the alternative oscillator strength scales (dominates the others).
The first three sources can be characterized by standard errors of the mean, because we are comparing multiple realizations of
ostensibly similar objects.  For the oscillator strength scales, we have only two examples, so
we quote the uncertainty as one-half the difference between the alternative results (equivalent to a 1\,$\sigma$ dispersion for a sample of two).  

In addition, we can characterize the further uncertainty that arises from the reality
that we do not know, {\em a priori,}\/ which is the more correct 3D model scenario: baseline, Goldilocks, or MAX.  This error must be treated as a standard deviation, because we are comparing ostensibly different
objects.  The derived ``scenario uncertainty'' typically is larger than any of the individual sources, and in fact is larger than all those
collectively (quadratic sum), except again in the case of isotopomer 27, for which the sample random error is large and comparable to the scenario component.

\noindent
We find: \\[4mm]

\begin{tabular}{rrrrrrrl}
$^{12}$C/$^{13}$C= & 91.4 & ${\pm}0.5$ & ${\pm}0.3$ & ${\pm}0.2$ & ${\pm}1.2$ & ${\pm}3.0$ & $({\pm}1.3,~{\pm}3.3)$\\ 
$^{16}$O/$^{18}$O= & 511 & ${\pm}3$ & ${\pm}3$ & ${\pm}1.3$ & ${\pm}9$ & ${\pm}21$ & $({\pm}10,~{\pm}23)$\\  
$^{16}$O/$^{17}$O= &2738 & ${\pm}16$ & ${\pm}105$ & ${\pm}7$ & ${\pm}50$ & ${\pm}110$ & $({\pm}118,~{\pm}161)$\\ 
$\epsilon_{\rm O}({\rm ppm})$= & 603 & ${\pm}2$ & ${\pm}1.3$ & ${\pm}2$ & ${\pm}8$ & ${\pm}28$ & $({\pm}9,~{\pm}29)$\\
\end{tabular}

The first error (1~s.e.) is that
associated with the uncertainty of the $\Delta v=1$ $\epsilon_{\rm O}$ applied in the isotopomer synthesis, resulting from
scaling the derived $\Delta v=2$ abundance (which has negligible random error owing to the
large number of contributing overtone transitions and their generally small scatter) according to the 
$\rho$ factor.  The $\langle{\sigma_{\rho}}\rangle\sim 1.2$\% for the ten $\Delta v=1$ hybrid lines (1~s.e.$\sim 0.38$\%) is amplified for the
isotopic ratios by a factor of $\sim 1.6$ (derived by numerical experiments).  The second uncertainty
(1~s.e.) is that due to the isotopomer sample random error (or the $\Delta v=2$ hybrid sample for $\epsilon_{\rm O}$).
The third (1~s.e.) is the snapshot variability, taking the dispersion of results
from the three independent reference snapshots, for the $\langle{f}\rangle$ scale, and dividing by $\sqrt{16}$ to estimate the s.e.\ over
the full 3D model.  The fourth (1\,$\sigma$) is from the $f$-values, as described above.  Finally, the fifth (1\,$\sigma$)
is the scenario uncertainty: i.e., which level of temperature correction, if any, should be applied
to the baseline model.  The leading error quoted in the trailing parentheses is the total uncertainty for the
Goldilocks option treated alone, which is the ``optimistic'' value if we believe that Goldilocks
is the most realistic of the scenarios.  The second parenthetical 
error includes the scenario component, and thus would fall at the conservative, pessimistic, end of the uncertainty spectrum.
Note that in all cases, the snapshot variability contribution is smaller than the total sample random error
(sum of first two terms in quadrature), demonstrating
that 16 snapshots is sufficient for the purpose.

Table~3 summarizes abundances and isotopic ratios derived with the three temperature correction scenarios applied to
the full 3D model, for the $\langle{f}\rangle$ scale; and the Goldilocks variants of the full model separately for the G94 and HR96 $f$-values.

For comparison, we carried out the isotopic analysis using the FAL-C 1D model illustrated earlier.  In this case, as with the 3D models, and for all three oscillator strength options, it also was necessary to impose a balance
temperature to bring the $\Delta v=1$ and $\Delta v=2$ $\epsilon_{\rm O}$'s into agreement.  This
$\Delta{T}$ was $\sim +100$~K for the $\langle{f}\rangle$ scale.  Note that because the
1D temperature stratification in a sense already is at the maximum level, owing to the UV/optical bias in its
construction, the additional positive temperature enhancement is counterintuitive (and plainly unphysical, to the extent that 1D models can claim to be physical
in the first place).  Nevertheless, the results from the 1D model follow the pattern seen in
3D: the Goldilocks version yields a higher oxygen abundance than the baseline model by
about 13\%, while the isotopic ratios also are higher, but only about half of the 
corresponding $\epsilon_{\rm O}$ rise.  

The results for the 1D Goldilocks scenario with the average $f$-values are as follows: $\epsilon_{\rm O}= 850$~ppm;
$^{12}$C/$^{13}$C=$78.2$; $^{16}$O/$^{18}$O=$420$; and $^{16}$O/$^{17}$O=$2195$.  Although not
directly comparable, these are 
similar to the 1D values obtained previously by APK, in the sense that the derived isotopic ratios all are significantly
lower than terrestrial, while the inferred oxygen abundance is much higher than any contemporary-accepted value.  The discouraging corollary is that all
the previous efforts devoted to deriving abundances and isotopic ratios from 1D synthesis of CO, and other molecules,
must be considered in vain, since it is clear that this particular problem is especially prone to the thermal differences between a 1D average
temperature stratification versus the fluctuations inherent in a dynamically convecting late-type photosphere.  

To be sure, the earlier 
Ayres et al.\ work made a concerted effort
to incorporate temperature perturbations in a schematic way by considering a set of 1D multicomponent models that were
scaled from a baseline thermal profile by depth-independent positive and negative increments in temperature, and weighted by
a Gaussian-like distribution to mimic observed continuum brightness histograms.  However, consideration of the different types of thermal structures populating a 3D
snapshot (e.g., Ayres 2008: his Fig.~2), shows that convection does not simply raise and lower the temperature
profile uniformly, but rather skews it so that those stratifications that are hot deep in tend to be cooler higher up, 
and vice versa (origin of the so-called ``reversed granulation'' pattern at high altitudes).
That skewness likely has a fundamental impact on the CO problem, since the continuum forms in deeper layers, while
the molecular absorptions form higher up.  Furthermore, thanks to the Wedemeyer-B{\"o}hm \& Rouppe van der Voort (2009)
{\em Hinode}\/ study, we now know that the amplitudes of thermal fluctuations in
the photosphere are significantly larger than assumed by APK, who had appealed to ground-based granulation measurements (that apparently were more affected by stray light and the telescope point response profile than realized).  Thus, the
APK multicomponent strategy was, in effect, much closer to 1D than 3D.

We also note, with respect to the earlier SAGS study, which derived rather lower equivalent values of $\epsilon_{\rm O}$ from CO\footnote{Although
the authors cast their results in terms of the carbon abundance, they did adopt the previous Asplund et al.\ (2004) low oxygen abundance,
and obtained a C/O ratio of about 0.5, consistent with what we use here.}, that the
3D snapshots utilized in that work apparently were somewhat cooler above $\tau_{\rm Ross}\sim 1$ than those employed here (see, e.g., Caffau et al.\ 2008, their Fig.~1).  The cooler temperature profile likely was a consequence of the fewer-frequency opacity binning used in the earlier models.  As we have seen here, cooler mid-photospheric temperatures tend to predict systematically lower oxygen abundances from CO and lower isotopic ratios, all else being equal (although, to be fair, the SAGS isotopic ratios are consistent with those derived here, given the large error bars of the earlier 3D study [Fig.~8]).

\section{Discussion and Conclusions}

The value $R_{68}= 511{\pm}$(10,\,23) [$\delta{^{18}\mbox{O}}= -23{\pm}22^{\circ}\!\!/\!_{\circ\circ}$] derived 
from the ATMOS spectra is somewhat smaller of a deficit than the {\em Genesis}\/ inferred photospheric value of 530 [$\delta{^{18}\mbox{O}}= -59$].  Our photospheric $R_{67}$ ratio implies a larger deficit [$\delta{^{17}\mbox{O}}= -42{\pm}45^{\circ}\!\!/\!_{\circ\circ}$], closer to the {\em Genesis}\/ result, although hardly any weight can be attached to this owing to the uncomfortably large measurement errors.  Therefore our results, while tantalizing, cannot confirm that the solar photosphere falls $\sim$28\,$^{\circ}\!\!/\!_{\circ\circ}$ below the terrestrial fractionation line, as determined by {\em Genesis}\/ (McKeegan et al.\ 2011). The value of $R_{23}= 91.4{\pm}$(1.3,\,3.3) [$\delta{^{13}\mbox{C}}= -25{\pm}15$] measured here is similar
to our $\delta{^{18}\mbox{O}}$ and is slightly higher than the terrestrial value of 89.2. The main conclusion --- given our intermediate modelization strategy and middle choice of
oscillator strength scales --- is that the photospheric isotopic ratios of
carbon and oxygen are slightly on the high side of the terrestrial values, although not to the extent
found by {\em Genesis}\/ for the heavy oxygen isotopes.  That the Sun likely is lighter than the Earth, isotopically speaking, means that exotic $\delta> 0$ fractionation mechanisms need not be invoked for the early solar system, although the {\em Genesis}\/ 
findings for $\delta^{18}\mbox{O}$ and $\delta^{17}\mbox{O}$ imply that whatever the $\delta<0$ process, it must have been mass-independent (McKeegan et al.\ 2011).

Taking some additional liberties, we can turn the problem around and ask what are the implications if {\em Genesis}\/ is correct and the actual
photospheric oxygen isotopic ratios are $\sim 60$\,$^{\circ}\!\!/\!_{\circ\circ}$ above the terrestrial standard values?  If we further require that a satisfactory
model must have $\rho\sim 1$, we see that this combination favors the HR96 $f$-value scale (Fig.~8 and Table~3), and by implication $\delta{^{13}\mbox{C}}\sim -50$, close to the {\em Genesis}\/ heavy oxygen deficits; and one step further, an
$\epsilon_{\rm O}\sim 620$~ppm well beyond the upper side of the ``low-O'' error flags, just at the lower edge of the seismic range.

In short, we reinforce the conclusions of Scott et al.\ 
that the solar CO problem is a poster child for
the importance of 3D effects: the 1D models simply fail to account for key aspects of the
highly temperature sensitive molecular formation, and provide completely misleading results in the end.  We
showed, moreover, that altering the thermal profiles of the 3D models slightly according to reasonable criteria can achieve a better match
to the empirical CO spectra, but with an accompanying significant
impact on especially the oxygen abundance.  Although we again caution that CO is a less than ideal abundance indicator, and uncertainties remain due to the discordance between the available oscillator strength scales, the derived $\epsilon_{\rm O}$'s for our favored ``Goldilocks'' scenario are in accord with the 
`high' range found in recent 3D studies of solar atomic
oxygen, and possibly within striking distance of a value that could be swallowed, without grimacing, by helioseismology.

\acknowledgments

This work was supported by NSF grant AST--0908293.  JRL thanks NASA Origins of Solar Systems and the Penn  
State Astrobiology Research Center.  EC and HGL acknowledge financial support by the Sonderforschungsbereich
SFB\,881 ``The Milky Way System'' (subproject A4) of the German Research
Foundation (DFG).

\begin{deluxetable}{rccrccc}
\tablenum{1}
\tablecaption{$^{12}$C$^{16}$O Hybrid Sample}
\tablewidth{0pt}
\tablecolumns{7}
\tablehead{
\colhead{Transition} &  \colhead{$\omega$} & \colhead{$E_{\rm low}$}  &\colhead{$gf$} & \colhead{$W_{\omega}$} & 
 \colhead{FWHM}  & \colhead{Flag} \\[3pt]                
\colhead{} &  \colhead{(cm$^{-1}$)} & \colhead{(cm$^{-1}$)}  & \colhead{} &  \colhead{($10^{-3}$ cm$^{-1}$)}& 
\colhead{(km~s$^{-1}$)}  
& \colhead{}}                
\startdata   
2--1~P92 & 1638.5309   &  17997.97  &  $ 1.716{\times}10^{-3 }$ & 
$ 2.788{\pm}0.010 $  &   4.44  &  n \\
2--1~P91 & 1644.9269   &  17666.30  &  $ 1.702{\times}10^{-3 }$ & 
$ 3.039{\pm}0.010 $  &   4.56  &  n \\
2--1~P90 & 1651.3006   &  17337.83  &  $ 1.686{\times}10^{-3 }$ & 
$ 3.178{\pm}0.011 $  &   4.50  &  n \\
2--1~P89 & 1657.6517   &  17012.59  &  $ 1.672{\times}10^{-3 }$ & 
$ 3.354{\pm}0.012 $  &   4.51  &  n \\
\hline
{\bf 1A26 } & {\bf 1648.8911 } & {\bf 17464.58 } & {\bf 
 $\mathbf{1.683{\times}10^{-3 }}$} & {\bf $\mathbf{ 3.072{\pm}0.005} $} & {\bf  4.54} \\
\hline\\[3pt]
1--0~P94 & 1648.8738   &  16683.95  &  $ 8.841{\times}10^{-4 }$ & 
$ 2.420{\pm}0.013 $  &   4.36  &  n \\
1--0~P93 & 1655.3485   &  16342.62  &  $ 8.768{\times}10^{-4 }$ & 
$ 2.614{\pm}0.010 $  &   4.41  &  l \\
1--0~P92 & 1661.8011   &  16004.50  &  $ 8.692{\times}10^{-4 }$ & 
$ 2.826{\pm}0.014 $  &   4.50  &  l \\
\hline
{\bf 1B26 } & {\bf 1655.7844 } & {\bf 16321.21 } & {\bf 
$\mathbf{ 8.733{\times}10^{-4 }}$} & {\bf $\mathbf{ 2.588{\pm}0.004 }$} & {\bf  4.44} \\
\hline\\[3pt]
1--0~P90 & 1674.6395   &  15337.95  &  $ 8.543{\times}10^{-4 }$ & 
$ 3.195{\pm}0.012 $  &   4.49  &  n \\
1--0~P89 & 1681.0251   &  15009.55  &  $ 8.468{\times}10^{-4 }$ & 
$ 3.363{\pm}0.014 $  &   4.55  &  l \\
1--0~P87 & 1693.7284   &  14362.57  &  $ 8.315{\times}10^{-4 }$ & 
$ 3.707{\pm}0.014 $  &   4.61  &  n \\
1--0~P86 & 1700.0458   &  14044.01  &  $ 8.239{\times}10^{-4 }$ & 
$ 3.948{\pm}0.019 $  &   4.61  &  l \\
\hline
{\bf 1C26 } & {\bf 1688.8924 } & {\bf 14613.11 } & {\bf 
$\mathbf{ 8.283{\times}10^{-4 }}$} & {\bf $\mathbf{ 3.538{\pm}0.005 }$} & {\bf  4.60} \\
\hline\\[3pt]
3--2~P82 & 1677.9226   &  16824.71  &  $ 2.316{\times}10^{-3 }$ & 
$ 4.219{\pm}0.016 $  &   4.75  &  n \\
4--3~P76 & 1690.6726   &  17085.25  &  $ 2.860{\times}10^{-3 }$ & 
$ 4.505{\pm}0.029 $  &   4.83  &  l \\
5--4~P71 & 1696.4065   &  17725.30  &  $ 3.330{\times}10^{-3 }$ & 
$ 4.237{\pm}0.021 $  &   4.74  &  r \\
5--4~P70 & 1702.2023   &  17470.95  &  $ 3.291{\times}10^{-3 }$ & 
$ 4.513{\pm}0.018 $  &   4.76  &  n \\
\hline
{\bf 1D26 } & {\bf 1692.2540 } & {\bf 17285.70 } & {\bf 
$\mathbf{  2.932{\times}10^{-3 }}$} & {\bf $\mathbf{  4.342{\pm}0.009} $} & {\bf  4.80} \\
\hline\\[3pt]
2--1~P85 & 1682.8292   &  15744.06  &  $ 1.612{\times}10^{-3 }$ & 
$ 4.189{\pm}0.018 $  &   4.68  &  n \\
2--1~P84 & 1689.0661   &  15435.10  &  $ 1.596{\times}10^{-3 }$ & 
$ 4.387{\pm}0.019 $  &   4.72  &  n \\
2--1~P83 & 1695.2798   &  15129.43  &  $ 1.581{\times}10^{-3 }$ & 
$ 4.458{\pm}0.020 $  &   4.78  &  r \\
2--1~P82 & 1701.4702   &  14827.07  &  $ 1.565{\times}10^{-3 }$ & 
$ 4.758{\pm}0.018 $  &   4.79  &  n \\
\hline
{\bf 1E26 } & {\bf 1692.8647 } & {\bf 15250.59 } & {\bf 
$ \mathbf{  1.579{\times}10^{-3 }}$} & {\bf $\mathbf{   4.426{\pm}0.009 }$} & {\bf  4.77} \\
\hline\\[3pt]
9--8~P28 & 1820.8624   &  17851.92  &  $ 2.428{\times}10^{-3 }$ & 
$ 4.273{\pm}0.017 $  &   4.65  &  n \\
9--8~P22 & 1847.1185   &  17308.93  &  $ 1.930{\times}10^{-3 }$ & 
$ 4.171{\pm}0.013 $  &   4.56  &  n \\
9--8~P20 & 1855.6181   &  17156.12  &  $ 1.760{\times}10^{-3 }$ & 
$ 4.183{\pm}0.017 $  &   4.61  &  l \\
\hline
{\bf 1F26 } & {\bf 1840.6961 } & {\bf 17451.14 } & {\bf 
$ \mathbf{ 2.030{\times}10^{-3 }}$} & {\bf $\mathbf{  4.183{\pm}0.006 }$} & {\bf  4.63} \\
\hline\\[3pt]
8--7~P8 & 1929.5857   &  14578.75  &  $ 6.506{\times}10^{-4 }$ & 
$ 3.973{\pm}0.011 $  &   4.47  &  n \\
8--7~P7 & 1933.4265   &  14549.96  &  $ 5.704{\times}10^{-4 }$ & 
$ 3.737{\pm}0.016 $  &   4.46  &  r \\
8--7~R5 & 1980.2139   &  14503.17  &  $ 4.996{\times}10^{-4 }$ & 
$ 3.628{\pm}0.018 $  &   4.41  &  n \\
8--7~R6 & 1983.5660   &  14524.77  &  $ 5.838{\times}10^{-4 }$ & 
$ 4.024{\pm}0.017 $  &   4.46  &  n \\
\hline
{\bf 1G26 } & {\bf 1955.6292 } & {\bf 14541.29 } & {\bf 
$\mathbf{  5.761{\times}10^{-4 }}$} & {\bf $\mathbf{  3.821{\pm}0.007 }$} & {\bf  4.47} \\
\hline\\[3pt]
7--6~P5 & 1966.8906   &  12518.28  &  $ 3.639{\times}10^{-4 }$ & 
$ 4.330{\pm}0.016 $  &   4.52  &  n \\
7--6~P4 & 1970.6641   &  12500.11  &  $ 2.916{\times}10^{-4 }$ & 
$ 3.797{\pm}0.010 $  &   4.40  &  n \\
7--6~R3 & 1999.6015   &  12485.57  &  $ 2.955{\times}10^{-4 }$ & 
$ 3.827{\pm}0.068 $  &   4.39  &  l \\
7--6~R4 & 2003.0602   &  12500.11  &  $ 3.701{\times}10^{-4 }$ & 
$ 4.553{\pm}0.020 $  &   4.53  &  b \\
\hline
{\bf 1H26 } & {\bf 1985.3169 } & {\bf 12501.79 } & {\bf 
$\mathbf{  3.302{\times}10^{-4 }}$} & {\bf $\mathbf{  4.115{\pm}0.006 }$} & {\bf  4.49} \\
\hline\\[3pt]
6--5~P2 & 2004.1715   &  10463.23  &  $ 1.275{\times}10^{-4 }$ & 
$ 3.493{\pm}0.017 $  &   4.37  &  n \\
5--4~R0 & 2041.4229   &   8414.46  &  $ 5.431{\times}10^{-5 }$ & 
$ 3.157{\pm}0.014 $  &   4.33  &  l \\
\hline
{\bf 1I26 } & {\bf 2021.0742 } & {\bf  9539.74 } & {\bf 
$\mathbf{  8.641{\times}10^{-5 }}$} & {\bf $\mathbf{  3.310{\pm}0.009} $} & {\bf  4.37} \\
\hline\\[3pt]
4--3~R0 & 2067.7353   &   6350.43  &  $ 4.415{\times}10^{-5 }$ & 
$ 4.399{\pm}0.023 $  &   4.49  &  n \\
\hline
{\bf 1J26 } & {\bf 2067.7353 } & {\bf  6350.43 } & {\bf 
$\mathbf{  4.415{\times}10^{-5 }}$} & {\bf $\mathbf{  4.399{\pm}0.023 }$} & {\bf  4.49} \\
\hline\\[3pt]
7--5~P63 & 3635.1206   &  17751.86  &  $ 9.009{\times}10^{-5 }$ & 
$ 1.067{\pm}0.007 $  &   4.35  &  n \\
7--5~P62 & 3642.8485   &  17526.74  &  $ 8.898{\times}10^{-5 }$ & 
$ 1.067{\pm}0.006 $  &   4.23  &  n \\
8--6~P54 & 3652.1604   &  17807.96  &  $ 1.074{\times}10^{-4 }$ & 
$ 1.162{\pm}0.009 $  &   4.21  &  r \\
9--7~P43 & 3676.6781   &  17832.96  &  $ 1.159{\times}10^{-4 }$ & 
$ 1.259{\pm}0.005 $  &   4.15  &  n \\
9--7~P42 & 3683.0854   &  17680.10  &  $ 1.137{\times}10^{-4 }$ & 
$ 1.397{\pm}0.010 $  &   4.36  &  l \\
\hline
{\bf 2A26 } & {\bf 3659.7804 } & {\bf 17723.43 } & {\bf 
$\mathbf{  1.031{\times}10^{-4 }}$} & {\bf $\mathbf{ 1.187{\pm}0.004 }$} & {\bf  4.28} \\
\hline\\[3pt]
8--6~P42 & 3733.6516   &  15726.29  &  $ 8.776{\times}10^{-5 }$ & 
$ 1.719{\pm}0.007 $  &   4.05  &  l \\
8--6~P41 & 3740.0303   &  15575.43  &  $ 8.605{\times}10^{-5 }$ & 
$ 1.933{\pm}0.007 $  &   4.24  &  r \\
9--7~P25 & 3782.1151   &  15616.62  &  $ 7.307{\times}10^{-5 }$ & 
$ 1.649{\pm}0.008 $  &   4.21  &  l \\
9--7~P22 & 3797.6212   &  15358.43  &  $ 6.524{\times}10^{-5 }$ & 
$ 1.716{\pm}0.008 $  &   4.34  &  r \\
\hline
{\bf 2B26 } & {\bf 3761.0750 } & {\bf 15576.80 } & {\bf 
$\mathbf{  7.796{\times}10^{-5 }}$} & {\bf $\mathbf{  1.761{\pm}0.004 }$} & {\bf  4.23} \\
\hline\\[3pt]
7--5~P34 & 3833.8782   &  12627.24  &  $ 5.481{\times}10^{-5 }$ & 
$ 2.928{\pm}0.006 $  &   4.25  &  n \\
7--5~P32 & 3845.6079   &  12383.19  &  $ 5.206{\times}10^{-5 }$ & 
$ 3.104{\pm}0.016 $  &   4.34  &  r \\
7--5~P29 & 3862.7120   &  12044.06  &  $ 4.786{\times}10^{-5 }$ & 
$ 3.146{\pm}0.006 $  &   4.21  &  n \\
7--5~P25 & 3884.5963   &  11642.40  &  $ 4.207{\times}10^{-5 }$ & 
$ 3.190{\pm}0.008 $  &   4.31  &  l \\
\hline
{\bf 2C26 } & {\bf 3857.0982 } & {\bf 12167.44 } & {\bf 
$\mathbf{  4.886{\times}10^{-5 }}$} & {\bf $\mathbf{  3.072{\pm}0.004 }$} & {\bf  4.28} \\
\hline\\[3pt]
6--4~P35 & 3879.0270   &  10738.94  &  $ 3.976{\times}10^{-5 }$ & 
$ 3.807{\pm}0.010 $  &   4.26  &  r \\
6--4~P34 & 3885.0247   &  10610.31  &  $ 3.881{\times}10^{-5 }$ & 
$ 3.937{\pm}0.006 $  &   4.27  &  n \\
6--4~P33 & 3890.9573   &  10485.30  &  $ 3.784{\times}10^{-5 }$ & 
$ 3.948{\pm}0.010 $  &   4.28  &  r \\
\hline
{\bf 2D26 } & {\bf 3885.0686 } & {\bf 10610.17 } & {\bf 
$\mathbf{  3.878{\times}10^{-5 }}$} & {\bf $\mathbf{  3.850{\pm}0.005 }$} & {\bf  4.25} \\
\hline\\[3pt]
9--7~R15 & 3940.7097   &  14880.83  &  $ 5.795{\times}10^{-5 }$ & 
$ 1.778{\pm}0.007 $  &   4.36  &  n \\
9--7~R17 & 3945.4187   &  14999.42  &  $ 6.592{\times}10^{-5 }$ & 
$ 1.916{\pm}0.006 $  &   4.31  &  b \\
9--7~R18 & 3947.6644   &  15064.07  &  $ 6.999{\times}10^{-5 }$ & 
$ 1.930{\pm}0.009 $  &   4.19  &  l \\
\hline
{\bf 2E26 } & {\bf 3944.7547 } & {\bf 14985.55 } & {\bf 
$\mathbf{  6.460{\times}10^{-5 }}$} & {\bf $\mathbf{  1.855{\pm}0.005 }$} & {\bf  4.27} \\
\hline\\[3pt]
5--3~P35 & 3930.2825   &   8696.96  &  $ 2.625{\times}10^{-5 }$ & 
$ 4.456{\pm}0.007 $  &   4.15  &  r \\
5--3~P34 & 3936.3153   &   8567.10  &  $ 2.563{\times}10^{-5 }$ & 
$ 4.771{\pm}0.008 $  &   4.26  &  l \\
5--3~P33 & 3942.2830   &   8440.90  &  $ 2.499{\times}10^{-5 }$ & 
$ 4.937{\pm}0.008 $  &   4.32  &  l \\
5--3~P30 & 3959.7946   &   8084.27  &  $ 2.305{\times}10^{-5 }$ & 
$ 5.472{\pm}0.014 $  &   4.49  &  l \\
5--3~P27 & 3976.7156   &   7760.68  &  $ 2.104{\times}10^{-5 }$ & 
$ 5.044{\pm}0.007 $  &   4.26  &  r \\
5--3~P26 & 3982.2240   &   7660.18  &  $ 2.037{\times}10^{-5 }$ & 
$ 5.060{\pm}0.023 $  &   4.25  &  b \\
\hline
{\bf 2F26 } & {\bf 3955.2330 } & {\bf  8189.89 } & {\bf 
$\mathbf{  2.337{\times}10^{-5 }}$} & {\bf $\mathbf{  4.940{\pm}0.006 }$} & {\bf  4.30} \\
\hline\\[3pt]
2--0~P50 & 3985.0994   &   4862.74  &  $ 3.360{\times}10^{-6 }$ & 
$ 2.182{\pm}0.008 $  &   4.23  &  n \\
2--0~P49 & 3992.1960   &   4673.53  &  $ 3.306{\times}10^{-6 }$ & 
$ 2.325{\pm}0.011 $  &   4.25  &  r \\
2--0~P48 & 3999.2296   &   4488.00  &  $ 3.253{\times}10^{-6 }$ & 
$ 2.415{\pm}0.006 $  &   4.26  &  r \\
\hline
{\bf 2G26 } & {\bf 3992.3851 } & {\bf  4669.23 } & {\bf 
$\mathbf{  3.303{\times}10^{-6 }}$} & {\bf $\mathbf{  2.331{\pm}0.007} $} & {\bf  4.29} \\
\hline\\[3pt]
3--1~P42 & 3988.6813   &   5563.81  &  $ 8.922{\times}10^{-6 }$ & 
$ 4.653{\pm}0.009 $  &   4.35  &  n \\
3--1~P40 & 4001.7259   &   5251.07  &  $ 8.576{\times}10^{-6 }$ & 
$ 4.575{\pm}0.008 $  &   4.20  &  l \\
3--1~P38 & 4014.5138   &   4953.09  &  $ 8.225{\times}10^{-6 }$ & 
$ 4.832{\pm}0.005 $  &   4.23  &  l \\
3--1~P36 & 4027.0438   &   4669.91  &  $ 7.863{\times}10^{-6 }$ & 
$ 4.974{\pm}0.013 $  &   4.21  &  l \\
\hline
{\bf 2H26 } & {\bf 4008.8637 } & {\bf  5089.47 } & {\bf 
$\mathbf{  8.349{\times}10^{-6 }}$} & {\bf $\mathbf{  4.701{\pm}0.004 }$} & {\bf  4.23} \\
\hline\\[3pt]
9--7~R29 & 3967.5315   &  16010.58  &  $ 1.177{\times}10^{-4 }$ & 
$ 2.349{\pm}0.009 $  &   4.14  &  r \\
9--7~R30 & 3968.8942   &  16117.92  &  $ 1.223{\times}10^{-4 }$ & 
$ 2.612{\pm}0.007 $  &   4.33  &  n \\
7--5~R61 & 4081.9755   &  17305.01  &  $ 1.724{\times}10^{-4 }$ & 
$ 2.587{\pm}0.009 $  &   4.28  &  l \\
7--5~R60 & 4082.9238   &  17086.66  &  $ 1.686{\times}10^{-4 }$ & 
$ 2.850{\pm}0.011 $  &   4.44  &  n \\
\hline
{\bf 2I26 } & {\bf 4025.4474 } & {\bf 16626.88 } & {\bf 
$ \mathbf{ 1.429{\times}10^{-4 }}$} & {\bf $\mathbf{  2.582{\pm}0.008 }$} & {\bf  4.30} \\
\hline\\[3pt]
7--5~R20 & 4057.6387   &  11221.85  &  $ 4.539{\times}10^{-5 }$ & 
$ 4.117{\pm}0.008 $  &   4.29  &  n \\
7--5~R22 & 4061.7586   &  11379.17  &  $ 5.031{\times}10^{-5 }$ & 
$ 4.375{\pm}0.016 $  &   4.38  &  l \\
7--5~R23 & 4063.7085   &  11463.29  &  $ 5.281{\times}10^{-5 }$ & 
$ 4.346{\pm}0.010 $  &   4.28  &  n \\
\hline
{\bf 2J26 } & {\bf 4061.1143 } & {\bf 11357.84 } & {\bf 
$\mathbf{  4.948{\times}10^{-5 }}$} & {\bf $\mathbf{  4.217{\pm}0.006} $} & {\bf  4.29} \\
\hline\\[3pt]
7--5~R29 & 4073.8593   &  12044.06  &  $ 6.842{\times}10^{-5 }$ & 
$ 4.921{\pm}0.021 $  &   4.54  &  r \\
7--5~R33 & 4079.1405   &  12503.42  &  $ 7.945{\times}10^{-5 }$ & 
$ 4.825{\pm}0.009 $  &   4.35  &  n \\
7--5~R37 & 4083.2217   &  13020.18  &  $ 9.100{\times}10^{-5 }$ & 
$ 4.831{\pm}0.009 $  &   4.37  &  n \\
\hline
{\bf 2K26 } & {\bf 4078.7026 } & {\bf 12518.27 } & {\bf 
$ \mathbf{ 7.899{\times}10^{-5 }}$} & {\bf $\mathbf{  4.816{\pm}0.012 }$} & {\bf  4.41} \\
\hline\\[3pt]
3--1~P25 & 4091.2978   &   3378.95  &  $ 5.768{\times}10^{-6 }$ & 
$ 5.746{\pm}0.011 $  &   4.34  &  n \\
3--1~P22 & 4107.4365   &   3105.64  &  $ 5.155{\times}10^{-6 }$ & 
$ 5.498{\pm}0.013 $  &   4.24  &  n \\
3--1~P18 & 4128.0192   &   2794.07  &  $ 4.309{\times}10^{-6 }$ & 
$ 5.280{\pm}0.012 $  &   4.26  &  l \\
3--1~P17 & 4132.9968   &   2725.63  &  $ 4.091{\times}10^{-6 }$ & 
$ 5.173{\pm}0.016 $  &   4.32  &  n \\
\hline
{\bf 2L26 } & {\bf 4114.0663 } & {\bf  3014.95 } & {\bf 
$\mathbf{  4.814{\times}10^{-6 }}$} & {\bf $\mathbf{  5.376{\pm}0.007} $} & {\bf  4.28} \\
\hline\\[3pt]
6--4~R72 & 4121.1779   &  17982.98  &  $ 1.557{\times}10^{-4 }$ & 
$ 1.969{\pm}0.008 $  &   4.27  &  n \\
6--4~R70 & 4124.6822   &  17470.95  &  $ 1.495{\times}10^{-4 }$ & 
$ 2.166{\pm}0.013 $  &   4.25  &  l \\
6--4~R67 & 4129.3348   &  16728.00  &  $ 1.404{\times}10^{-4 }$ & 
$ 2.508{\pm}0.016 $  &   4.30  &  l \\
\hline
{\bf 2M26 } & {\bf 4125.4780 } & {\bf 17331.22 } & {\bf 
$\mathbf{  1.466{\times}10^{-4 }}$} & {\bf $\mathbf{  2.199{\pm}0.006 }$} & {\bf  4.27} \\
\hline\\[3pt]
6--4~R26 & 4122.3991   &   9711.92  &  $ 4.304{\times}10^{-5 }$ & 
$ 5.943{\pm}0.022 $  &   4.30  &  l \\
6--4~R28 & 4125.7047   &   9914.68  &  $ 4.680{\times}10^{-5 }$ & 
$ 6.099{\pm}0.012 $  &   4.33  &  n \\
6--4~R30 & 4128.7133   &  10132.02  &  $ 5.063{\times}10^{-5 }$ & 
$ 6.116{\pm}0.008 $  &   4.30  &  n \\
6--4~R31 & 4130.1058   &  10246.15  &  $ 5.260{\times}10^{-5 }$ & 
$ 6.211{\pm}0.008 $  &   4.34  &  n \\
\hline
{\bf 2N26 } & {\bf 4126.7659 } & {\bf 10003.51 } & {\bf 
$\mathbf{  4.816{\times}10^{-5 }}$} & {\bf $\mathbf{  6.046{\pm}0.006}$} & {\bf  4.31} \\
\hline\\[3pt]
6--4~R36 & 4135.9451   &  10871.18  &  $ 6.273{\times}10^{-5 }$ & 
$ 6.106{\pm}0.008 $  &   4.34  &  n \\
6--4~R37 & 4136.8872   &  11007.03  &  $ 6.483{\times}10^{-5 }$ & 
$ 6.110{\pm}0.013 $  &   4.36  &  r \\
6--4~R39 & 4138.5446   &  11289.53  &  $ 6.909{\times}10^{-5 }$ & 
$ 5.957{\pm}0.010 $  &   4.33  &  n \\
6--4~R40 & 4139.2596   &  11436.17  &  $ 7.126{\times}10^{-5 }$ & 
$ 5.839{\pm}0.008 $  &   4.31  &  n \\
\hline
{\bf 2O26 } & {\bf 4137.6319 } & {\bf 11146.33 } & {\bf 
$\mathbf{  6.681{\times}10^{-5 }}$} & {\bf $\mathbf{  5.984{\pm}0.008 }$} & {\bf  4.34} \\
\hline\\[3pt]
5--3~R27 & 4177.5502   &   7760.68  &  $ 2.979{\times}10^{-5 }$ & 
$ 7.307{\pm}0.008 $  &   4.32  &  n \\
5--3~R28 & 4179.2008   &   7864.86  &  $ 3.105{\times}10^{-5 }$ & 
$ 7.494{\pm}0.017 $  &   4.39  &  l \\
5--3~R29 & 4180.7772   &   7972.73  &  $ 3.232{\times}10^{-5 }$ & 
$ 7.415{\pm}0.005 $  &   4.32  &  n \\
5--3~R30 & 4182.2791   &   8084.27  &  $ 3.361{\times}10^{-5 }$ & 
$ 7.483{\pm}0.008 $  &   4.33  &  l \\
5--3~R33 & 4186.3371   &   8440.90  &  $ 3.756{\times}10^{-5 }$ & 
$ 7.563{\pm}0.009 $  &   4.37  &  r \\
5--3~R35 & 4188.6677   &   8696.96  &  $ 4.028{\times}10^{-5 }$ & 
$ 7.272{\pm}0.007 $  &   4.31  &  n \\
\hline
{\bf 2P26 } & {\bf 4182.4733 } & {\bf  8136.71 } & {\bf 
$\mathbf{  3.391{\times}10^{-5 }}$} & {\bf $\mathbf{  7.398{\pm}0.006} $} & {\bf  4.34} \\
\hline\\[3pt]
5--3~R69 & 4181.2480   &  15240.46  &  $ 9.776{\times}10^{-5 }$ & 
$ 2.856{\pm}0.009 $  &   4.39  &  n \\
5--3~R68 & 4182.7632   &  14990.39  &  $ 9.573{\times}10^{-5 }$ & 
$ 2.963{\pm}0.008 $  &   4.29  &  n \\
5--3~R66 & 4185.5532   &  14500.46  &  $ 9.172{\times}10^{-5 }$ & 
$ 3.252{\pm}0.007 $  &   4.31  &  n \\
5--3~R65 & 4186.8282   &  14260.61  &  $ 8.976{\times}10^{-5 }$ & 
$ 3.487{\pm}0.020 $  &   4.37  &  l \\
\hline
{\bf 2Q26 } & {\bf 4184.2968 } & {\bf 14713.50 } & {\bf 
$\mathbf{  9.304{\times}10^{-5 }}$} & {\bf $\mathbf{  3.127{\pm}0.008 }$} & {\bf  4.34} \\
\hline\\[3pt]
4--2~R73 & 4229.6013   &  14278.71  &  $ 6.369{\times}10^{-5 }$ & 
$ 2.624{\pm}0.014 $  &   4.49  &  l \\
4--2~R72 & 4231.4032   &  14012.58  &  $ 6.240{\times}10^{-5 }$ & 
$ 2.617{\pm}0.007 $  &   4.23  &  l \\
4--2~R69 & 4236.3246   &  13234.62  &  $ 5.864{\times}10^{-5 }$ & 
$ 3.034{\pm}0.007 $  &   4.20  &  r \\
4--2~R68 & 4237.8042   &  12982.14  &  $ 5.742{\times}10^{-5 }$ & 
$ 3.417{\pm}0.018 $  &   4.43  &  l \\
\hline
{\bf 2R26 } & {\bf 4234.2155 } & {\bf 13559.28 } & {\bf 
$\mathbf{  5.967{\times}10^{-5 }}$} & {\bf $\mathbf{  2.900{\pm}0.008 }$} & {\bf  4.32} \\
\hline\\[3pt]
4--2~R27 & 4231.1521   &   5683.53  &  $ 1.777{\times}10^{-5 }$ & 
$ 8.344{\pm}0.013 $  &   4.35  &  l \\
4--2~R28 & 4232.8376   &   5788.69  &  $ 1.852{\times}10^{-5 }$ & 
$ 8.422{\pm}0.012 $  &   4.36  &  r \\
4--2~R31 & 4237.4479   &   6126.46  &  $ 2.083{\times}10^{-5 }$ & 
$ 8.760{\pm}0.014 $  &   4.43  &  n \\
4--2~R32 & 4238.8355   &   6246.46  &  $ 2.162{\times}10^{-5 }$ & 
$ 8.445{\pm}0.008 $  &   4.32  &  n \\
\hline
{\bf 2S26 } & {\bf 4235.0904 } & {\bf  5962.72 } & {\bf 
$\mathbf{  1.963{\times}10^{-5 }}$} & {\bf $\mathbf{  8.472{\pm}0.010 }$} & {\bf  4.37} \\
\hline\\[3pt]
4--2~R66 & 4240.5231   &  12487.48  &  $ 5.499{\times}10^{-5 }$ & 
$ 3.959{\pm}0.019 $  &   4.62  &  l \\
4--2~R65 & 4241.7626   &  12245.32  &  $ 5.381{\times}10^{-5 }$ & 
$ 3.920{\pm}0.014 $  &   4.40  &  r \\
4--2~R64 & 4242.9224   &  12006.63  &  $ 5.263{\times}10^{-5 }$ & 
$ 4.051{\pm}0.008 $  &   4.32  &  n \\
4--2~R63 & 4244.0026   &  11771.41  &  $ 5.147{\times}10^{-5 }$ & 
$ 4.180{\pm}0.008 $  &   4.28  &  n \\
4--2~R62 & 4245.0034   &  11539.67  &  $ 5.033{\times}10^{-5 }$ & 
$ 4.347{\pm}0.006 $  &   4.26  &  n \\
\hline
{\bf 2T26 } & {\bf 4242.9628 } & {\bf 11984.87 } & {\bf 
$ \mathbf{ 5.235{\times}10^{-5 }}$} & {\bf $\mathbf{  4.065{\pm}0.009} $} & {\bf  4.36} \\
\hline\\[3pt]
4--2~R35 & 4242.5487   &   6628.63  &  $ 2.405{\times}10^{-5 }$ & 
$ 8.591{\pm}0.015 $  &   4.39  &  l \\
4--2~R36 & 4243.6361   &   6763.39  &  $ 2.488{\times}10^{-5 }$ & 
$ 8.394{\pm}0.008 $  &   4.35  &  n \\
4--2~R37 & 4244.6480   &   6901.83  &  $ 2.571{\times}10^{-5 }$ & 
$ 8.295{\pm}0.008 $  &   4.34  &  l \\
4--2~R39 & 4246.4450   &   7189.72  &  $ 2.742{\times}10^{-5 }$ & 
$ 8.309{\pm}0.025 $  &   4.43  &  l \\
4--2~R40 & 4247.2298   &   7339.16  &  $ 2.830{\times}10^{-5 }$ & 
$ 7.966{\pm}0.025 $  &   4.31  &  r \\
4--2~R41 & 4247.9386   &   7492.25  &  $ 2.918{\times}10^{-5 }$ & 
$ 7.966{\pm}0.008 $  &   4.34  &  n \\
\hline
{\bf 2U26 } & {\bf 4245.3507 } & {\bf  7043.10 } & {\bf 
$\mathbf{  2.646{\times}10^{-5 }}$} & {\bf $\mathbf{  8.246{\pm}0.008 }$} & {\bf  4.37} \\
\hline\\[3pt]
3--1~R17 & 4263.6356   &   2725.63  &  $ 5.325{\times}10^{-6 }$ & 
$ 6.667{\pm}0.007 $  &   4.27  &  r \\
3--1~R18 & 4266.0904   &   2794.07  &  $ 5.657{\times}10^{-6 }$ & 
$ 6.900{\pm}0.011 $  &   4.29  &  n \\
3--1~R19 & 4268.4725   &   2866.29  &  $ 5.993{\times}10^{-6 }$ & 
$ 7.101{\pm}0.009 $  &   4.29  &  n \\
3--1~R20 & 4270.7816   &   2942.30  &  $ 6.332{\times}10^{-6 }$ & 
$ 7.308{\pm}0.011 $  &   4.28  &  n \\
3--1~R21 & 4273.0177   &   3022.08  &  $ 6.675{\times}10^{-6 }$ & 
$ 7.582{\pm}0.017 $  &   4.35  &  l \\
3--1~R22 & 4275.1805   &   3105.64  &  $ 7.025{\times}10^{-6 }$ & 
$ 7.641{\pm}0.007 $  &   4.31  &  n \\
3--1~R23 & 4277.2699   &   3192.98  &  $ 7.377{\times}10^{-6 }$ & 
$ 7.742{\pm}0.008 $  &   4.31  &  l \\
\hline
{\bf 2V26 } & {\bf 4270.9033 } & {\bf  2958.84 } & {\bf 
$\mathbf{  6.333{\times}10^{-6 }}$} & {\bf $\mathbf{  7.242{\pm}0.006} $} & {\bf  4.30} \\
\hline\\[3pt]
3--1~R82 & 4265.4063   &  14827.07  &  $ 3.812{\times}10^{-5 }$ & 
$ 1.379{\pm}0.008 $  &   4.36  &  n \\
3--1~R81 & 4267.9059   &  14528.02  &  $ 3.740{\times}10^{-5 }$ & 
$ 1.408{\pm}0.009 $  &   4.23  &  l \\
3--1~R80 & 4270.3235   &  14232.31  &  $ 3.669{\times}10^{-5 }$ & 
$ 1.409{\pm}0.006 $  &   4.03  &  r \\
3--1~R78 & 4274.9127   &  13650.91  &  $ 3.528{\times}10^{-5 }$ & 
$ 1.706{\pm}0.012 $  &   4.30  &  l \\
3--1~R77 & 4277.0847   &  13365.25  &  $ 3.459{\times}10^{-5 }$ & 
$ 1.767{\pm}0.029 $  &   4.13  &  r \\
\hline
{\bf 2W26 } & {\bf 4271.7245 } & {\bf 14039.34 } & {\bf 
$\mathbf{  3.574{\times}10^{-5 }}$} & {\bf $ \mathbf{ 1.515{\pm}0.006} $} & {\bf  4.19} \\
\hline\\[3pt]
3--1~R66 & 4295.6311   &  10448.09  &  $ 2.752{\times}10^{-5 }$ & 
$ 3.472{\pm}0.007 $  &   4.26  &  n \\
3--1~R65 & 4296.8351   &  10203.62  &  $ 2.692{\times}10^{-5 }$ & 
$ 3.859{\pm}0.010 $  &   4.28  &  n \\
3--1~R64 & 4297.9594   &   9962.65  &  $ 2.633{\times}10^{-5 }$ & 
$ 3.759{\pm}0.006 $  &   4.23  &  b \\
\hline
{\bf 2X26 } & {\bf 4296.8517 } & {\bf 10195.79 } & {\bf 
$\mathbf{  2.687{\times}10^{-5 }}$} & {\bf $ \mathbf{ 3.699{\pm}0.008 }$} & {\bf  4.27} \\
\hline\\[3pt]
2--0~R13 & 4306.4749   &    349.69  &  $ 1.331{\times}10^{-6 }$ & 
$ 3.975{\pm}0.008 $  &   4.25  &  n \\
2--0~R14 & 4309.2544   &    403.46  &  $ 1.436{\times}10^{-6 }$ & 
$ 4.265{\pm}0.016 $  &   4.35  &  l \\
2--0~R16 & 4314.5966   &    522.47  &  $ 1.648{\times}10^{-6 }$ & 
$ 4.638{\pm}0.007 $  &   4.27  &  n \\
2--0~R17 & 4317.1590   &    587.72  &  $ 1.757{\times}10^{-6 }$ & 
$ 4.774{\pm}0.013 $  &   4.30  &  r \\
\hline
{\bf 2Y26 } & {\bf 4312.2073 } & {\bf   473.25 } & {\bf 
$ \mathbf{ 1.542{\times}10^{-6 }}$} & {\bf $\mathbf{  4.387{\pm}0.008} $} & {\bf  4.29} \\
\hline\\[3pt]
2--0~R20 & 4324.4098   &    806.38  &  $ 2.089{\times}10^{-6 }$ & 
$ 5.294{\pm}0.009 $  &   4.30  &  n \\
2--0~R21 & 4326.6808   &    886.90  &  $ 2.203{\times}10^{-6 }$ & 
$ 5.450{\pm}0.015 $  &   4.36  &  r \\
2--0~R22 & 4328.8785   &    971.23  &  $ 2.318{\times}10^{-6 }$ & 
$ 5.422{\pm}0.006 $  &   4.27  &  n \\
2--0~R23 & 4331.0029   &   1059.37  &  $ 2.435{\times}10^{-6 }$ & 
$ 5.508{\pm}0.008 $  &   4.26  &  n \\
2--0~R24 & 4333.0537   &   1151.31  &  $ 2.554{\times}10^{-6 }$ & 
$ 5.647{\pm}0.013 $  &   4.29  &  l \\
2--0~R25 & 4335.0309   &   1247.05  &  $ 2.673{\times}10^{-6 }$ & 
$ 5.670{\pm}0.007 $  &   4.25  &  n \\
\hline
{\bf 2Z26 } & {\bf 4329.9746 } & {\bf  1025.69 } & {\bf 
$\mathbf{  2.376{\times}10^{-6 }}$} & {\bf $\mathbf{  5.478{\pm}0.006} $} & {\bf  4.29} \\
\hline\\[3pt]
\enddata
\tablecomments{~Line center frequencies, $\omega$, and lower level energies, $E_{\rm low}$, are from
G94;
$gf$-values are averages of G94 and HR96.  Uncertainties on equivalent widths, $W_{\omega}$, 
are 1\,$\sigma$, based on Lenz \& Ayres (1992), with an empirical estimate of the photometric noise.  Rightmost column (``Flag'') indicates how profile was trimmed before the coaddition
step: ``n''-- no trimming; ``l''-- left wing trimmed; ``r''-- right wing; ``b''-- both wings.}
\end{deluxetable}

\begin{deluxetable}{rccrccc}
\tablenum{2}
\tablecaption{Isotopomer Hybrid Sample}
\tablewidth{0pt}
\tablecolumns{7}
\tablehead{
\colhead{Transition} &  \colhead{$\omega$} & \colhead{$E_{\rm low}$}  &\colhead{$gf$} & \colhead{$W_{\omega}$} & 
 \colhead{FWHM}  & \colhead{Flag} \\[3pt]                
\colhead{} &  \colhead{(cm$^{-1}$)} & \colhead{(cm$^{-1}$)}  & \colhead{} &  \colhead{($10^{-3}$ cm$^{-1}$)}& 
\colhead{(km~s$^{-1}$)}  
& \colhead{}}                
\startdata   
\cutinhead{$^{13}$C$^{16}$O}
2--1~P53 & 1835.8986   &   7263.82  &  $ 1.035{\times}10^{-3 }$ & 
$ 0.919{\pm}0.008 $  &   3.97  &  l \\
2--1~P52 & 1841.0593   &   7074.05  &  $ 1.017{\times}10^{-3 }$ & 
$ 1.013{\pm}0.007 $  &   4.21  &  n \\
2--1~P51 & 1846.1940   &   6887.74  &  $ 9.999{\times}10^{-4 }$ & 
$ 0.986{\pm}0.006 $  &   4.02  &  n \\
\hline
{\bf 1A36 } & {\bf 1841.2046 } & {\bf  7069.78 } & {\bf 
$\mathbf{  1.016{\times}10^{-3 }}$} & {\bf $\mathbf{  0.968{\pm}0.003 }$} & {\bf  4.14} \\
\hline\\[3pt]
5--4~P36 & 1847.9960   &  10583.74  &  $ 1.745{\times}10^{-3 }$ & 
$ 0.589{\pm}0.007 $  &   4.08  &  n \\
5--4~P32 & 1866.2495   &  10098.20  &  $ 1.564{\times}10^{-3 }$ & 
$ 0.613{\pm}0.011 $  &   4.32  &  r \\
5--4~P29 & 1879.6415   &   9770.51  &  $ 1.425{\times}10^{-3 }$ & 
$ 0.650{\pm}0.006 $  &   4.18  &  l \\
5--4~P28 & 1884.0480   &   9668.24  &  $ 1.379{\times}10^{-3 }$ & 
$ 0.631{\pm}0.006 $  &   4.08  &  n \\
\hline
{\bf 1B36 } & {\bf 1869.8756 } & {\bf 10022.02 } & {\bf 
$\mathbf{  1.518{\times}10^{-3 }}$} & {\bf $\mathbf{  0.613{\pm}0.004 }$} & {\bf  4.14} \\
\hline\\[3pt]
2--1~P45 & 1876.4530   &   5842.88  &  $ 8.935{\times}10^{-4 }$ & 
$ 1.299{\pm}0.010 $  &   4.16  &  r \\
2--1~P44 & 1881.4034   &   5680.97  &  $ 8.756{\times}10^{-4 }$ & 
$ 1.272{\pm}0.012 $  &   4.03  &  l \\
\hline
{\bf 1C36 } & {\bf 1878.9695 } & {\bf  5760.63 } & {\bf 
$\mathbf{  8.842{\times}10^{-4 }}$} & {\bf $\mathbf{  1.279{\pm}0.005 }$} & {\bf  4.14} \\
\hline\\[3pt]
2--1~P40 & 1900.9348   &   5068.49  &  $ 8.025{\times}10^{-4 }$ & 
$ 1.508{\pm}0.009 $  &   4.25  &  r \\
2--1~P37 & 1915.2963   &   4646.21  &  $ 7.468{\times}10^{-4 }$ & 
$ 1.542{\pm}0.011 $  &   4.06  &  l \\
\hline
{\bf 1D36 } & {\bf 1908.3619 } & {\bf  4850.57 } & {\bf 
$\mathbf{  7.729{\times}10^{-4 }}$} & {\bf $\mathbf{  1.514{\pm}0.006} $} & {\bf  4.14} \\
\hline\\[3pt]
4--3~P30 & 1899.3493   &   7871.16  &  $ 1.195{\times}10^{-3 }$ & 
$ 0.934{\pm}0.007 $  &   4.04  &  l \\
4--3~P26 & 1917.0482   &   7465.37  &  $ 1.044{\times}10^{-3 }$ & 
$ 0.936{\pm}0.010 $  &   4.01  &  r \\
4--3~P23 & 1930.0159   &   7198.08  &  $ 9.289{\times}10^{-4 }$ & 
$ 0.947{\pm}0.006 $  &   4.07  &  n \\
4--3~P22 & 1934.2794   &   7116.07  &  $ 8.902{\times}10^{-4 }$ & 
$ 0.957{\pm}0.010 $  &   4.18  &  n \\
\hline
{\bf 1E36 } & {\bf 1919.9709 } & {\bf  7418.37 } & {\bf 
$\mathbf{  1.010{\times}10^{-3 }}$} & {\bf $\mathbf{  0.940{\pm}0.003} $} & {\bf  4.14} \\
\hline\\[3pt]
2--1~P34 & 1929.4078   &   4255.87  &  $ 6.907{\times}10^{-4 }$ & 
$ 1.646{\pm}0.010 $  &   4.09  &  l \\
2--1~P33 & 1934.0555   &   4132.88  &  $ 6.718{\times}10^{-4 }$ & 
$ 1.654{\pm}0.008 $  &   4.05  &  n \\
2--1~P30 & 1947.8280   &   3785.33  &  $ 6.145{\times}10^{-4 }$ & 
$ 1.707{\pm}0.009 $  &   4.14  &  n \\
2--1~P28 & 1956.8665   &   3571.53  &  $ 5.759{\times}10^{-4 }$ & 
$ 1.787{\pm}0.009 $  &   4.19  &  n \\
2--1~P27 & 1961.3423   &   3470.01  &  $ 5.564{\times}10^{-4 }$ & 
$ 1.719{\pm}0.010 $  &   4.14  &  n \\
\hline
{\bf 1F36 } & {\bf 1946.1313 } & {\bf  3838.53 } & {\bf 
$\mathbf{  6.188{\times}10^{-4 }}$} & {\bf $\mathbf{  1.693{\pm}0.005 }$} & {\bf  4.14} \\
\hline\\[3pt]
5--4~R27 & 2080.6760   &   9569.47  &  $ 1.522{\times}10^{-3 }$ & 
$ 0.853{\pm}0.007 $  &   4.09  &  l \\
5--4~R30 & 2088.1925   &   9876.26  &  $ 1.692{\times}10^{-3 }$ & 
$ 0.863{\pm}0.009 $  &   4.07  &  r \\
5--4~R35 & 2099.9808   &  10457.16  &  $ 1.979{\times}10^{-3 }$ & 
$ 0.886{\pm}0.011 $  &   4.26  &  l \\
5--4~R37 & 2104.4344   &  10713.78  &  $ 2.095{\times}10^{-3 }$ & 
$ 0.865{\pm}0.008 $  &   4.17  &  l \\
\hline
{\bf 1G36 } & {\bf 2093.1762 } & {\bf 10145.43 } & {\bf 
$\mathbf{  1.803{\times}10^{-3 }}$} & {\bf $\mathbf{  0.850{\pm}0.004} $} & {\bf  4.14} \\
\hline\\[3pt]
1--0~R12 & 2140.8278   &    286.58  &  $ 1.469{\times}10^{-4 }$ & 
$ 1.498{\pm}0.036 $  &   3.99  &  r \\
1--0~R14 & 2147.2045   &    385.72  &  $ 1.702{\times}10^{-4 }$ & 
$ 1.883{\pm}0.010 $  &   4.27  &  l \\
1--0~R17 & 2156.5089   &    561.89  &  $ 2.053{\times}10^{-4 }$ & 
$ 1.938{\pm}0.029 $  &   4.17  &  l \\
1--0~R18 & 2159.5403   &    627.93  &  $ 2.171{\times}10^{-4 }$ & 
$ 1.928{\pm}0.042 $  &   4.06  &  l \\
\hline
{\bf 1H36 } & {\bf 2151.8492 } & {\bf   480.27 } & {\bf 
$ \mathbf{ 1.847{\times}10^{-4 }}$} & {\bf $\mathbf{  1.804{\pm}0.005}$} & {\bf  4.14} \\
\hline\\[3pt]
1--0~R42 & 2221.4090   &   3301.16  &  $ 5.117{\times}10^{-4 }$ & 
$ 2.117{\pm}0.009 $  &   4.15  &  n \\
1--0~R43 & 2223.5190   &   3457.45  &  $ 5.245{\times}10^{-4 }$ & 
$ 2.109{\pm}0.020 $  &   4.18  &  r \\
1--0~R44 & 2225.5905   &   3617.29  &  $ 5.373{\times}10^{-4 }$ & 
$ 2.047{\pm}0.013 $  &   4.15  &  r \\
\hline
{\bf 1I36 } & {\bf 2223.4707 } & {\bf  3455.82 } & {\bf 
$\mathbf{  5.241{\times}10^{-4 }}$} & {\bf $\mathbf{  2.077{\pm}0.010 }$} & {\bf  4.14} \\
\hline\\[3pt]
\cutinhead{$^{12}$C$^{18}$O}
3--2~P53 & 1809.5808   &   9260.84  &  $ 1.527{\times}10^{-3 }$ & 
$ 0.167{\pm}0.006 $  &   4.27  &  r \\
3--2~P50 & 1824.8215   &   8709.06  &  $ 1.449{\times}10^{-3 }$ & 
$ 0.181{\pm}0.006 $  &   4.36  &  n \\
3--2~P49 & 1829.8500   &   8531.98  &  $ 1.424{\times}10^{-3 }$ & 
$ 0.178{\pm}0.006 $  &   4.86  &  l \\
4--3~P42 & 1840.5850   &   9401.32  &  $ 1.628{\times}10^{-3 }$ & 
$ 0.152{\pm}0.008 $  &   3.74  &  n \\
4--3~P36 & 1868.8777   &   8564.81  &  $ 1.412{\times}10^{-3 }$ & 
$ 0.169{\pm}0.011 $  &   4.12  &  l \\
\hline
{\bf 1A28 } & {\bf 1835.4927 } & {\bf  8869.81 } & {\bf 
$\mathbf{  1.478{\times}10^{-3 }}$} & {\bf $\mathbf{  0.162{\pm}0.002 }$} & {\bf  4.14} \\
\hline\\[3pt]
1--0~P15 & 2033.8515   &    439.11  &  $ 1.600{\times}10^{-4 }$ & 
$ 0.300{\pm}0.009 $  &   4.36  &  l \\
1--0~P11 & 2050.0805   &    241.59  &  $ 1.182{\times}10^{-4 }$ & 
$ 0.238{\pm}0.005 $  &   4.16  &  n \\
1--0~R9 & 2126.9299   &    164.74  &  $ 1.120{\times}10^{-4 }$ & 
$ 0.237{\pm}0.005 $  &   3.80  &  r \\
1--0~R11 & 2133.4895   &    241.59  &  $ 1.348{\times}10^{-4 }$ & 
$ 0.332{\pm}0.005 $  &   4.49  &  r \\
\hline
{\bf 1B28 } & {\bf 2084.7945 } & {\bf   281.80 } & {\bf 
$\mathbf{  1.312{\times}10^{-4 }}$} & {\bf $\mathbf{  0.268{\pm}0.002} $} & {\bf  4.14} \\
\hline\\[3pt]
1--0~R22 & 2167.0986   &    925.05  &  $ 2.637{\times}10^{-4 }$ & 
$ 0.453{\pm}0.006 $  &   3.91  &  b \\
1--0~R26 & 2178.2619   &   1282.61  &  $ 3.117{\times}10^{-4 }$ & 
$ 0.512{\pm}0.005 $  &   4.07  &  r \\
1--0~R29 & 2186.2556   &   1588.75  &  $ 3.481{\times}10^{-4 }$ & 
$ 0.449{\pm}0.010 $  &   3.82  &  l \\
\hline
{\bf 1C28 } & {\bf 2177.4391 } & {\bf  1272.49 } & {\bf 
$\mathbf{  3.067{\times}10^{-4 }}$} & {\bf $\mathbf{  0.474{\pm}0.004 }$} & {\bf  4.14} \\
\hline\\[3pt]
2--1~R37 & 2179.5029   &   4632.65  &  $ 8.799{\times}10^{-4 }$ & 
$ 0.456{\pm}0.012 $  &   4.15  &  r \\
1--0~R48 & 2229.1062   &   4275.82  &  $ 5.864{\times}10^{-4 }$ & 
$ 0.376{\pm}0.005 $  &   4.19  &  n \\
1--0~R49 & 2230.9793   &   4452.65  &  $ 5.994{\times}10^{-4 }$ & 
$ 0.384{\pm}0.005 $  &   4.33  &  n \\
\hline
{\bf 1D28 } & {\bf 2209.7509 } & {\bf  4472.02 } & {\bf 
$\mathbf{  6.878{\times}10^{-4 }}$} & {\bf $\mathbf{  0.391{\pm}0.002} $} & {\bf  4.14} \\
\hline\\[3pt]
\cutinhead{$^{12}$C$^{17}$O}
1--0~P42 & 1931.5996   &   3365.44  &  $ 4.333{\times}10^{-4 }$ & 
$ 0.056{\pm}0.048 $  &   4.38  &  r \\
1--0~P35 & 1966.0712   &   2351.97  &  $ 3.664{\times}10^{-4 }$ & 
$ 0.082{\pm}0.007 $  &   6.71  &  n \\
1--0~P34 & 1970.8816   &   2221.79  &  $ 3.569{\times}10^{-4 }$ & 
$ 0.072{\pm}0.008 $  &   3.49  &  l \\
\hline
{\bf 1A27 } & {\bf 1957.6087 } & {\bf  2607.09 } & {\bf 
$ \mathbf{ 3.804{\times}10^{-4 }}$} & {\bf $\mathbf{  0.058{\pm}0.002} $} & {\bf  4.14} \\
\hline\\[3pt]
1--0~P50 & 1890.5340   &   4740.88  &  $ 5.074{\times}10^{-4 }$ & 
$ 0.087{\pm}0.006 $  &   4.88  &  n \\
1--0~P49 & 1895.7624   &   4556.38  &  $ 4.981{\times}10^{-4 }$ & 
$ 0.027{\pm}0.005 $  &   2.42  &  n \\
2--1~P37 & 1931.8009   &   4715.92  &  $ 7.615{\times}10^{-4 }$ & 
$ 0.055{\pm}0.007 $  &   4.02  &  l \\
2--1~P34 & 1946.2171   &   4318.04  &  $ 7.040{\times}10^{-4 }$ & 
$ 0.085{\pm}0.004 $  &   3.56  &  n \\
2--1~P33 & 1950.9647   &   4192.67  &  $ 6.850{\times}10^{-4 }$ & 
$ 0.068{\pm}0.010 $  &   2.53  &  r \\
3--2~P20 & 1984.9124   &   4978.67  &  $ 6.301{\times}10^{-4 }$ & 
$ 0.049{\pm}0.005 $  &   3.49  &  l \\
3--2~P16 & 2001.9015   &   4706.90  &  $ 5.081{\times}10^{-4 }$ & 
$ 0.046{\pm}0.006 $  &   4.38  &  l \\
2--1~R37 & 2205.3932   &   4715.92  &  $ 9.013{\times}10^{-4 }$ & 
$ 0.063{\pm}0.005 $  &   2.70  &  l \\
\hline
{\bf 1B27 } & {\bf 1991.2737 } & {\bf  4592.26 } & {\bf 
$\mathbf{  6.475{\times}10^{-4 }}$} & {\bf $\mathbf{  0.064{\pm}0.002} $} & {\bf  4.14} \\
\hline\\[3pt]
2--1~P47 & 1881.9118   &   6276.42  &  $ 9.470{\times}10^{-4 }$ & 
$ 0.111{\pm}0.011 $  &   6.46  &  l \\
2--1~P42 & 1907.2048   &   5451.32  &  $ 8.554{\times}10^{-4 }$ & 
$ 0.038{\pm}0.005 $  &   2.31  &  l \\
3--2~P33 & 1926.3298   &   6264.26  &  $ 1.013{\times}10^{-3 }$ & 
$ 0.080{\pm}0.007 $  &   8.32  &  r \\
3--2~P21 & 1980.5875   &   5055.74  &  $ 6.603{\times}10^{-4 }$ & 
$ 0.038{\pm}0.005 $  &   2.30  &  n \\
2--1~R42 & 2216.4885   &   5451.32  &  $ 1.028{\times}10^{-3 }$ & 
$ 0.130{\pm}0.011 $  &   5.24  &  b \\
\hline
{\bf 1C27 } & {\bf 2002.3517 } & {\bf  5677.56 } & {\bf 
$\mathbf{  8.903{\times}10^{-4 }}$} & {\bf $\mathbf{  0.058{\pm}0.002} $} & {\bf  4.14} \\
\hline\\[3pt]
1--0~P29 & 1994.4960   &   1625.94  &  $ 3.076{\times}10^{-4 }$ & 
$ 0.066{\pm}0.007 $  &   3.60  &  b \\
1--0~P24 & 2017.3692   &   1122.28  &  $ 2.572{\times}10^{-4 }$ & 
$ 0.120{\pm}0.005 $  &   5.44  &  l \\
1--0~P21 & 2030.7301   &    864.53  &  $ 2.265{\times}10^{-4 }$ & 
$ 0.073{\pm}0.005 $  &   3.67  &  n \\
1--0~P15 & 2056.6177   &    449.41  &  $ 1.638{\times}10^{-4 }$ & 
$ 0.097{\pm}0.026 $  &   4.88  &  l \\
1--0~R7 & 2145.0538   &    104.92  &  $ 9.140{\times}10^{-5 }$ & 
$ 0.065{\pm}0.005 $  &   5.48  &  l \\
1--0~R9 & 2151.8983   &    168.60  &  $ 1.147{\times}10^{-4 }$ & 
$ 0.067{\pm}0.006 $  &   4.25  &  l \\
\hline
{\bf 1D27 } & {\bf 2052.3259 } & {\bf   844.97 } & {\bf 
$\mathbf{  1.907{\times}10^{-4 }}$} & {\bf $\mathbf{  0.067{\pm}0.003} $} & {\bf  4.14} \\
\hline\\[3pt]
\enddata
\tablecomments{~Line center frequencies, $\omega$, and lower level energies, $E_{\rm low}$, are from
G94;
$gf$-values are averages of G94 and HR96.  Uncertainties on equivalent widths, $W_{\omega}$, 
are 1\,$\sigma$, based on Lenz \& Ayres (1992), with an empirical estimate of the photometric noise.  Rightmost column (``Flag'') indicates how profile was trimmed before the coaddition
step: ``n''-- no trimming; ``l''-- left wing trimmed; ``r''-- right wing; ``b''-- both wings.  A uniform FWHM= 4.14~km s$^{-1}$ was imposed in the Gaussian fitting procedure.}
\end{deluxetable}

\begin{deluxetable}{crlcllll}
\rotate
\tablenum{3}
\tablecaption{Oxygen Abundances and Isotopic Ratios for Full 3D Model}
\tablewidth{0pt}
\tablecolumns{8}
\tablehead{
\colhead{Scenario} & \colhead{$\Delta{T}$} &  \colhead{$\epsilon_{\rm O}$} &  \colhead{$c_{1}$} & \colhead{$\rho$}  &\colhead{$R_{23}$} & \colhead{$R_{68}$} &  
\colhead{$R_{67}$}  }                
\startdata   
\cutinhead{Average $f$-values}
B  &        0  & 572 ${\pm}$  6 &  $+7$  &  0.972 ${\pm}$ 0.013 & 88.9 ${\pm}$ 0.7\,(0.2) & 493 ${\pm}$  7\,( 3) & 2642 ${\pm}$ 204\,(102) \\
G  &       +34 & 603 ${\pm}$  6 &  $-1$  &  1.000 ${\pm}$ 0.011 & 91.4 ${\pm}$ 0.7\,(0.2) & 511 ${\pm}$  9\,( 4) & 2738 ${\pm}$ 209\,(104) \\   
M  &       +90 & 644 ${\pm}$  9 & $-15$  &  1.050 ${\pm}$ 0.017 & 96.2 ${\pm}$ 0.9\,(0.3) & 544 ${\pm}$ 10\,( 5) & 2915 ${\pm}$ 203\,(101) \\[3pt]
\cutinhead{G94 $f$-values}
G  &       +17 &  587 ${\pm}$ 7  &  $+7$  &  1.000 ${\pm}$ 0.014 & 89.3 ${\pm}$ 0.9\,(0.3) & 496 ${\pm}$  7\,( 3) & 2657 ${\pm}$ 209\,(104) \\[3pt]
\cutinhead{HR96 $f$-values}
G  &       +51 &  620 ${\pm}$ 7  &  $-10$  &  1.000 ${\pm}$ 0.010 & 93.7 ${\pm}$ 0.8\,(0.3) & 528 ${\pm}$  11\,( 5) & 2829 ${\pm}$ 210\,(105) 
\enddata
\tablecomments{~``Scenario'' corresponds to one of three temperature correction schemes,
parametrized by $\Delta{T}$ (in K): ``B''= baseline; ``G''= Goldilocks; and ``M''= MAX.  Abundance gradient, $c_{1}= d\,\epsilon_{\rm O}\,/\,d\,E_{\rm low}$, is in units of
ppm per $10^4$ cm$^{-1}$.  In $\epsilon_{\rm O}$ (ppm) and $c_{1}$ columns, values refer to $\Delta v=2$ parent sample;
corresponding $\Delta v=1$ values can be obtained by multiplying by $\rho$ factor.  In $\epsilon_{\rm O}$ and $\rho$ columns, cited uncertainties are 1\,$\sigma$ dispersions.  In $R_{\rm ISO}$ columns,
first uncertainty is a 1\,$\sigma$ dispersion, second is 1\,s.e.}
\end{deluxetable}

\begin{figure}
\figurenum{1}
\caption[]{{\em (a)}\/ Comparison of two independent oscillator strength scales used in this study: G94 (solid curves) and HR96
(small crosses).  The $f$-values display smooth behavior with $J_{\rm low}$ (negative values refer to P branch, and positive to R), and systematically
increase with $v_{\rm low}$.  Truncation of higher $\Delta v = 1$ bands is due to $\omega> 1600$~cm$^{-1}$ cutoff
applied to the line samples, corresponding to highest quality ATMOS spectra.  {\em (b)}\/ Similar to {\em (a)}\/ but showing ratios of G94 to HR96 
oscillator strengths on a magnified
scale.  Slight ``noise'' in the ratios is due to finite precision in tabular values of G94 oscillator strengths;
HR96 values were calculated from smooth analytic functions.  G94 fundamental bands ($\Delta v = 1$) are systematically lower by a few percent, without much dependence on
$v_{\rm low}$ or $J_{\rm low}$.  In contrast, G94 overtone bands mostly are higher, and the deviations depend strongly on both $v_{\rm low}$ and $J_{\rm low}$.
{\em (c)}\/ Similar to {\em (b)}\/ but showing ratios of isotopomer 36 to parent 26 for both oscillator strength
scales.  Offsets are identical for G94 and HR96, and nearly independent of $v_{\rm low}$ and $J_{\rm low}$.  Similar behavior is seen for the other isotopomers: 28 systematically lower than 26 by about 4\%, 27 lower than 26 by about 2\%.}
\end{figure}

\clearpage
\begin{figure}
\figurenum{1a}
\hskip  -5mm
\includegraphics[scale=0.75,angle=0]{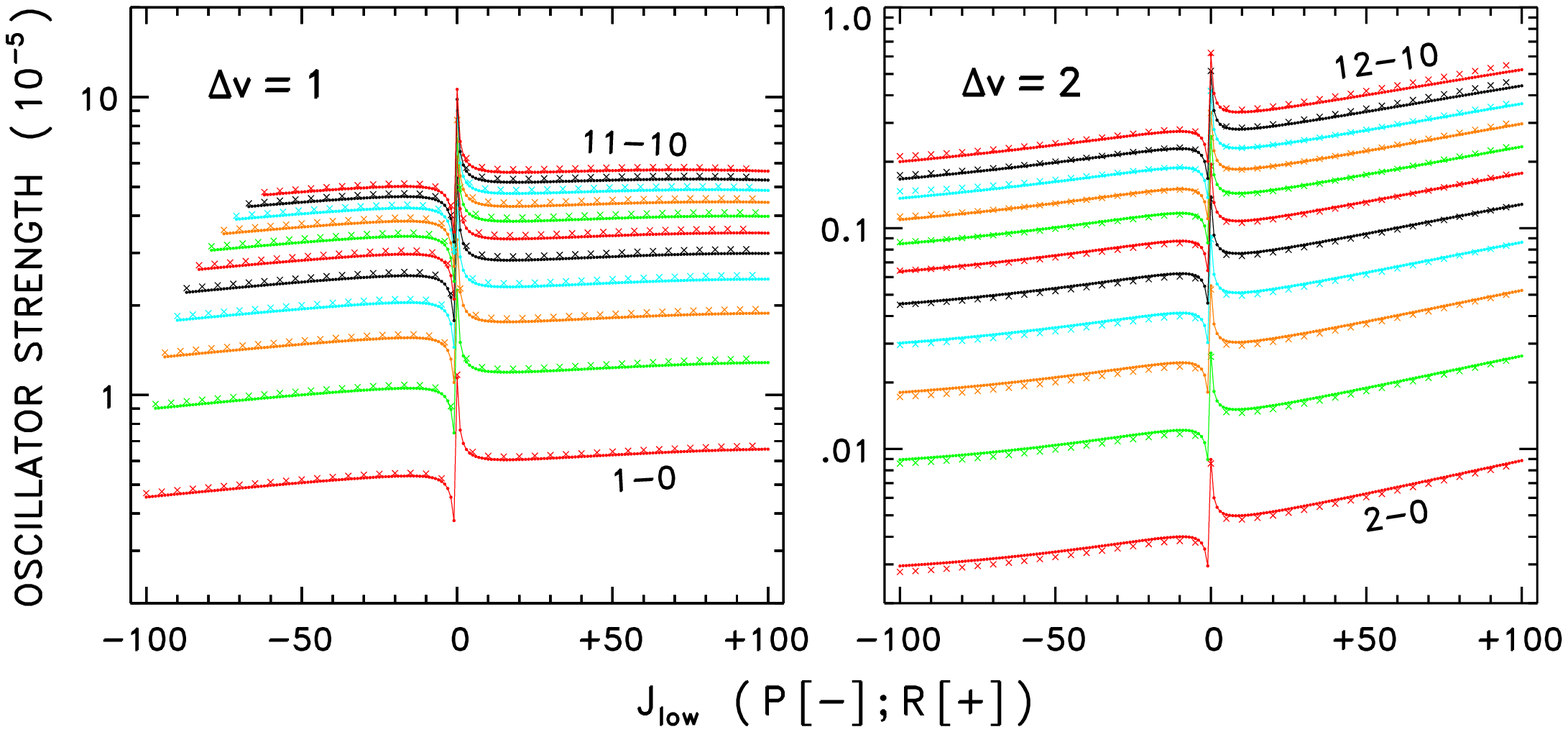} 
\caption[]{}
\end{figure}

\begin{figure}
\figurenum{1b}
\hskip  -5mm
\includegraphics[scale=0.75,angle=0]{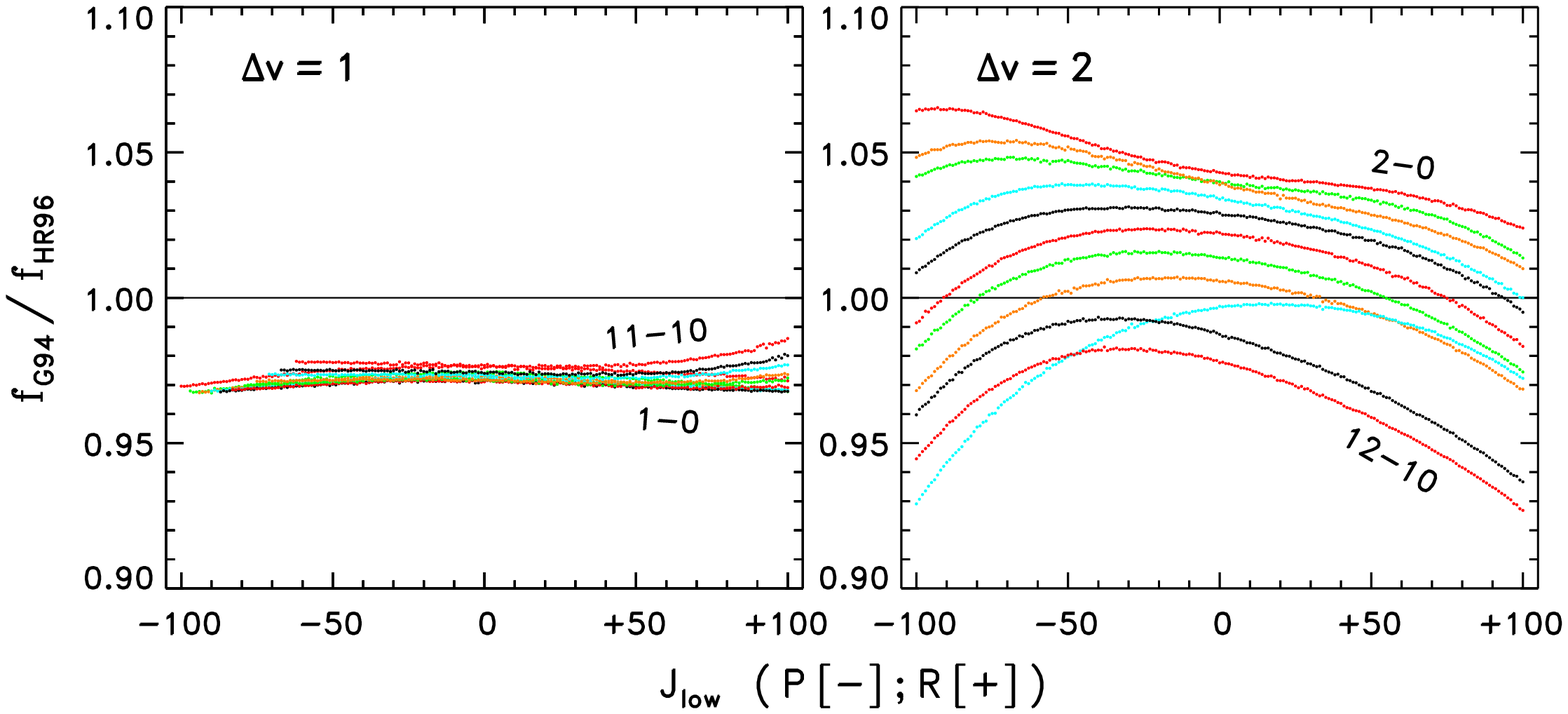} 
\caption[]{}
\end{figure}

\begin{figure}
\figurenum{1c}
\hskip  -5mm
\includegraphics[scale=0.75,angle=0]{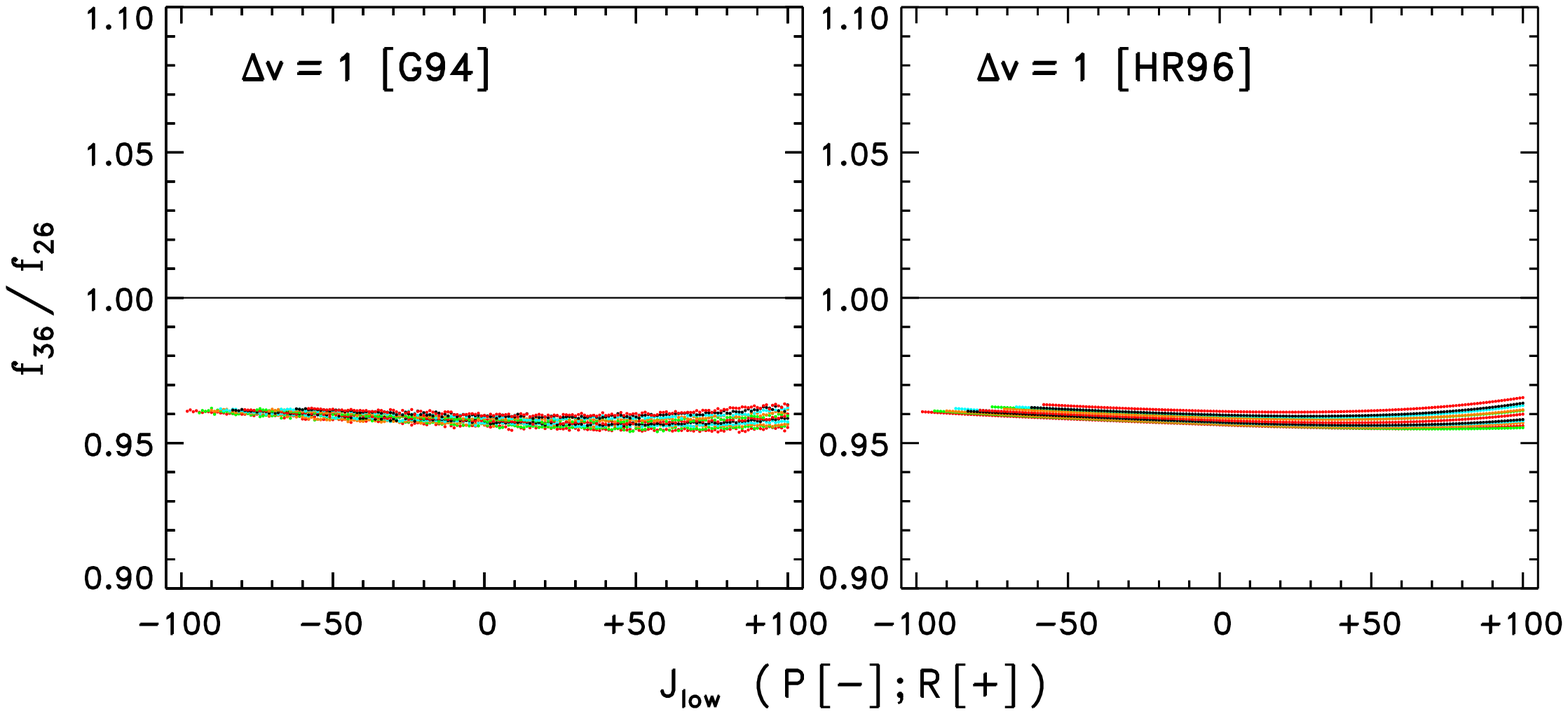} 
\caption[]{}
\end{figure}

\clearpage
\begin{figure}
\figurenum{2}
\caption[]{{\em (a)}\/ Hybridization of $^{12}$C$^{16}$O rovibrational lines from ATMOS spectra, for four representative $\Delta v = 1$
transitions.  In upper right hand corner
of each panel, leading numerical designation in blue is for $\Delta v$ (1= fundamental; 2= first overtone) followed by letter indicating the transition (cross-referenced to Table~1).  Trailing number in red is the isotopomer designation.  Below the hybrid
line specification is a list of individual transitions that were coadded.  These are depicted in the spectral sub-panel
at left by thin curves.  If colored red, that part of the line profile was ignored in the coaddition.  Sum is indicated by solid dots, and underlying thick blue curve is a similar coaddition of isotopomer 26 synthetic spectra calculated from
a 1D model but with line oscillator strengths adjusted to compensate for the large gradient in $\epsilon_{\rm O}$ with
$E_{\rm low}$ found with such models.  The synthetic curves were combined without trimming to illustrate 
corruption to the hybrid line shape that could have occurred.  In the lower parts of the panels, temperature-dependent factors for the line opacities (not including $n_{\rm CO}$) are shown by thin curves (of various colors), and the average at discrete temperatures by large blue dots.  Mean of the G94 and HR96 $f$-values was used for the input line strengths.  Red dot-dashed curve is a fit
that optimizes values of $E_{\rm low}$ and $gf$, for the weighted transition frequency, $<\omega>$.  {\em (b)}\/ Same as {\em (a)}\/ but for selected $\Delta v = 2$
transitions.  {\em (c)}\/ Same as {\em (a)}\/ but for selected $\Delta v = 1$ $^{13}$C$^{16}$O 
transitions.  Thick green curves are synthetic spectra for this isotopomer (36), and (wing) blends due to isotopomer 26 (blue) are present in some cases.  Negative
slope for low-excitation transition 1H36 is caused by the stimulated emission factor.  {\em (d)}\/ Same as {\em (a)}\/ for $\Delta v = 1$ $^{12}$C$^{18}$O 
transitions. Thick red curves are synthetic spectra for this isotopomer (28), and (wing) contributions
from other isotopomers can be seen as well.  {\em (e)}\/ Same as {\em (a)}\/ for $\Delta v = 1$ $^{12}$C$^{17}$O 
transitions. Thick orange curves are synthetic spectra for this isotopomer (27).  Blending by other
isotopomers is particularly conspicuous since the $^{12}$C$^{17}$O line depths are only a few tenths of a
percent at best.}
\end{figure}

\clearpage
\begin{figure}
\figurenum{2a}
\hskip  -10mm
\includegraphics[scale=0.800,angle=90]{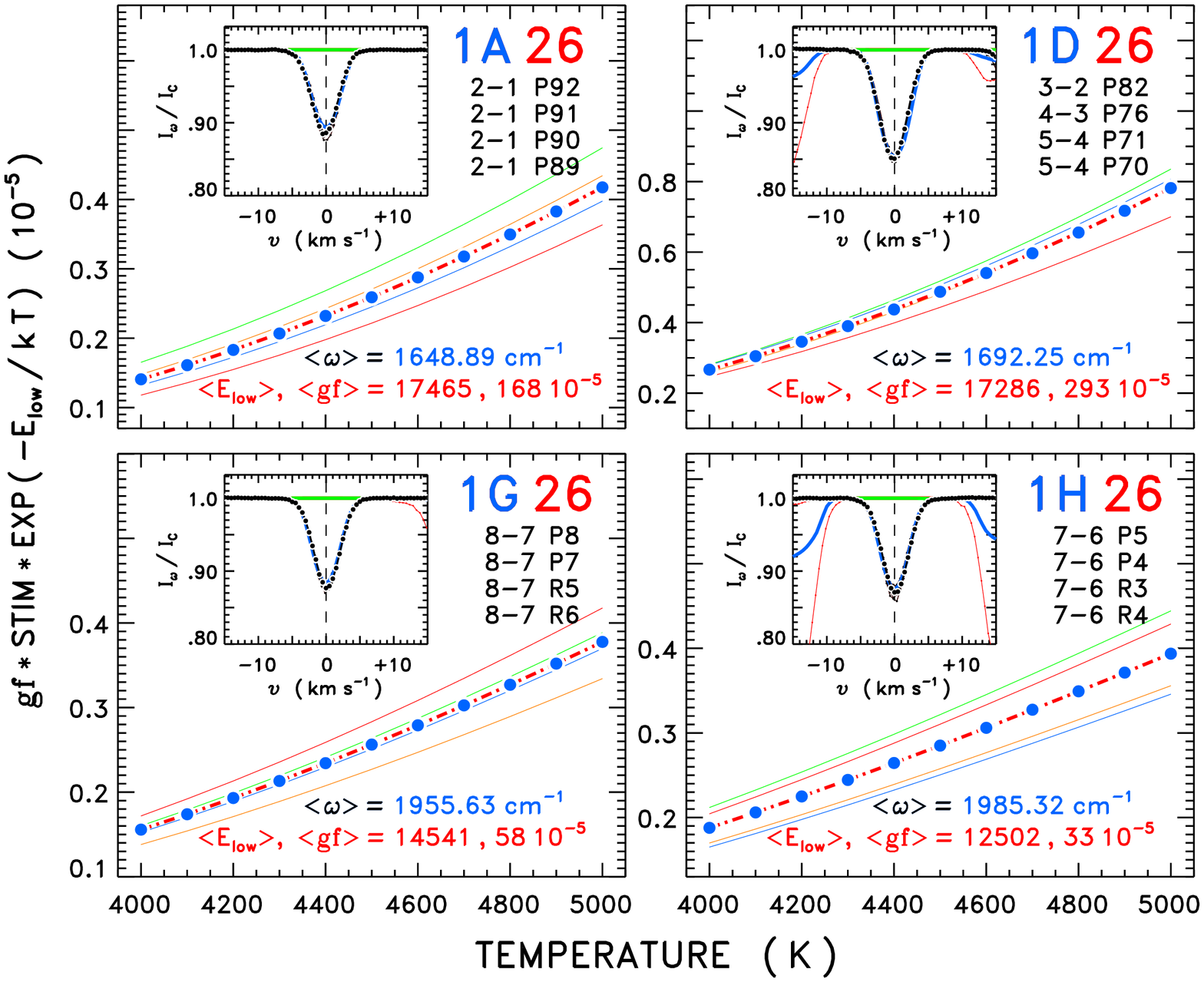} 
\caption[]{}
\end{figure}

\clearpage
\begin{figure}
\figurenum{2b}
\hskip  -10mm
\includegraphics[scale=0.800,angle=90]{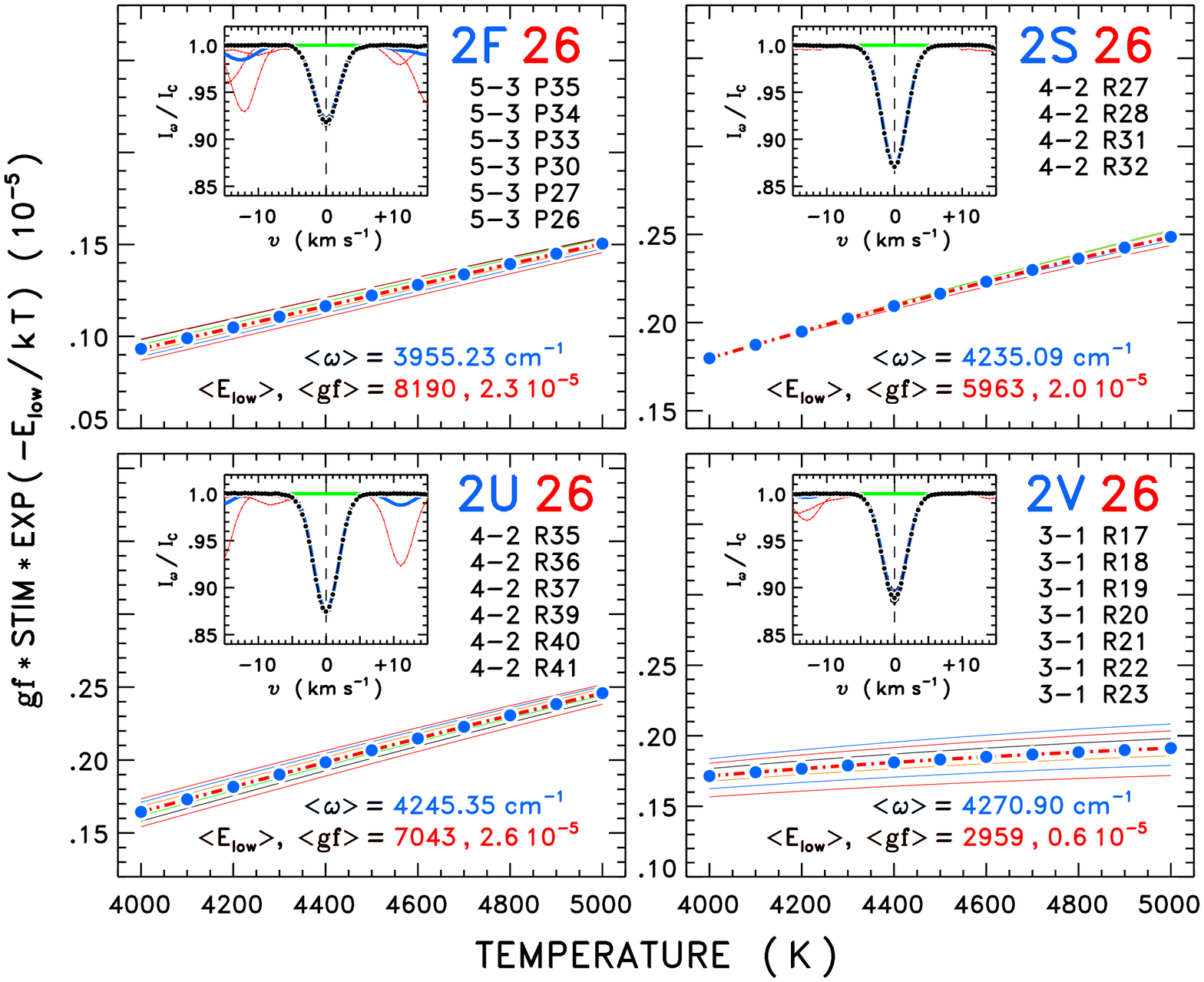} 
\caption[]{}
\end{figure}

\clearpage
\begin{figure}
\figurenum{2c}
\hskip  -10mm
\includegraphics[scale=0.800,angle=90]{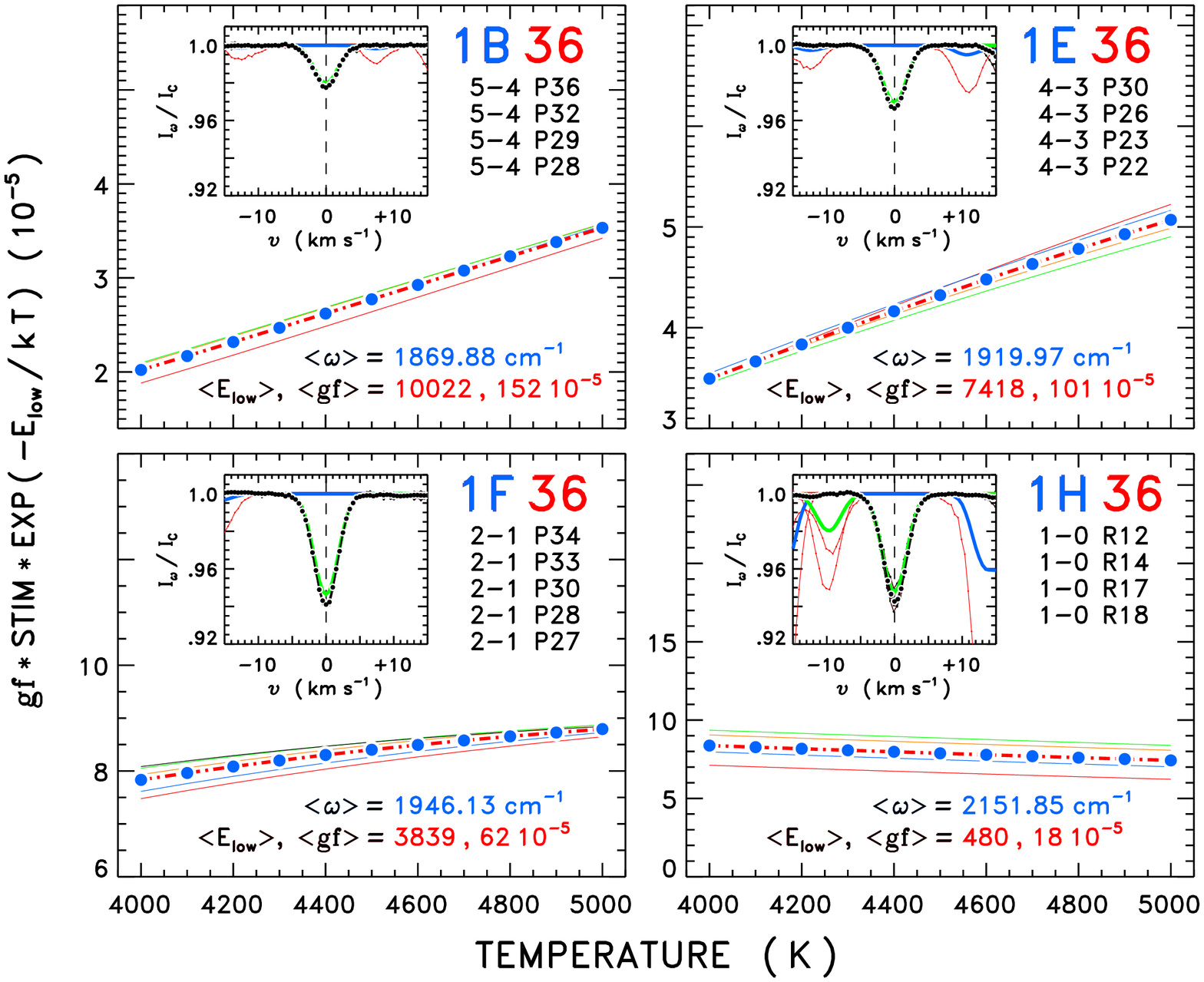} 
\caption[]{}
\end{figure}

\clearpage
\begin{figure}
\figurenum{2d}
\hskip  -10mm
\includegraphics[scale=0.800,angle=90]{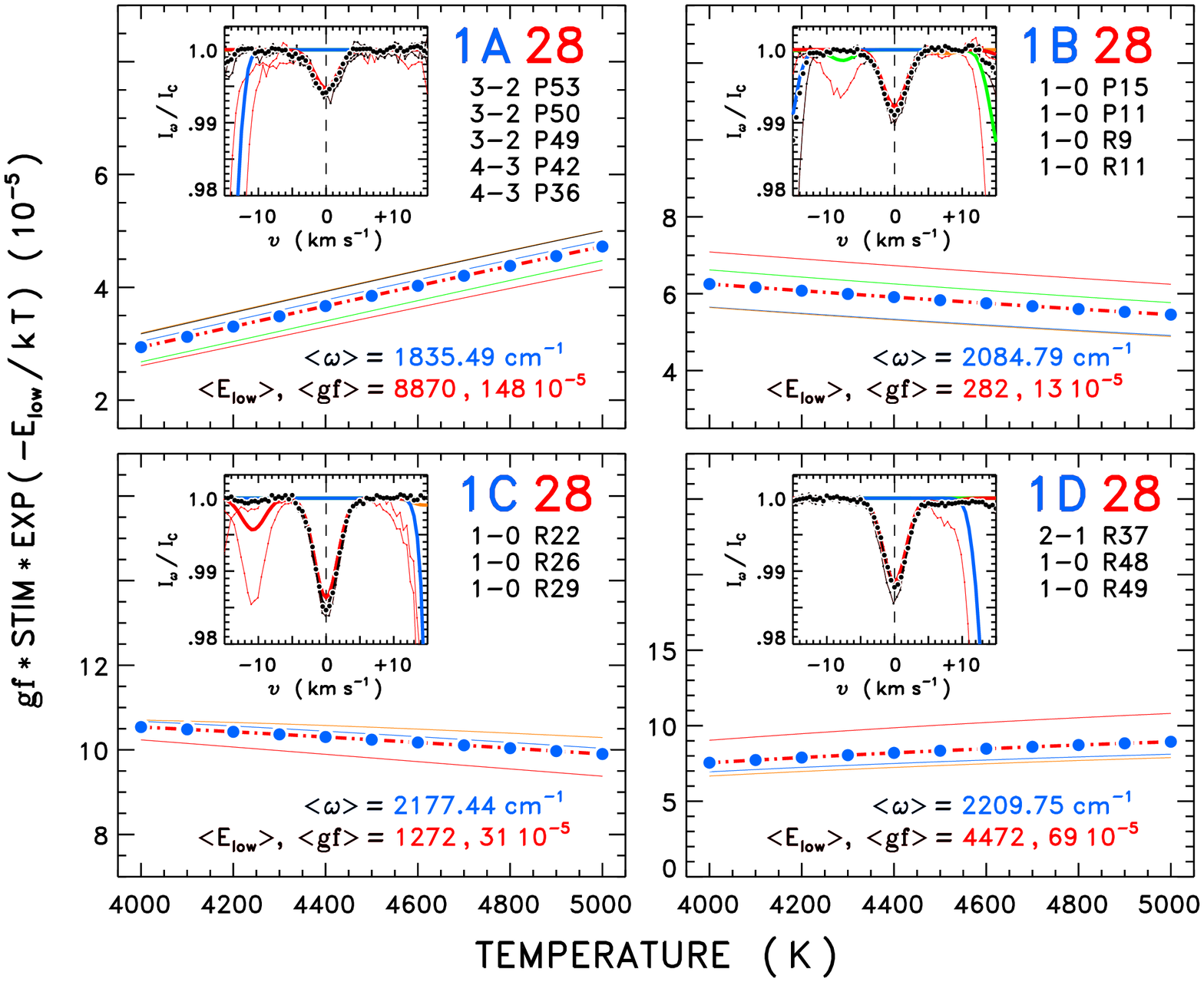} 
\caption[]{}
\end{figure}

\clearpage
\begin{figure}
\figurenum{2e}
\hskip  -10mm
\includegraphics[scale=0.800,angle=90]{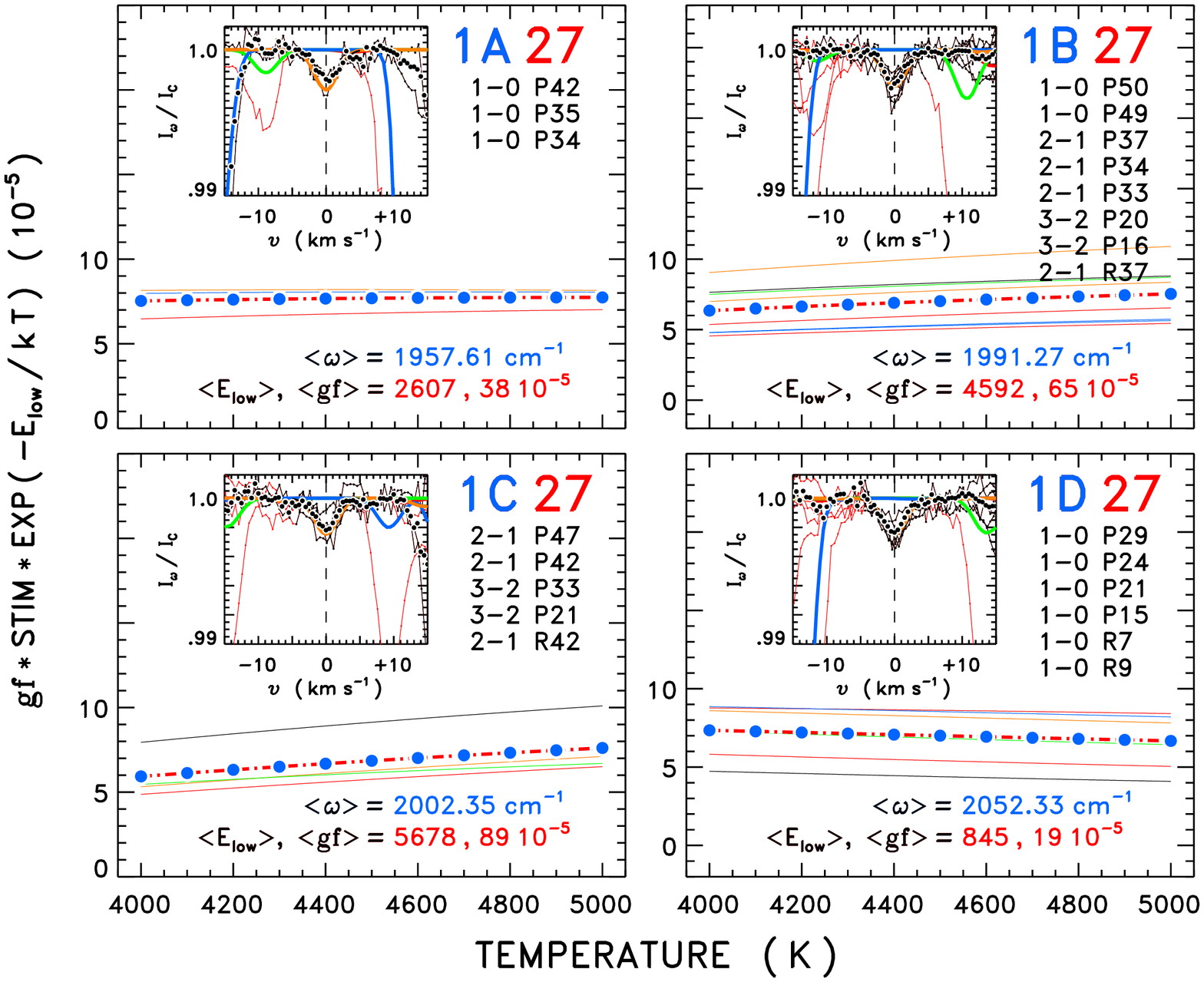} 
\caption[]{}
\end{figure}

\clearpage
\begin{figure}
\figurenum{3}
\caption[]{{\em (a)}\/ Schematic maps of velocity (left) and temperature (right) for sixteen CO5BOLD
snapshots utilized in this study.  Cuts are at constant height (median of $\log{p}= 5$ pressure surface).  In the velocity map, 
lighter areas are upflows and darker areas are downflows, and red ticks indicate horizontal velocity components.  The gray scale saturates at ${\pm}$ 2 km~s$^{-1}$.  In the temperature map, lighter areas are warmer and darker areas are
cooler.  The gray scale covers the range 5000--7000~K.  Three snapshots highlighted in yellow were considered separately, and in more detail, to gauge systematic errors in the isotopic analysis.  {\em (b)}\/ Left panel depicts probability density map of temperature as a function of
pressure from the full 3D model.  Dashed blue curve is mean temperature profile (averaged over constant pressure
surfaces).  Right panel compares mean $T( p )$ (again blue dashed) with FAL-C semi-empirical 1D model
(black dot-dashed).  Thick
red dashed profile was obtained by shifting temperatures of the 3D model uniformly by 90~K
in the outer layers (``MAX'' scenario), so that the mean stratification 
matched the 1D model in the middle photosphere where the CO molecules are most abundant.  Boosting
function is depicted, exaggerated, in upper part of the panel.  Large dots indicate
average pressures of continuum optical depth unity calculated with the
full 3D model for 0.5~$\mu$m (green: visible), 2.3~$\mu$m (orange: CO $\Delta v= 2$), and 4.6~$\mu$m (red: CO $\Delta v= 1$).  Shaded areas are relative CO number densities for the 1D model (black), baseline 3D snapshot (blue), and 
MAX modified 3D model (red) calculated
using the same $\epsilon_{\rm O}=~600$~ppm.  Even though the mean temperature profiles of 3D MAX 
and 1D FAL-C are similar, the former apparently can produce substantially more CO,
thanks to the large 3D fluctuations acting on the strong low-temperature bias of molecular formation.
Also note that the IR continuua arise in deeper layers than where the bulk of CO
resides.}
\end{figure}

\clearpage
\begin{figure}
\figurenum{3a}
\hskip -10mm
\includegraphics[scale=0.675,angle=90]{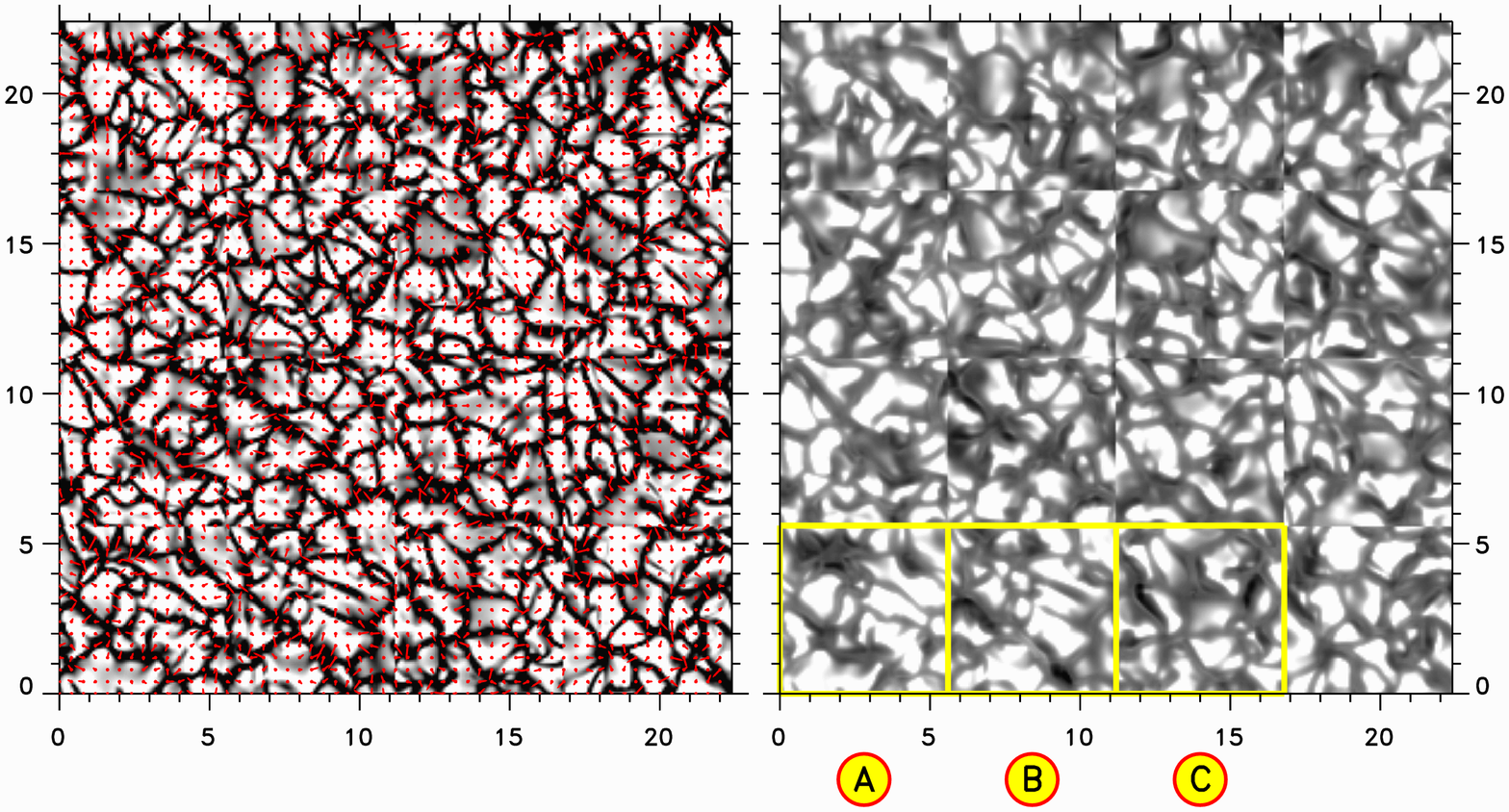} 
\caption[]{}
\end{figure}

\begin{figure}
\figurenum{3b}
\hskip -10mm
\includegraphics[scale=0.800,angle=90]{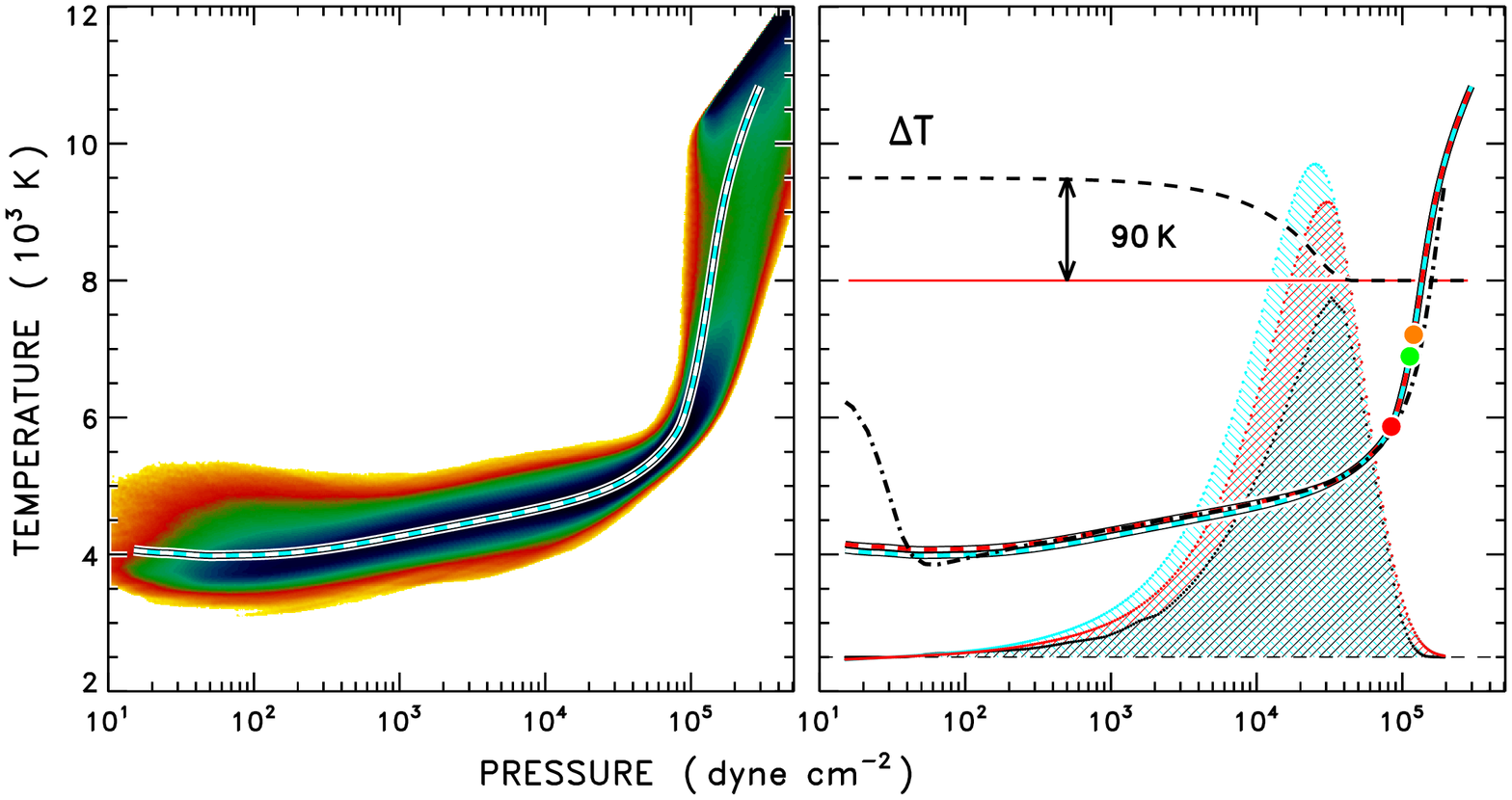} 
\caption[]{}
\end{figure}

\clearpage
\begin{figure}
\figurenum{4}
\hskip   10mm
\includegraphics[scale=0.85,angle=0]{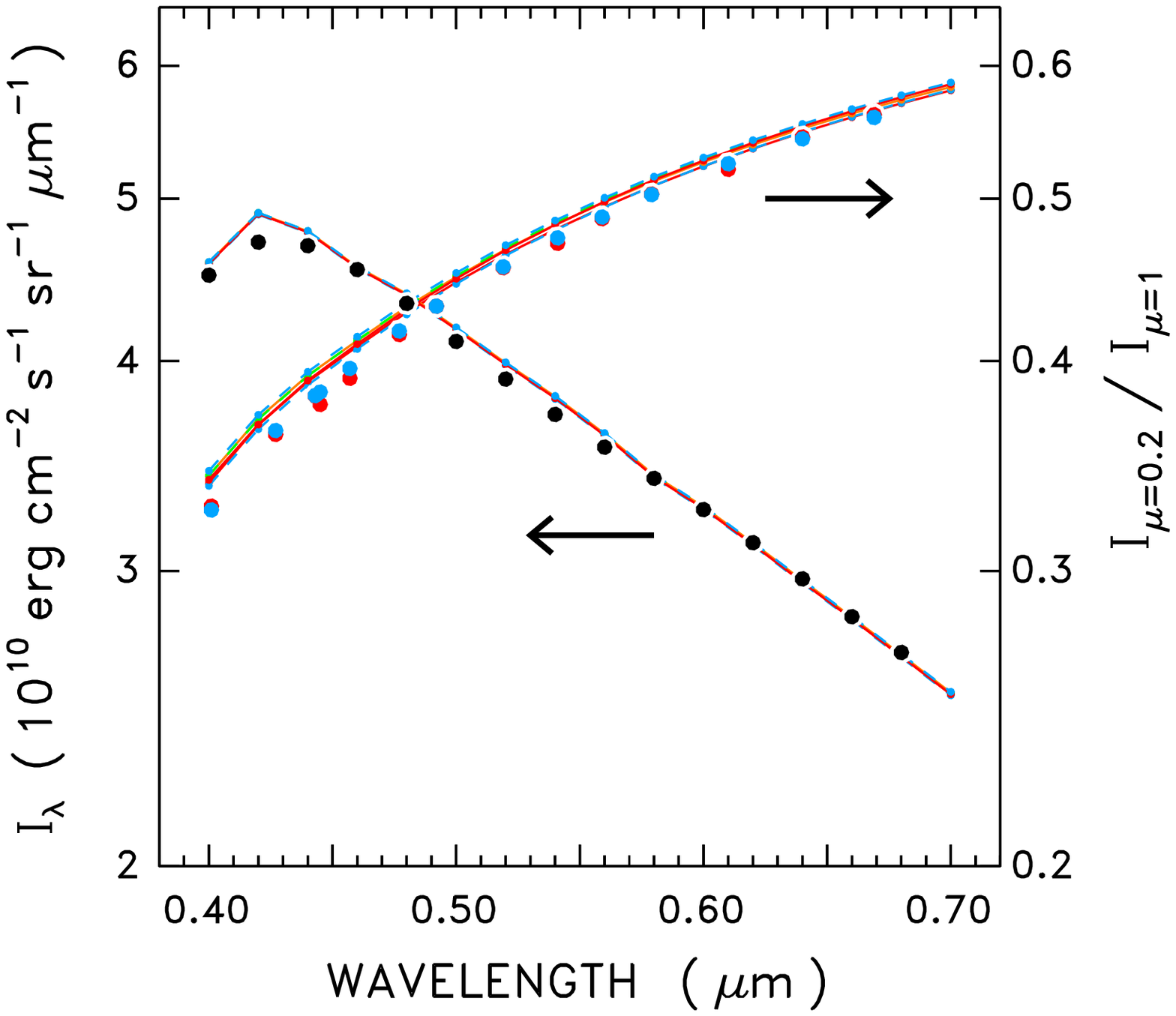} 
\caption[]{Comparison of absolute continuum intensities (left hand scale) and center-to-limb
behavior (right hand scale: $\mu= 0.2$) at visible wavelengths
predicted by the three baseline 3D snapshots, and their MAX counterparts.  This is a fundamental consistency test
for a model that seeks to mimic the solar photosphere.  Large dots refer to measurements (two
independent sets for the limb ratios); curves (and smaller dots) to simulations.  Observational uncertainties are
comparable to the symbol sizes.  Baseline continuum intensities differ slightly from each other (although too small to be seen at this scale), owing to the
stochastic nature of the convective process when viewed over the snapshot-size spatial area, but
average is very close (actually 0.6--0.9\% higher, by chance, for the three reference snapshots) to observed intensities (by design: the full 3D model was adjusted slightly in pressure to
force agreement with these absolute measurements).  The MAX perturbed cases also
fall on top of the baseline models (the temperature enhancement is too shallow to
affect the deep-seated visible continuum).  There is more of a separation in the center-to-limb curves, with
the baseline models predicting the lower curves (better matches to observations), while MAX
models predict slightly higher limb intensities.}
\end{figure}

\clearpage
\begin{figure}
\figurenum{5}
\caption[]{{\em (a)}\/ Montage of $^{12}$C$^{16}$O hybrid profiles (larger dark dots) and synthesized
line shapes (smaller red dots: smooth curves are Gaussian fits) for full 3D baseline model and four values of the oxygen abundance.  Average of G94 and HR96
oscillator strengths was used.  Line
designator and isotopomer (all are 26 here) are listed in upper part of each panel; 
line $<\omega>$ and $E_{\rm low}$ (both in cm$^{-1}$) are listed below each panel.  Note slight convective blueshift
of each line (amounting to about 300 m s$^{-1}$ for $\Delta v=1$, and about 380 m s$^{-1}$ for $\Delta v=2$).  {\em (b)}\/ Calculated equivalent widths of hybrid lines
of {\em (a)}\/ for the four discrete oxygen abundances.  Curves are parabolic fits to the points; 
horizontal lines indicate observed $W_{\omega}$ (error bars too small to be seen). Vertical ticks mark $\epsilon_{\rm O}$'s that correspond to observed $W_{\omega}$'s. {\em (c)}\/ Oxygen abundances
of {\em (b)}\/ as functions of $E_{\rm low}$ (upper panel) and $W_{\omega}$ (lower panel), separated by $\Delta v$: the
more numerous overtone transitions are blue symbols; fundamental lines are
red.  Full 3D baseline model occupies lower part
of each panel, while upper points are for 1D FAL-C.   Green dashed lines are least-squares fits to the overtone sample;  orange dashed
lines are a forced fit to the sparser fundamental sample, with same slope.  For the 3D model, 
$\epsilon_{\rm O}= 572\,+\,7\,(E_{\rm low} / {10^4})$ from first-overtone lines, and $\rho= 0.972{\pm}0.013$ is the
offset of fundamental relative to
first overtone.  Note that the $\epsilon_{\rm O} / E_{\rm low}$ slopes are very different in 1D and 3D,
and the 1D model yields much higher derived oxygen abundances.  In the lower panel, lack of conspicuous trends between derived 3D $\epsilon_{\rm O}$ and $W_{\omega}$
suggests that the sample has successfully avoided saturation effects. {\em (d)}\/ Similar to {\em (c)}, but now
on an expanded $\epsilon_{\rm O}$ scale and including MAX (small squares and orange lines) and Goldilocks
($\rho\equiv 1$: large diamonds and black lines) temperature enhanced versions of baseline snapshot, again using the $\langle{f}\rangle$ scale.}
\end{figure}

\clearpage
\begin{figure}
\figurenum{5a}
\vskip  -5mm
\hskip  -8mm
\includegraphics[scale=0.875,angle=0]{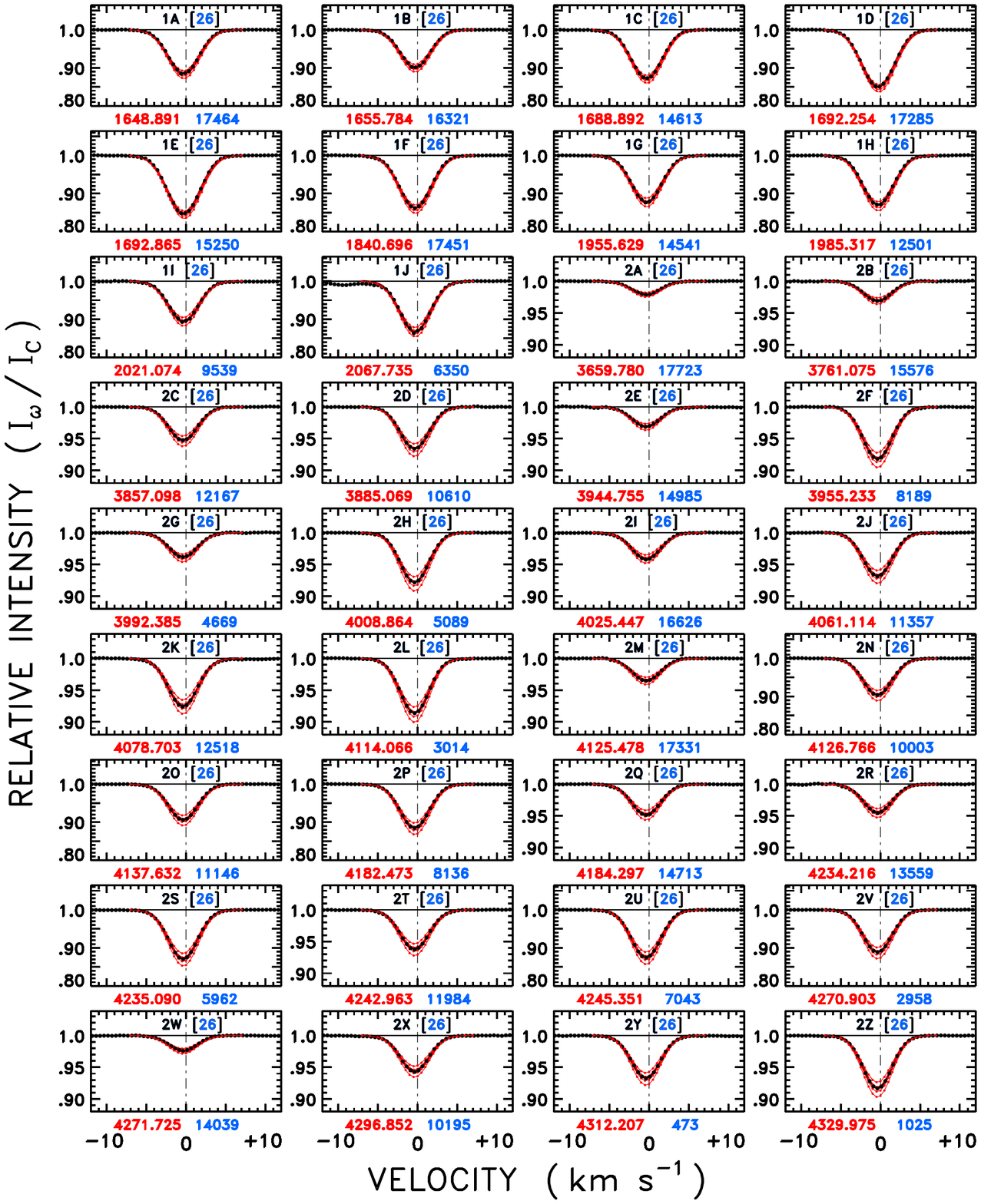} 
\caption[]{}
\end{figure}

\clearpage
\begin{figure}
\figurenum{5b}
\vskip  -5mm
\hskip  0mm
\includegraphics[scale=0.875,angle=0]{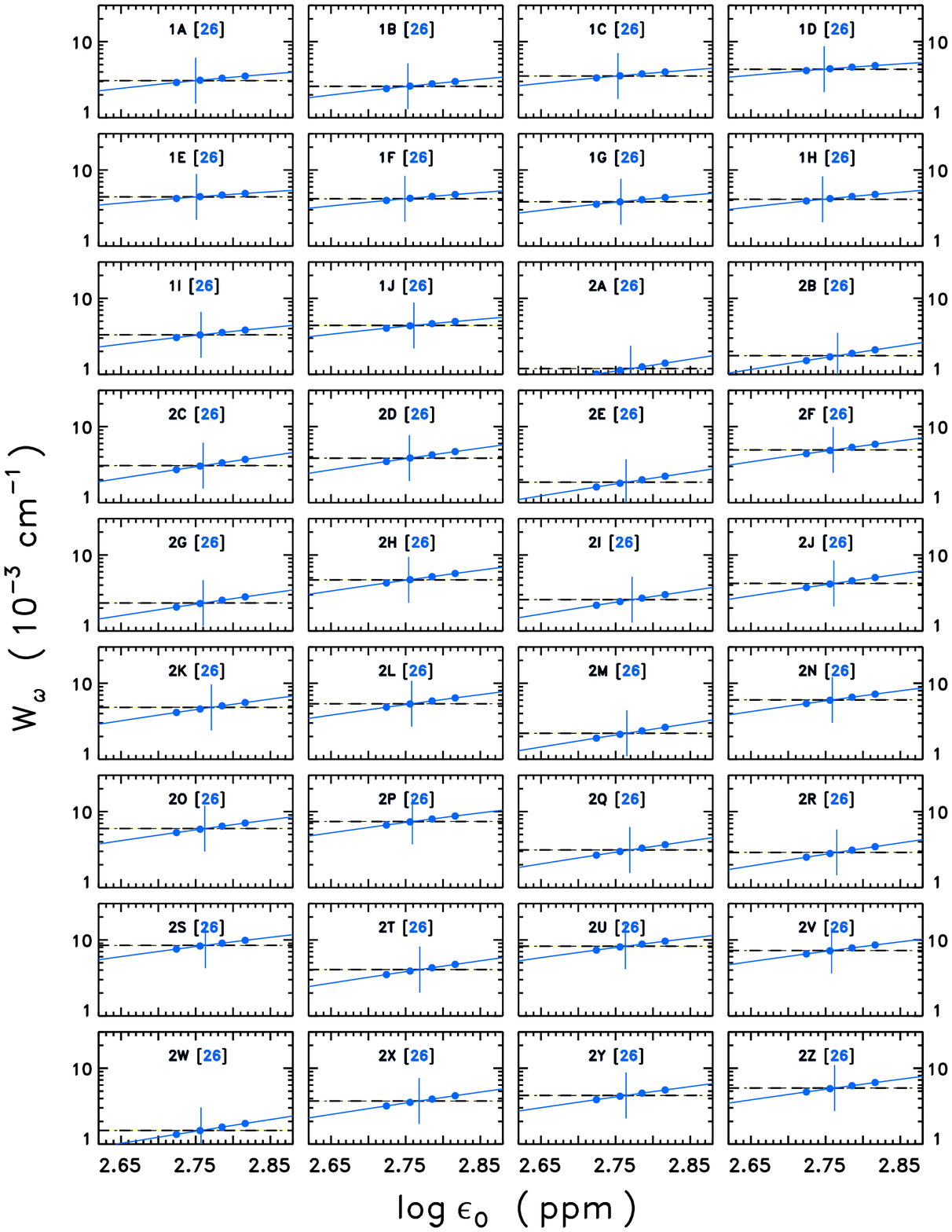} 
\caption[]{}
\end{figure}

\clearpage
\begin{figure}
\figurenum{5c}
\hskip  0mm
\includegraphics[scale=0.95,angle=0]{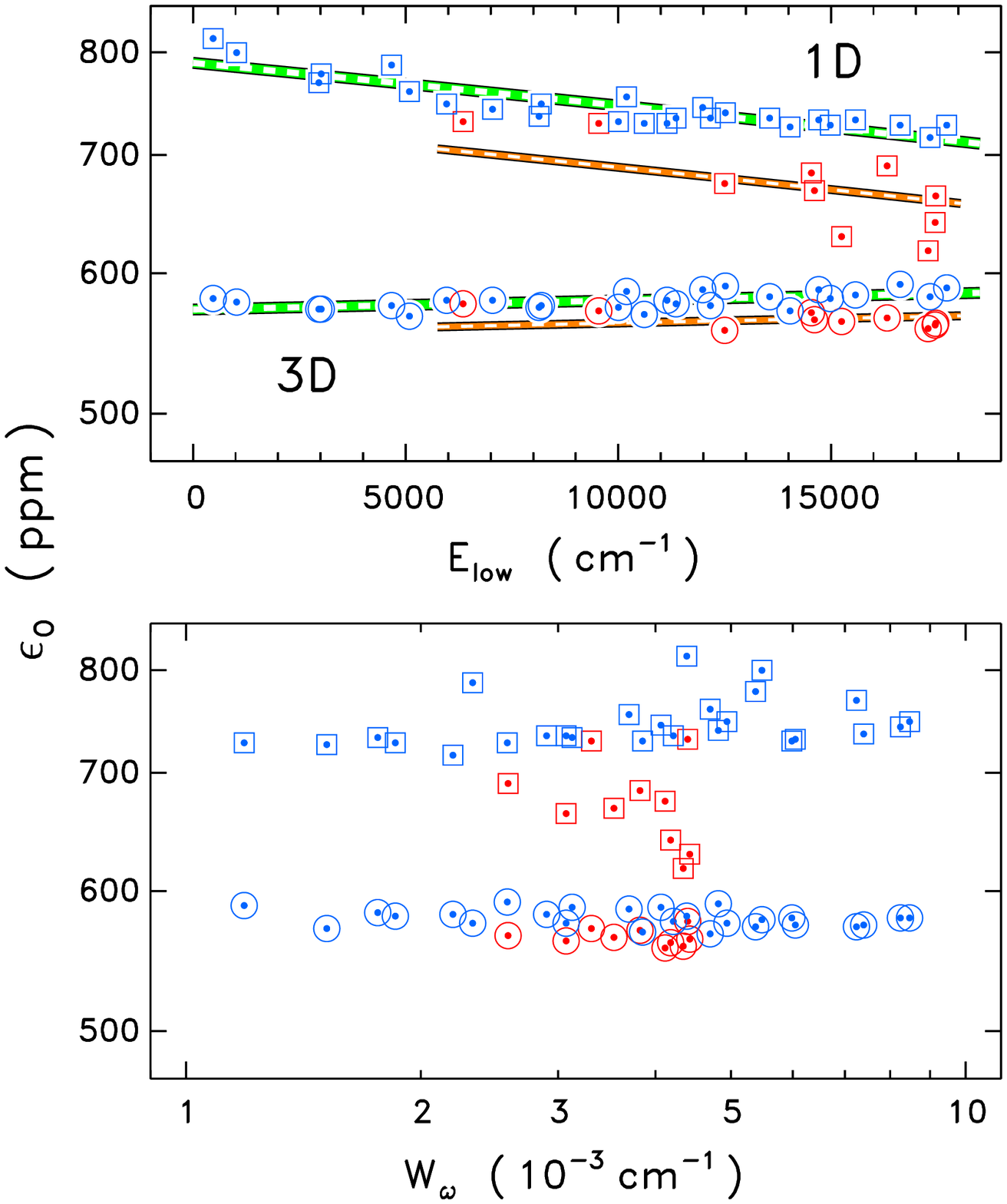} 
\caption[]{}
\end{figure}

\clearpage
\begin{figure}
\figurenum{5d}
\hskip  0mm
\includegraphics[scale=0.95,angle=0]{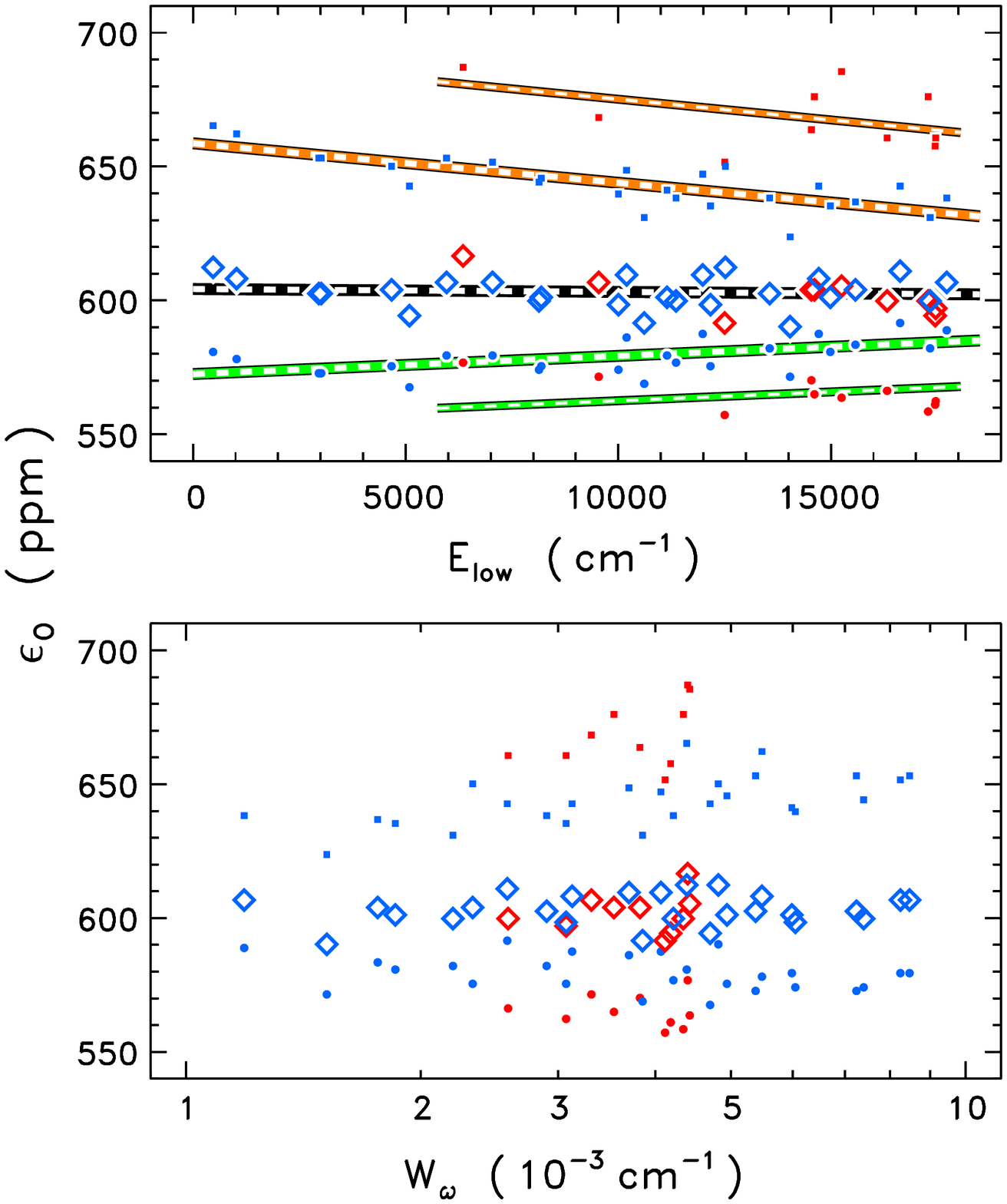} 
\caption[]{}
\end{figure}

\clearpage
\begin{figure}
\figurenum{6}
\caption[]{Summary of 3D abundance experiments utilizing the parent hybrid sample. Derived oxygen abundances
on $y$-axis are compared to $\epsilon_{\rm O}$/$E_{\rm low}$ slopes on $x$-axis.  ``Bow-tie''
symbols each refer to one of the three reference 3D snapshots (white shading for G94 $f$-values, green for HR96); or full 3D model, in yellow for the average $f$-values.
Right edge of the bow-tie is for the baseline model: bottom value is $\Delta v = 1$ (marked by red dot); top
value is $\Delta v = 2$.  Left edge represents MAX-perturbed models: higher $\epsilon_{\rm O}$,
and $\Delta v = 1$ values now above $\Delta v = 2$.  Center of bow-tie is for Goldilocks option,
where $\Delta v = 1$ and 2 coincide.  FAL-C 1D model is represented by an analogous figure (yellow shaded
for the $\langle{f}\rangle$ scale), 
although only a triangle because the MAX option is the 1D baseline model by definition.  Lower dark hatched area indicates currently recommended
oxygen abundance of Grevesse et al.\ (2010: GASS); upper thinner hatched zone is ``seismic'' value preferred
by solar interior modelers (thick dashed line was derived from the favored metallicity $Z$ parsed into an $\epsilon_{\rm O}$ according to Grevesse \& Savaul [1998] relative abundances; lower edge of hatched zone results if 
GASS abundance ratios are used instead).  Intermediate red shading refers to Caffau et al.\ (2008)
determination from atomic oxygen lines.}
\end{figure}

\clearpage
\begin{figure}
\figurenum{6}
\vskip +5mm
\includegraphics[scale=0.75,angle=0]{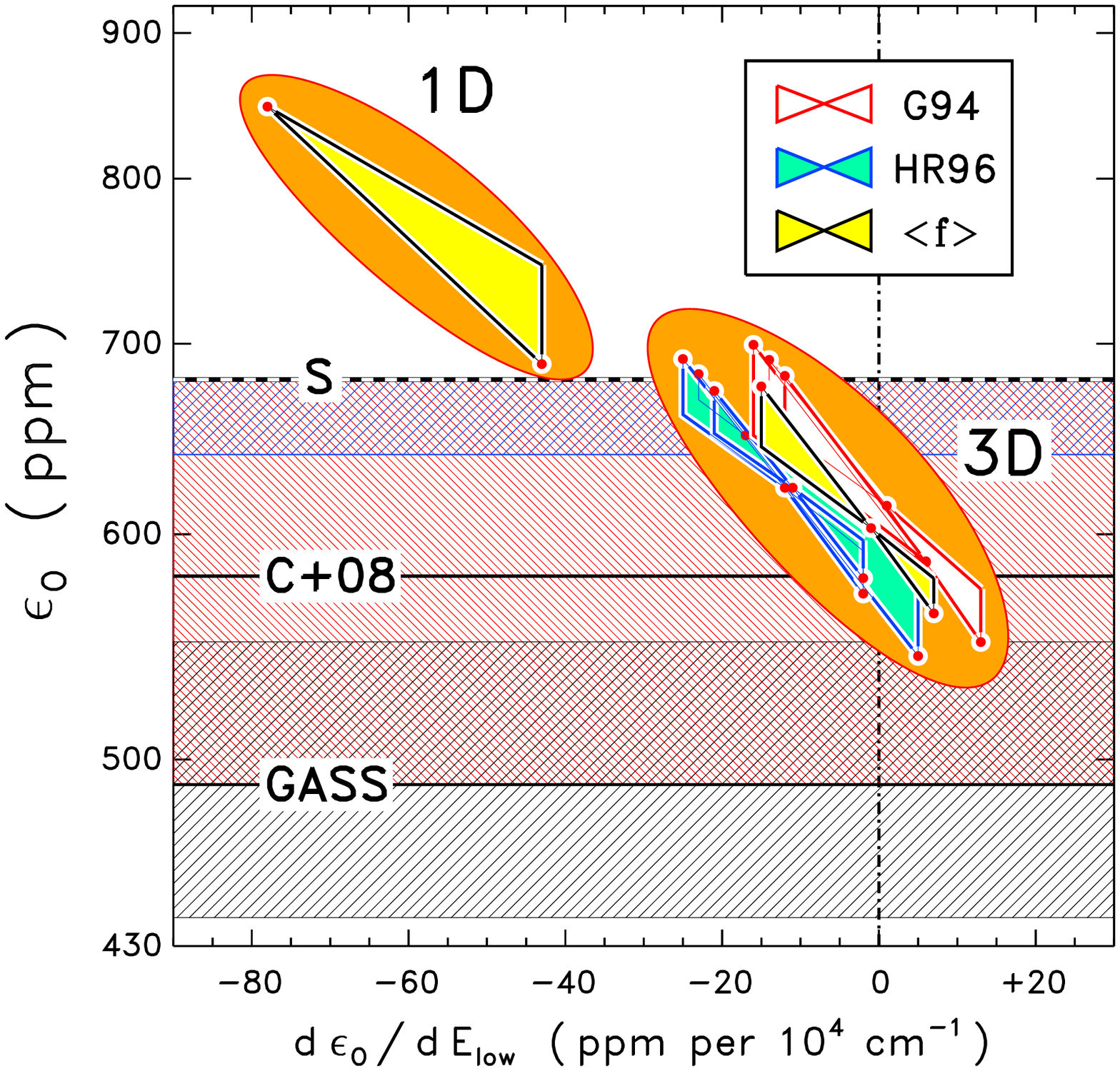} 
\end{figure}

\clearpage
\begin{figure}
\figurenum{7}
\hskip  0mm
\caption[]{{\em (a)}\/ Similar to Fig.~5a, but now montage of isotopomer hybrid profiles compared to
synthesized line shapes for six multiples of standard isotopic ratios, assuming the $\langle{f}\rangle$ scale.  {\em (b)}\/ Similar to Fig.~5b, to derive individual isotopic ratio scale factors, relative to adopted standard $(R_{\rm ISO})_{\rm STD}$ values,
from observed isotopomer $W_{\omega}$.  Widths of the horizontal lines [$(W_{\omega})_{\rm obs}$]
reflect measurement errors (only visible, and then just barely, for isotopomer 27).
{\em (c)}\/ Similar to Fig.~5c, but now showing the derived scale factors
for the three isotopic combinations: 23 (red), 68 (blue), and 67 (green).  The
23 and 68 values are tightly clustered about the means (solid lines), with small s.e.'s (dot-dashed lines);
67 displays larger spread consistent with its much smaller, and therefore less certain, equivalent widths.}
\end{figure}

\clearpage
\begin{figure}
\figurenum{7a}
\vskip  -35mm
\hskip  -8mm
\includegraphics[scale=1.0,angle=0]{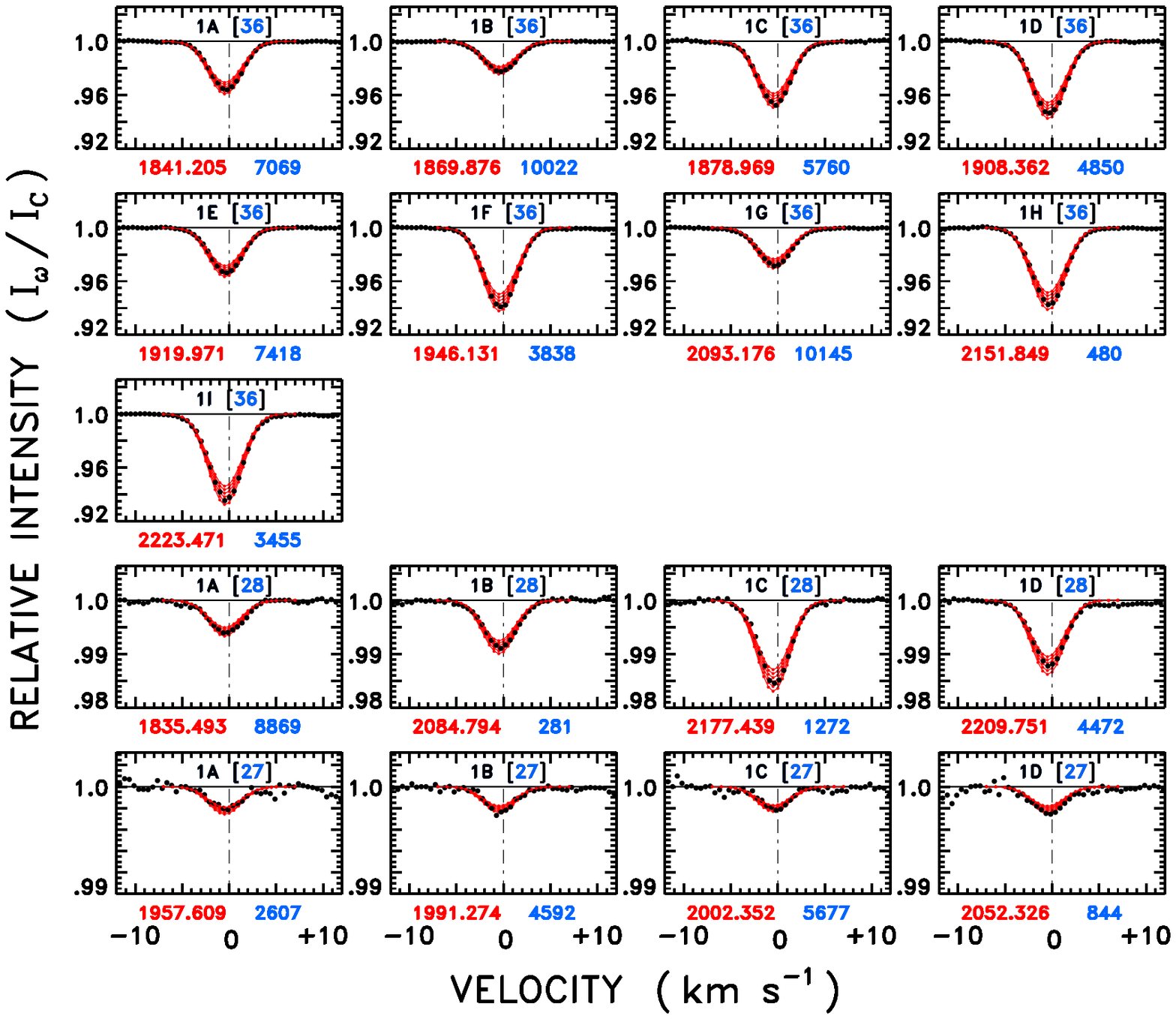} 
\caption[]{}
\end{figure}

\clearpage
\begin{figure}
\figurenum{7b}
\vskip  -35mm
\hskip  -8mm
\includegraphics[scale=1.0,angle=0]{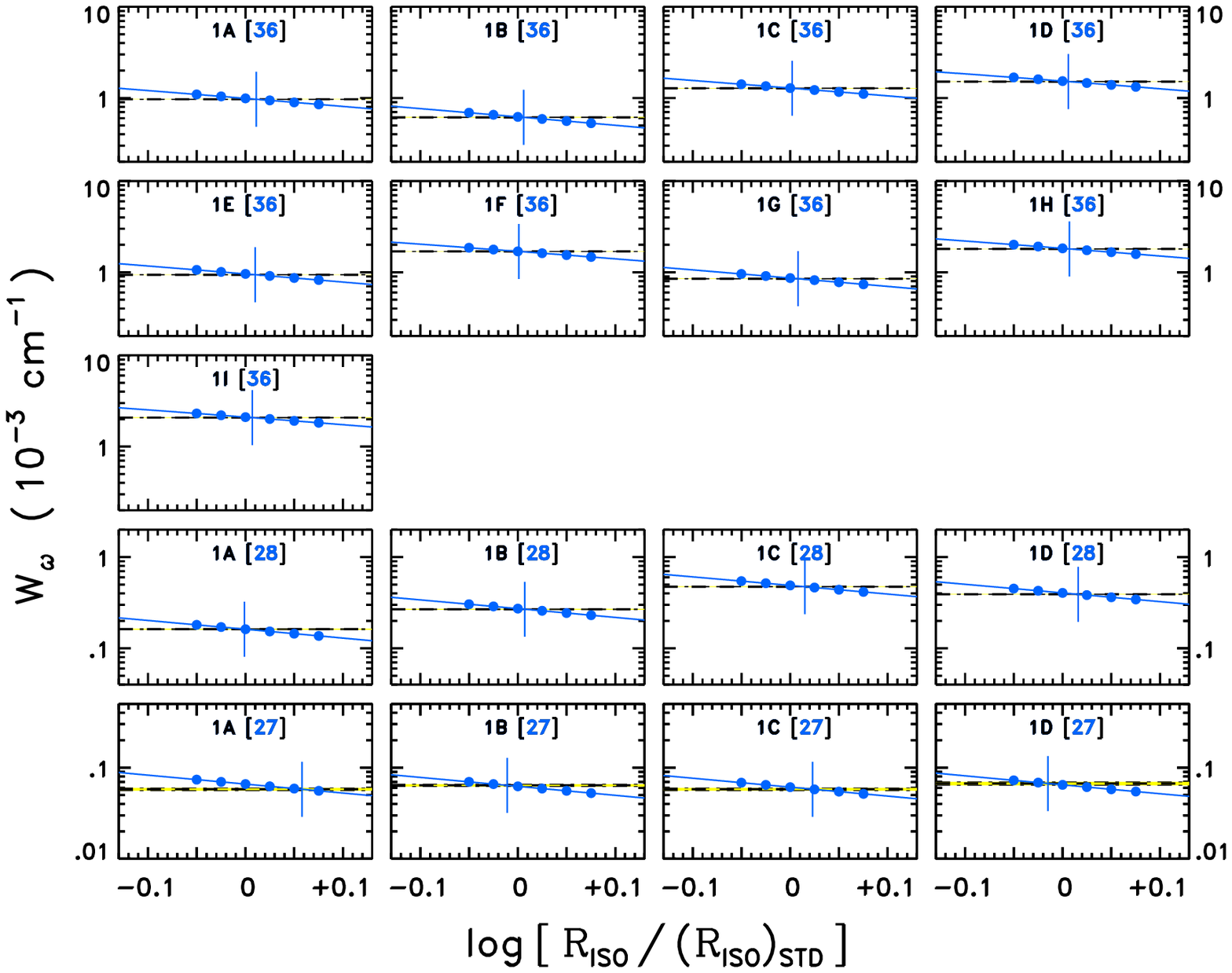} 
\caption[]{}
\end{figure}

\clearpage
\begin{figure}
\figurenum{7c}
\hskip  0mm
\includegraphics[scale=0.95,angle=0]{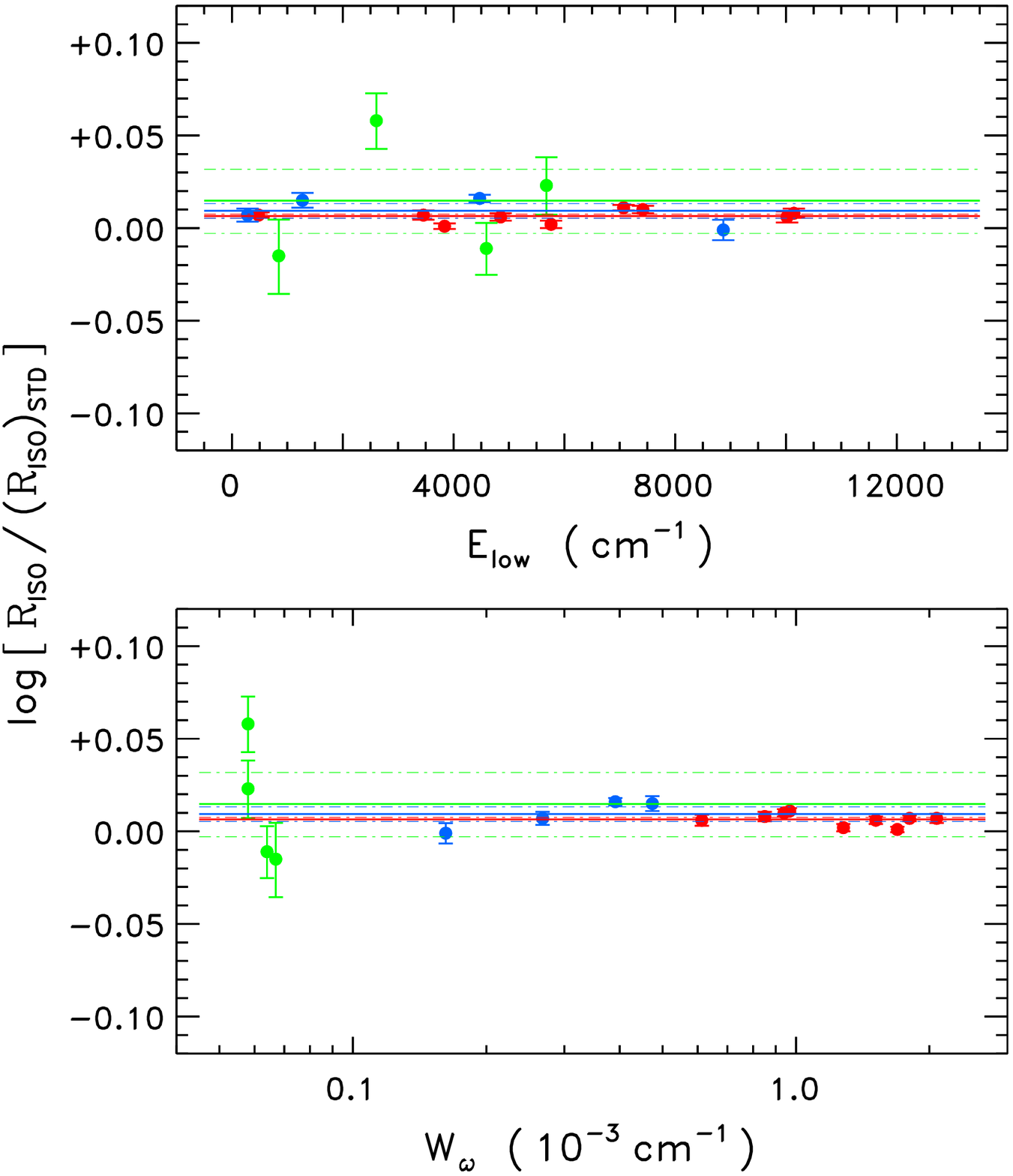} 
\caption[]{}
\end{figure}

\clearpage
\begin{figure}
\figurenum{8}
\vskip  -15mm
\hskip -13mm
\includegraphics[scale=0.785,angle=90]{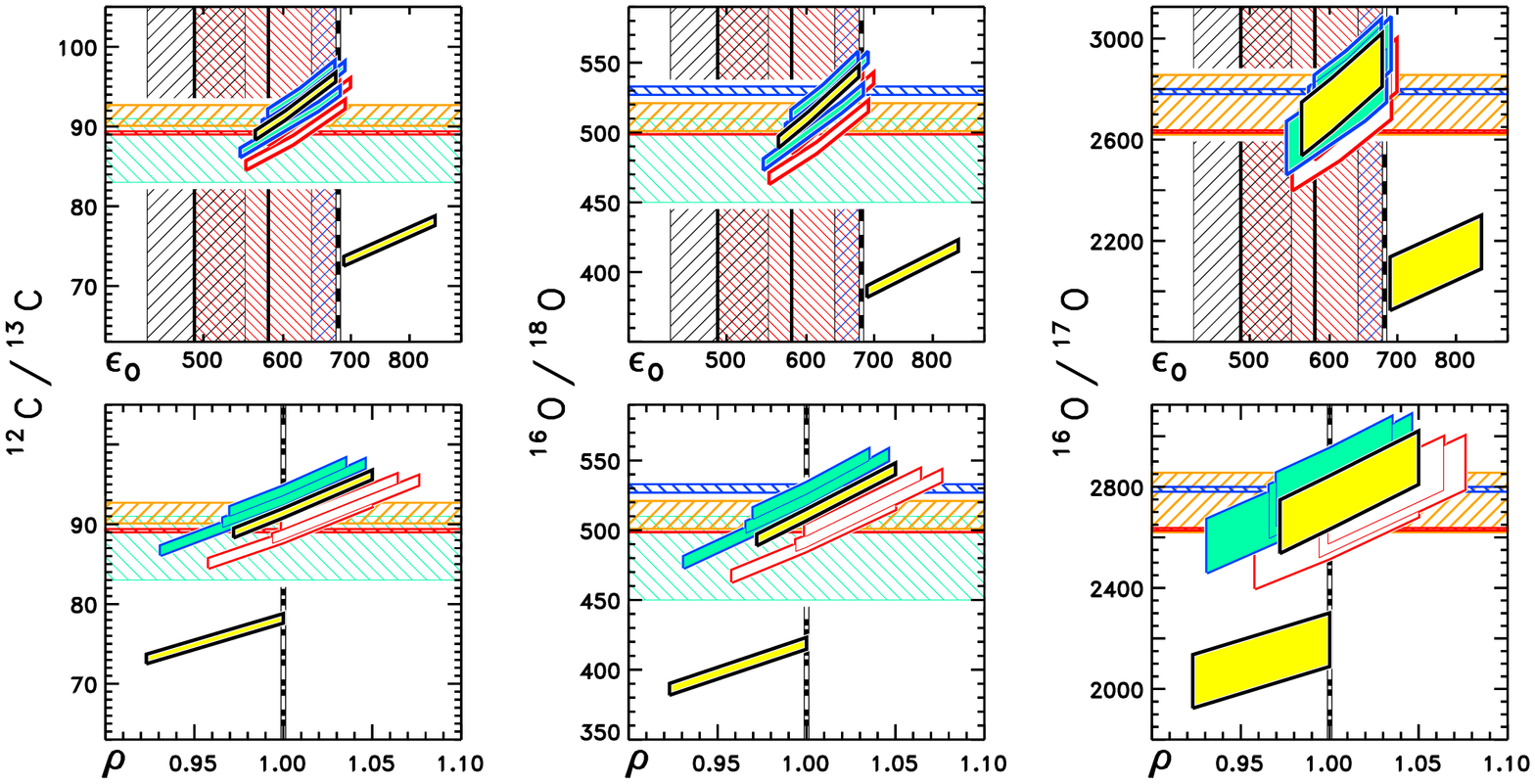} 
\caption[]{Summary of isotopomer experiments.  Derived isotopic ratios are illustrated versus $\epsilon_{\rm O}$ (from
$\Delta v= 1$ sample) in upper panels, and versus the $\rho$ value in lower panels.  Included are the three
reference snapshots, and the three temperature perturbation scenarios for each snapshot, for
both the G94 (white/red) and HR96 (blue/green) $f$-value scales; the full 3D model
and its three scenarios, for the average $f$-values (yellow/black); and the FAL-C 1D model,
also for the $\langle{f}\rangle$ scale.  The scenarios define
quasi-rectangular areas: vertical extent is average 1\,s.e.\ statistical measurement error due to the
hybrid sample.  Baseline models are at lower left edges; MAX variants are at upper right edges; Goldilocks in the middle
(or closer to the left for two of the G94 snapshots).
Thin red horizontal hatched lines refer to terrestrial standard values; upper blue lines are for {\em Genesis}\/
oxygen ($\delta\sim -60$\,$^{\circ}\!\!/\!_{\circ\circ}$); intermediate orange
hatched areas are our preferred Goldilocks results based on the $\langle{f}\rangle$ scale, and with ``optimistic'' error bars; and green shading indicates the earlier 3D results of SAGS
for $^{13}$C and $^{18}$O.  Vertical shaded bands in
the upper panels refer to the same oxygen abundance ranges illustrated in Fig.~6.}
\end{figure}

\end{document}